\pdfoutput=1
\documentclass[cernpreprint, coverpage=false, atlasdraft=false, texlive=2016, UKenglish, texmf]{atlasdoc}
 
\usepackage{atlaspackage}
\usepackage{atlasbiblatex}
 
\usepackage{rotating}
\usepackage{mathtools}
\usepackage{subfig}
 
\usepackage{atlasphysics}
\usepackage{atlasmisc}
\usepackage{atlasjetetmiss}
\usepackage{atlasprocess}
 
\graphicspath{{logos/}{figures/}}
 
\usepackage{ANA-EXOT-2020-11-PAPER-defs}

\AtlasTitle{Search for invisible Higgs-boson decays in events with vector-boson fusion signatures using $139\,\text{fb}^{-1}$ of proton--proton data recorded by the ATLAS experiment}
 
\AtlasAbstract{
A direct search for Higgs bosons produced via vector-boson fusion and subsequently decaying into invisible particles is reported.
The analysis uses $139\,\text{fb}^{-1}$ of $pp$ collision data at a centre-of-mass energy of $\sqrt{s}{\,=\,}13\,\text{TeV}$ recorded by the ATLAS detector at the LHC.
The  observed numbers of events are found to be in agreement with the background expectation from Standard Model processes.
For a scalar Higgs boson with a mass of $125\,\GeV$ and a Standard Model production cross section, an observed upper limit of $0.145$ is placed on the branching fraction of its decay into invisible particles at $95\%$ confidence level, with an expected limit of $0.103$.
These results are interpreted in the context of models where the Higgs boson acts as a portal to dark matter, and limits are set on the scattering cross section of weakly interacting massive particles and nucleons.
Invisible decays of additional scalar bosons with masses from $50\,\text{GeV}$ to $2\,\text{TeV}$ are also studied, and the derived upper limits on the cross section times branching fraction decrease with increasing mass from $1.0\,\text{pb}$ for a scalar boson mass of $50\,\text{GeV}$ to $0.1\,\text{pb}$ at a mass of $2\,\text{TeV}$.
}
 
\author{The ATLAS Collaboration}
 
\AtlasRefCode{EXOT-2020-11}

\AtlasJournal{JHEP}
\AtlasJournalRef{JHEP 08 (2022) 104}
\AtlasDOI{10.1007/JHEP08(2022)104}
 
\PreprintIdNumber{CERN-EP-2021-258}

\hypersetup{pdftitle={ATLAS document},pdfauthor={The ATLAS Collaboration}}
 
\begin{document}
 
\maketitle
 
\tableofcontents
 
\section{Introduction}
% The next lines are included from the .//sections/intro.tex input file
This paper presents a direct search for the decay of the Higgs boson~\cite{HIGG-2012-27,CMS-HIG-12-028} into invisible particles.
While the expectation for the branching fraction of invisible decays ($\mathcal{B}_\text{inv}$) from the Standard Model (SM) is $0.12\%$ due to the decay $H{\,\to\,}ZZ^\ast{\,\to\,}\nu\bar{\nu}\nu\bar{\nu}$, several scenarios of physics beyond the Standard Model (BSM) allow for anomalous and sizeable $\mathcal{O}(10\%)$ values.
In one class of models under consideration~\cite{Shrock:1982kd,Choudhury:1993hv,ArkaniHamed:1998vp,Eboli:2000ze,Godbole:2003it,Antoniadis:2004se,Davoudiasl:2004aj,Patt:2006fw,Kanemura:2010sh,Djouadi:2011aa,Ghosh:2012ep,Belanger:2013kya,Curtin:2013fra} the Higgs boson might decay into $\chi\chi$, a pair of weakly interacting massive particles (WIMPs)~\cite{Goldberg:1983nd,Ellis:1983ew}, which may explain the existence of dark matter (DM)~\cite{Zwicky:1937zza,Rubin:1970zza,Clowe:2006eq}.
Such models, in which the DM candidate is a singlet under the gauge symmetries of the SM and the Higgs boson acts as a mediator to the DM, are called Higgs-portal models.
They represent a simple extension of the SM to provide a DM candidate and are able to predict the observed relic DM density via $s$-channel $\chi\chi{\,\to\,}f\bar{f}$ annihilation processes~\cite{LopezHonorez:2012kv}, where $f$ denotes a fermion.
The phenomenology of such models is very rich, and depends strongly on the spin of the DM candidate.
 
A long list of searches for invisible Higgs boson decays that target various final states have been performed by the ATLAS and CMS experiments~\cite{CMS-HIG-13-030,HIGG-2014-07,CMS-EXO-12-055,HIGG-2013-16,EXOT-2016-23,EXOT-2016-37,CMS:2020ulv,CMS:2021far,cmscollaboration2022search,ATLAS:2021gcn} at the LHC~\cite{Evans:2008zzb}. 
The absence of an excess of such events is interpreted as an upper limit on $\mathcal{B}_\text{inv}$, assuming the SM production cross sections~\cite{deFlorian:2227475} of the Higgs boson.
Statistical combinations of search results have been performed, for which ATLAS reported an observed (expected) limit of $0.26$ ($0.17$)~\cite{HIGG-2018-54} and CMS reported $0.19$ ($0.15$)~\cite{CMS-HIG-17-023} at $95\%$ confidence level (CL), both using Run-1 and up to $36\,\ifb$ of Run-2 data.
The results with the highest expected sensitivity are from the vector-boson fusion (VBF) channel, for which ATLAS reported an upper limit of $0.37$ ($0.28$) using $36\,\ifb$ of Run-2 data~\cite{EXOT-2016-37}, while CMS reported $0.18$ ($0.10$) using a combination of the full Run-1 and Run-2 data~\cite{cmscollaboration2022search}.
 
Global fits to the measurements of visible decay channels of the Higgs boson place indirect constraints on the decay branching fraction $\mathcal{B}_{\text{BSM}}$ into final states which are either invisible ($\mathcal{B}_\text{inv}$) or not included in the combination of visible decays ($\mathcal{B}_\text{undet}$), such as a BSM Higgs boson that decays into light quarks.
A combination of ATLAS and CMS results, using only the cross-section measurements of the visible decay channels from Run-1 data and under the assumption that the couplings between the Higgs boson and the vector bosons are not larger than in the SM, results in an observed (expected) upper limit on $\mathcal{B}_\text{BSM}$ of $0.34$ ($0.39$)~\cite{HIGG-2015-07}.
Direct searches for invisible Higgs boson decays are complementary to indirect constraints from measurements of visible decays.
A simultaneous excess in the two approaches would confirm a signal, while a non-zero branching fraction $\mathcal{B}_\text{BSM}$ in the global fit, but no excess in the direct searches, would point towards other undetected decays.
In combinations of direct and indirect constraints, CMS reported a limit on $\mathcal{B}_\text{inv}$ of $0.22$ from $36\,\ifb$ of Run-2 data~\cite{CMS-HIG-17-031}, and ATLAS a limit of $0.30$ using up to $80\,\ifb$ of Run-2 data~\cite{HIGG-2018-57}.
 
The data sample used for the analysis reported in this paper corresponds to an integrated luminosity of $139\,\ifb$ of proton--proton ($pp$) collisions at $\sqrt{s}{\,=\,}13\,\TeV$ recorded by the ATLAS detector.
This search targets the VBF production process~\cite{Dittmaier:1318996,Dittmaier:1416519} with its distinct signature of a pair of energetic quark-induced jets with a wide gap in pseudorapidity (\detajj) resulting in a large invariant mass (\mjj).
Feynman diagrams for the signal and dominant background processes are shown in Figure~\ref{fig:Feynman}.
The characteristic VBF topology provides a powerful way to reject backgrounds from single vector-boson production in association with two jets from QCD radiation, which is the main background after the large backgrounds from multijet processes are suppressed by requiring a large amount of missing transverse momentum (\MET).
In this analysis, Higgs boson production via the gluon--gluon fusion (ggF) mechanism and in association with a vector boson ($VH$, where $V$ stands for $W$ or $Z$) are also considered as signal, but their contributions are small compared to the VBF process after the selection.
 
\begin{figure}[htbp]
\centering
\subfloat[Signal process]{
\parbox{0.45\textwidth}{
\centering
\includegraphics[width=.3\textwidth]{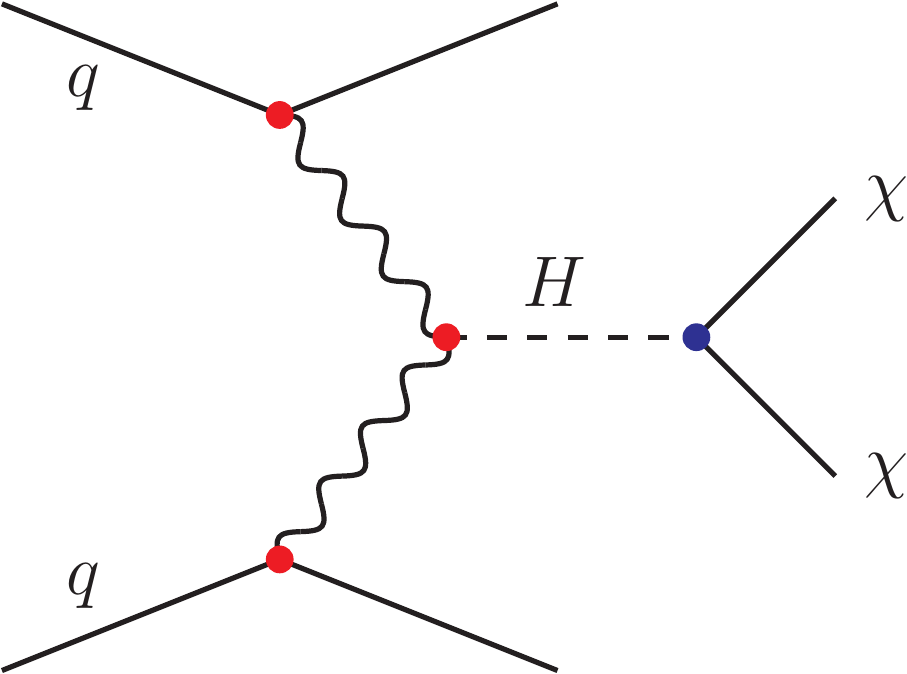}
}
\label{fig:Feynman_a}
}
\hspace{0.05\textwidth}
\subfloat[Strong $Z\text{+\,jets}$ background]{
\parbox{0.45\textwidth}{
\centering
\includegraphics[width=.3\textwidth]{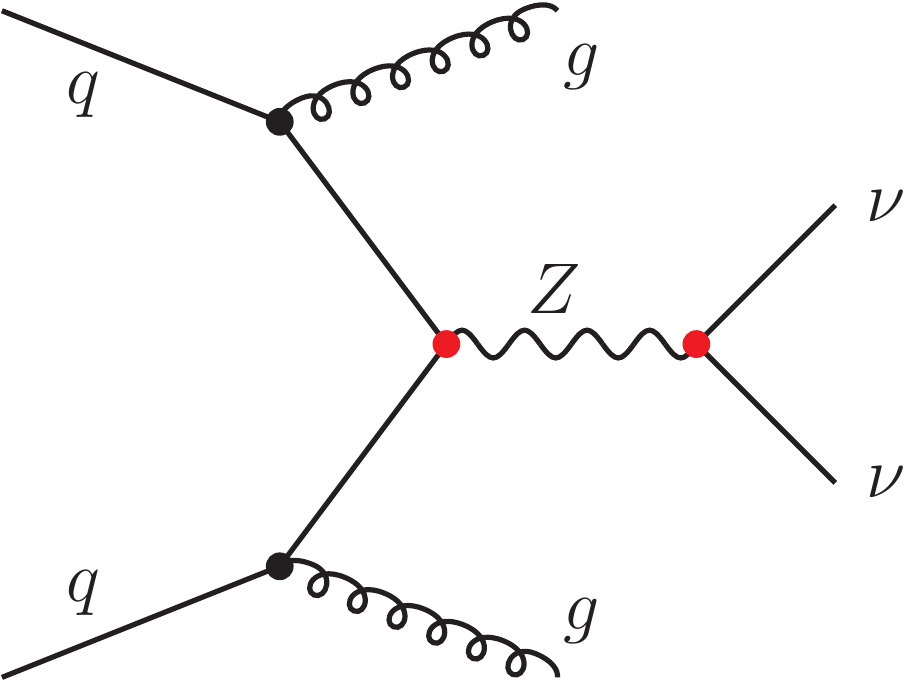}
}
\label{fig:Feynman_b}
}
 
\subfloat[Electroweak $Z\text{+\,jets}$ background]{
\parbox{0.45\textwidth}{
\centering
\includegraphics[width=.3\textwidth]{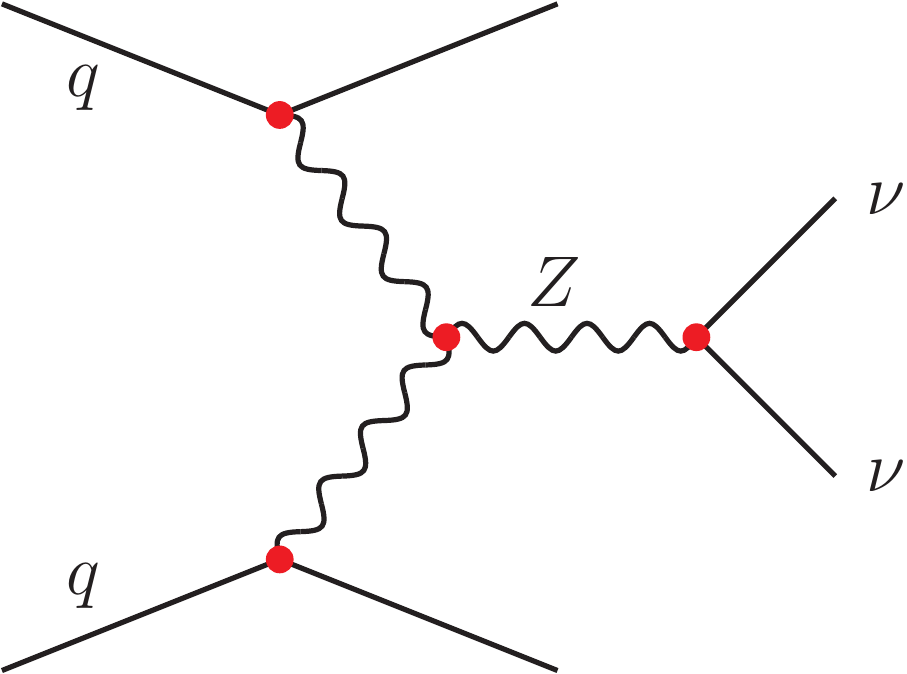}
}
\label{fig:Feynman_c}
}
\hspace{0.05\textwidth}
\subfloat[Diboson ($ZV$) background]{
\parbox{0.45\textwidth}{
\centering
\includegraphics[width=.3\textwidth]{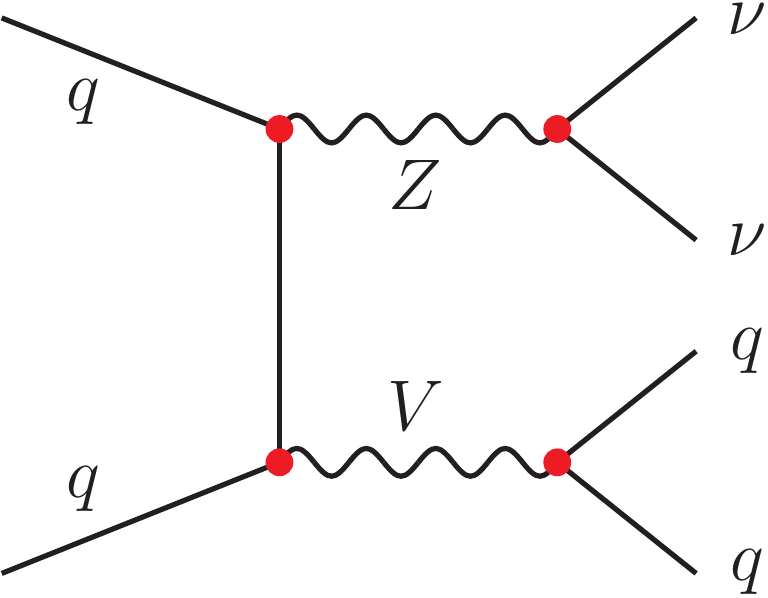}
}
\label{fig:Feynman_d}
}
\caption{Representative Feynman diagrams for \protect\subref{fig:Feynman_a} the signal and for the three dominant background processes:
\protect\subref{fig:Feynman_b} strong $Z\text{+\,jets}$, \protect\subref{fig:Feynman_c} electroweak $Z\text{+\,jets}$, and \protect\subref{fig:Feynman_d} diboson production.
The latter two processes are collectively referred to as `electroweak' $Z$ production.
Electroweak vertices are shown as red markers.
For simplicity, no distinction is made between particles and antiparticles, or between different quark flavours.
}
\label{fig:Feynman}
\end{figure}
 
In this analysis, several changes and improvements are made relative to the previous ATLAS paper on this topic~\cite{EXOT-2016-37}.
The selection criteria for \mjj, \detajj and the azimuthal angle between the two highest-$\pT$ jets (\dphijj) are relaxed, while the \MET requirement is increased slightly.
While the previous analysis rejected events with a third jet, events with three or four jets are accepted in this analysis if these jets are compatible with the hypothesis that they originate from final-state radiation.
To profit from the larger data set, a finer binning in \mjj and for the first time a binning in \dphijj ($\dphijj{\,<\,}1$ and $1{\,\le\,}\dphijj{\,<\,}2$) and jet multiplicity are used.
This results in improved performance since it better captures the characteristic shape of the signal, which yields higher signal-to-background ratios in kinematic regions with larger values of \mjj and at smaller values of \dphijj.
 
The analysis extracts the signal yield using a binned likelihood fit to 16 search bins 
and corresponding $Z_{\ell\ell}$ and $W_{\ell\nu}$ ($\ell=e,\mu$) control region bins to estimate the main backgrounds from the \Znnjets and \Wlnjets processes.
A recent theoretical calculation~\cite{Lindert:2021xxx}, performed in the phase space relevant for this analysis, is used to constrain the ratio of \Zjets to \Wjets events.
This allows a reduction of the statistical uncertainties in the estimate of the background from \Znnjets events, by also using the $W_{\ell\nu}$ control region.
For the estimation of the multijet background, two independent data-driven techniques are used, and this reduces the overall uncertainties relative to the previously published result.
The signal yield extracted from the likelihood fit is used to determine an upper limit on $\mathcal{B}_\text{inv}$. 
The background predictions are validated in a separate kinematic region with two jets and $2{\,<\,}\dphijj{\,<\,}2.5$. 
% End of text imported from the .//sections/intro.tex input file
 
\section{ATLAS detector}
 
% The next lines are included from the .//sections/experiment.tex input file
The ATLAS detector~\cite{PERF-2007-01} at the LHC covers nearly the entire solid angle around the collision point.\footnote{
ATLAS uses a right-handed coordinate system with its origin at the nominal interaction point (IP) in the centre of the detector and the $z$-axis along the beam pipe.
The $x$-axis points from the IP to the centre of the LHC ring, and the $y$-axis points upward.
Cylindrical coordinates $(r,\phi)$ are used in the transverse plane, $\phi$ being the azimuthal angle around the $z$-axis.
The transverse momentum of an object is defined as $\pT{\,=\,}\sqrt{p_x^2+p_y^2}$.
The pseudorapidity is defined in terms of the polar angle $\theta$ as $\eta = -\ln\tan(\theta/2)$ and the rapidity is given by $y = (1/2)\ln[(E+p_z)/(E-p_z)]$.
Angular separation between two objects is quantified by the distance metric $\Delta R=\sqrt{(\Delta y)^2+(\Delta\phi)^2}$.
}
It consists of an inner tracking detector (ID) surrounded by a thin superconducting solenoid, electromagnetic (EM) and hadronic calorimeters, and a muon spectrometer incorporating three large superconducting toroidal magnets.
 
The ID system is immersed in a \SI{2}{\tesla} axial magnetic field and provides charged-particle tracking in the range \(|\eta|{\,<\,}2.5\).
The high-granularity silicon pixel detector covers the vertex region and typically provides four measurements per track, the first hit normally being in the insertable B-layer installed before the start of Run~2~\cite{ATLAS-TDR-2010-19,PIX-2018-001}.
It is followed by the silicon microstrip tracker which usually provides eight measurements per track.
These silicon detectors are complemented by the transition radiation tracker, which enables radially extended track reconstruction up to \(|\eta|{\,=\,}2.0\).
This detector also provides electron identification information based on the fraction of hits (typically 30 in total) above a higher energy-deposit threshold corresponding to transition radiation.
 
The calorimeter system covers the pseudorapidity range \(|\eta|{\,<\,}4.9\).
Within the region \(|\eta|{\,<\,} 3.2\), EM calorimetry is provided by barrel and endcap high-granularity lead/liquid-argon (LAr) calorimeters,
with an additional thin LAr presampler covering \(|\eta|{\,<\,}1.8\) to correct for energy loss in material upstream of the calorimeters.
Hadronic calorimetry is provided by the steel/scintillator-tile calorimeter, segmented into three barrel structures within \(|\eta|{\,<\,}1.7\), and two copper/LAr hadronic endcap calorimeters.
The solid angle coverage is completed with forward copper/LAr and tungsten/LAr calorimeter modules optimised for EM and hadronic measurements respectively.
 
The muon spectrometer (MS) comprises separate trigger and high-precision tracking chambers measuring the deflection of muons in a magnetic field generated by the superconducting air-core toroids.
The field integral of the toroids ranges between \num{2.0} and \SI{6.0}{\tesla\metre} across most of the detector.
A set of precision chambers covers the region \(|\eta|{\,<\,}2.7\) with three layers of monitored drift tubes, complemented by cathode-strip chambers in the forward region, where the background is highest.
The muon trigger system covers the range \(|\eta|{\,<\,}2.4\) with resistive-plate chambers in the barrel, and thin-gap chambers in the endcap regions.
 
Interesting events are selected to be recorded by the first-level (L1) trigger system implemented in custom hardware, followed by selections made by algorithms implemented in software in the high-level trigger (HLT)~\cite{TRIG-2016-01}.
The L1 trigger accepts events from the \SI{40}{\MHz} bunch crossings at a rate below \SI{100}{\kHz}, which the HLT reduces in order to record events to disk at about \SI{1}{\kHz}.
 
An extensive software suite~\cite{ATL-SOFT-PUB-2021-001} is used in the reconstruction and analysis of real and simulated data, in detector operations, and in the trigger and data acquisition systems of the experiment.
% End of text imported from the .//sections/experiment.tex input file
 
\section{Data and simulation}
% The next lines are included from the .//sections/samples.tex input file
\subsection{Data and triggers}
 
The analysis is performed using $pp$ collisions at $\sqrt{s}{\,=\,}13\,\TeV$ collected by ATLAS in 2015–2018 during stable beam conditions and with all subsystems operational~\cite{DAPR-2018-01}.
A total integrated luminosity of $139\,\ifb{}$ is used, with an uncertainty of 1.7\%~\cite{ATLAS-CONF-2019-021}, obtained using the LUCID-2 detector~\cite{LUCID2} for the primary luminosity measurements.
 
The data used for this analysis were recorded using several event triggers, as detailed in the following.
Signal event candidates were collected with unprescaled \met{} triggers~\cite{TRIG-2019-01} having the lowest threshold. 
In 2015, a threshold of 70\,\GeV was used, which was subsequently raised multiple times to cope with increasing effects from multiple interactions in the same and neighbouring bunch crossings (pile-up) and it reached $110\,\GeV$ during the 2017--2018 data-taking period.
Control samples for $W$ and $Z$ backgrounds were recorded with the unprescaled single-lepton and dilepton triggers having the lowest \pT{} thresholds~\cite{TRIG-2018-05,TRIG-2018-01}.
The thresholds for single-lepton triggers ranged from 20 to 26\,\GeV{} for the tightest lepton identification, while for dileptons the thresholds were typically lower, even though the lepton identification was looser.
The \MET{} trigger was also used for the muon control regions, and raised the efficiency for muons by 12\%--28\% depending on the detector region.
The \MET in the trigger is based only on calorimetric measurements and does not include any reconstructed muons, so the muons behave similarly to invisible particles in this trigger.
One of the techniques to estimate the multijet background uses inclusive jet-triggered data recorded with a set of single-jet triggers with jet \pT{} thresholds ranging from $15\,\GeV{}$ to ${\sim}400\,\GeV{}$~\cite{TRIG-2012-01}.
The single-jet triggers with low \pt{} thresholds have significant prescale factors that reduce the data collection rate.
To compensate for these factors, which can be as large as $\mathcal{O}(10^6$), each event is weighted appropriately.
After this treatment, a smooth jet \pt{} spectrum is obtained.
 
\subsection{Event simulation}
 
\begin{table}[]
\caption{Summary of generators used for simulation. The details and the corresponding references are provided in the body of the text.}
\label{tab:samples}
\centering
\resizebox{\textwidth}{!}{
\def\arraystretch{1.5}
\newcolumntype{L}[1]{>{\raggedright\let\newline\\\arraybackslash}p{#1}}
\begin{tabular}{L{0.175\textwidth}L{0.25\textwidth}L{0.18\textwidth}L{0.175\textwidth}L{0.15\textwidth}L{0.1\textwidth}}
\toprule
Process & Generator & ME order & PDF & Parton shower & Tune \\
\midrule
Strong \Vjets & \Sherpa[2.2.1] and \Sherpa[2.2.7] (\mjj-filtered) & NLO (up to 2 jets), \newline LO (up to 4 jets) & \NNPDF[3.0nnlo] & \Sherpa \MEPSatNLO & \Sherpa \\
Electroweak \Vjets & \Herwig[7.2.1]  & NLO (for 2 jets) & \MMHT[nlo] & \Herwig angular-order and \Pythia[8] dipole recoil & \Herwig[7.2] \\
\Vjets $\alphaewk^3$ interference & \MGNLO & LO & \PDFforLHC[15] & \Pythia[8] & \\
Strong \VVjets (including $gg \to \VVjets$) & \Sherpa[2.2.1] or \Sherpa[2.2.2] & NLO (up to 1 jet), LO (up to 3 jets) & \NNPDF[3.0nnlo] & \Sherpa \MEPSatNLO & \Sherpa \\
Electroweak \VVjets & \Sherpa[2.2.1] or \Sherpa[2.2.2] & LO & \NNPDF[3.0nnlo] & \Sherpa \MEPSatLO & \Sherpa \\
\ttbar & \POWHEGBOX[v2] & NLO & \NNPDF[3.0nlo] & \Pythia[8] & A14 \\
QCD multijet & \Pythia[8.230] & LO & \NNPDF[2.3lo] & \Pythia[8] & A14 \\
ggF Higgs & \POWHEG~NNLOPS & NNLO & \PDFforLHC[15nnlo] & \Pythia[8] & AZNLO \\
VBF Higgs & \POWHEG & NLO & \PDFforLHC[15] & \Pythia[8] dipole recoil & AZNLO \\
$VH$ Higgs & \POWHEGBOX[v2] & NLO & \PDFforLHC[15] & \Pythia[8] & AZNLO \\
\bottomrule
\end{tabular}
}
\end{table}
 
Simulated event samples are used to model both the signal and background processes.
The full set of simulated samples is summarised in Table~\ref{tab:samples}.
The generated events were processed through a simulation~\cite{SOFT-2010-01} of the ATLAS detector geometry and response using \GEANT~\cite{Agostinelli:2002hh}, and through the same reconstruction software as the data.
The effect of pile-up was modelled by adding detector signals from simulated inelastic $pp$ events to the original hard-scattering event.
Those were generated with \Pythia[8.186]~\cite{Sjostrand:2014zea} using the \NNPDF[2.3lo] set of parton distribution functions (PDFs)~\cite{Ball:2012cx} and the A3 set of tuned parameters (A3 tune)~\cite{ATL-PHYS-PUB-2016-017}.
The lepton energy scales and resolutions, as well as the lepton reconstruction and identification efficiencies, and the trigger efficiencies in the simulation are corrected to match those measured in data.
 
\subsubsection{\Vjets and \VVjets}
 
The dominant backgrounds in this analysis arise from single weak boson production in association with jets, denoted by \Vjets, where the boson decays via \Znn{} or \Wln{}.
The \Zll{} decay modes ($\ell{\,=\,}e$ or $\mu$) are also important in constraining the \Znn{} contribution. 
The \Vjets production processes defined by $\mathcal{O}(\alphaewk^2)$ diagrams, such as in Figure~\ref{fig:Feynman_b}, are referred to as `strong', while `electroweak' $V$ production is defined by $\mathcal{O}(\alphaewk^4)$ diagrams.
The latter is divided into diboson and single boson production.
Diboson processes (Figure~\ref{fig:Feynman_d}) can mimic the signal when one of the two vector bosons decays into jets and the other decays into leptons, which are either neutrinos or `lost' during reconstruction.
Electroweak single boson production includes the VBF diagram (Figure~\ref{fig:Feynman_c}), where the jets originate from the initial-state quarks that radiated the vector bosons.
 
Similar simulation configurations were used for strong production of single-boson \Vjets and for strong and electroweak production of diboson \VVjets.
These are based on the \Sherpa[2.2] Monte Carlo (MC) generator~\cite{Bothmann:2019yzt} with the \NNPDF[3.0nnlo] set of PDFs~\cite{Ball:2014uwa}.
Parton-shower matching~\cite{Schumann:2007mg} was performed using either the \MEPSatLO~\cite{Catani:2001cc} or \MEPSatNLO~\cite{Hoeche:2011fd,Hoeche:2012yf,Catani:2001cc,Hoeche:2009rj} prescription with associated parameters tuned by the \Sherpa authors.
 
Strong production of \Vjets was simulated using next-to-leading-order (NLO) matrix elements for up to two partons, and leading-order (LO) matrix elements for up to four partons, calculated with the Comix~\cite{Gleisberg:2008fv} and \OPENLOOPS~\cite{Cascioli:2011va,Denner:2016kdg} libraries and the \MEPSatNLO prescription.
The samples were normalised to a next-to-next-to-leading-order (NNLO) prediction~\cite{Anastasiou:2003ds}.
The simulated \Zjets events were further weighted as function of \mjj{}, such that the ratio of the strong \Wjets and \Zjets cross sections matches a dedicated NLO QCD and NLO electroweak calculation in the phase space of a VBF-like signature~\cite{Lindert:2021xxx}.
 
The strong \Vjets samples were filtered in dijet mass at the matrix-element level to enhance the statistical precision in the region $\mjj{\,>\,}0.8\,\TeV{}$, while generating fewer events in the lower $\mjj$ regions.
With the default CSS merging criterion~\cite{Hoeche:2009rj} used in \Sherpa, the filter was found to be inefficient due to the possibility of selecting different objects before and after the parton shower.
To improve the efficiency of the filter, the \kt algorithm~\cite{Ellis:1993tq,Catani:1993hr} was used as the jet criterion when matching the matrix element to the parton shower.
In comparison with the default CSS criterion, the \kt jet criterion classifies more of the forward partons as being above the threshold to be modelled by the matrix element, giving better agreement between the pre- and post-shower \mjj values.
 
Electroweak VBF production of \Vjets at NLO was modelled using \HERWIG[7.2.1]~\cite{Bahr:2008pv,Bellm:2019zci} and the \VBFNLO~\cite{Arnold:2008rz} external matrix element.
The renormalisation and factorisation scales were set to the parton-level \HT defined as the scalar sum of the \pT of all parton-level jets with $\pT{\,>\,}20\,\GeV{}$.
The \MMHT[nlo]~\cite{Harland-Lang:2014zoa} PDF set was used for the matrix-element calculation.
Two sets of samples were generated, one using the default parton shower with angular ordering, and the other using the dipole shower as an alternative.
 
Additionally, there is a term of $\mathcal{O}(\alphaewk^3)$ from interference between the strong and electroweak production modes.
This was modelled independently using \MGNLO~\cite{Alwall:2014hca}.
In the phase space of this search, its contribution was found to be negligible at LO and is therefore not considered further in the background estimation.
 
Diboson final states ($VV$) were simulated at NLO accuracy in QCD for up to one additional parton and at LO accuracy for up to three additional parton emissions.
Loop-induced $gg{\,\to\,}VV$ processes were generated using LO-accurate matrix elements for up to one additional parton emission.
In both cases, the matrix-element calculations were matched and merged with the parton shower using the \MEPSatNLO prescription.
Electroweak diboson production in association with two jets ($VVjj$) was simulated at LO accuracy with the \MEPSatLO prescription.
The diboson samples include only contributions from doubly resonant diboson diagrams, where either one or both bosons decay leptonically.
The off-shell component is assumed to be negligible.
 
\subsubsection{$\ttbar$, $tW$, and single top}
 
The production of \ttbar, the production of a top quark in association with a $W$ boson ($tW$), and single-top $s$-channel and $t$-channel production were modelled using the \POWHEGBOX~\cite{Frixione:2007nw,Nason:2004rx,Frixione:2007vw,Alioli:2010xd}~v2 generator at NLO with the \NNPDF[3.0nlo] PDF set.
For \ttbar\ production, the \hdamp\ parameter\footnote{
The \hdamp\ parameter is a resummation damping factor and one of the parameters that controls the matching of \powheg matrix elements to the parton shower and thus effectively regulates the high-\pt\ radiation against which the \ttbar\ system recoils.}
was set to $1.5\,\mtop$~\cite{ATL-PHYS-PUB-2016-020}.
The events were interfaced to \PYTHIA[8.230] to model the parton shower, hadronisation, and underlying event, with parameters set according to the A14 tune~\cite{ATL-PHYS-PUB-2014-021} and using the \NNPDF[2.3lo] set of PDFs.
The decays of bottom and charm hadrons were performed by \EVTGEN[1.6.0]~\cite{Lange:2001uf}.
 
\subsubsection{QCD multijet}
 
QCD multijet production was modelled using \Pythia[8.230] with leading-order matrix elements for dijet production interfaced to
a \pt-ordered parton shower.
The squared renormalisation and factorisation scales were set to the geometric mean of the squared transverse masses of the two outgoing particles,
$\sqrt{(p_{\mathrm{T},1}^2+m_1^2)(p_{\mathrm{T},2}^2+m_2^2)}$.
The \NNPDF[2.3lo] PDF set was used in the matrix-element generation, the shower, and in the simulation of the multi-parton interactions.
The A14 tune was used. Perturbative uncertainties are estimated through event weights~\cite{Mrenna:2016sih} that encompass
variations of the scales at which the strong coupling constant is evaluated for the initial- and final-state showers as well as the effect of
the PDF uncertainty on the shower and the non-singular part of the splitting functions.
 
\subsubsection{Signal}
 
This analysis targets VBF Higgs boson production.
The small contributions from the ggF and $VH$ Higgs boson production processes passing the analysis selection are also treated as signal.
The invisible decay of the Higgs boson for ggF, $VH$, and VBF was simulated using the SM $H{\,\to\,}ZZ^\ast{\,\to\,}\nu\bar{\nu}\nu\bar{\nu}$ decay, assuming a branching fraction into this final state of 100\%.
The difference between this and decays into new invisible particles is negligible, as the Higgs boson is always produced on-shell for any of the scenarios this analysis is sensitive to.
Events produced via VBF $H{\,\to\,}W^\ast W$ also enter the control regions and were modelled using that same set of generators.
 
Higgs boson production via gluon--gluon fusion was simulated at NNLO accuracy in QCD using the \POWHEG NNLOPS program~\cite{Hamilton:2013fea,Hamilton:2015nsa,Alioli:2010xd,Nason:2004rx,Frixione:2007vw}.
The simulation achieves NNLO accuracy for arbitrary inclusive $gg{\,\to\,}H$ observables by reweighting the Higgs boson rapidity spectrum from HJ-\MINLO~\cite{Hamilton:2012np,Campbell:2012am,Hamilton:2012rf} to that from HNNLO~\cite{Catani:2007vq}.
The \PDFforLHC{15nnlo} PDF set~\cite{Butterworth:2015oua} and the AZNLO tune of \Pythia[8] were used.
The gluon--gluon fusion prediction from the Monte Carlo samples was normalised to the next-to-NNLO cross section in QCD plus NLO electroweak corrections~\cite{deFlorian:2227475,Anastasiou:2016cez,Anastasiou:2015ema,Dulat:2018rbf,Harlander:2009mq,Harlander:2009bw,Harlander:2009my,Pak:2009dg,Actis:2008ug,Actis:2008ts,Bonetti:2018ukf}.
 
The $qq{\,\to\,}VH$ production processes were simulated to NLO accuracy in QCD using the \POWHEGBOX[v2] generator with the \PDFforLHC[15] set of PDFs.
 
Higgs boson production via VBF processes was modelled with \POWHEG~\cite{Nason:2009ai,Alioli:2010xd,Nason:2004rx,Frixione:2007vw} and interfaced with \Pythia[8] for parton showering, using the dipole recoil scheme, and non-perturbative effects.
The \powheg prediction is accurate to NLO and tuned to match calculations with effects due to finite heavy-quark masses and soft-gluon resummations up to next-to-next-to-leading-logarithm accuracy.
The \PDFforLHC[15] PDF set and the AZNLO tune of \Pythia[8] were used.
The MC prediction was normalised to an approximate-NNLO QCD cross section~\cite{Ciccolini:2007jr,Ciccolini:2007ec,Bolzoni:2010xr}. 
NLO electroweak (EWK) corrections to VBF Higgs boson ($m_H{\,=\,}125\,\GeV{}$) production were computed separately using \hawk~\cite{Denner:2014cla}.
This was achieved by applying a reweighing factor of $1-\alpha_{\text{NLO EWK}}^{\text{HAWK}}$ to the \POWHEG events described above, where $\alpha_{\text{NLO EWK}}^{\text{HAWK}}{\,=\,}-0.000350\,\GeV^{-1}\cdot\pT^{\text{Higgs}}-0.0430$.
This correction was derived in a kinematic region defined by mimicking the analysis selection requirements on jet \pT{}, \detajj{}, and \mjj{}.
 
Additional VBF samples with lighter or heavier Higgs bosons were generated in a mass range from $50\,\GeV{}$ to $2\,\TeV$ to allow interpretation of the results in scenarios with massive scalar mediators.
The parameter values used in the production of these samples are exactly the same as those for the main signal sample; the only difference is the Higgs boson mass.
The samples with alternative Higgs boson masses were processed with a fast simulation~\cite{SOFT-2010-01} which relies on a parameterisation of the calorimeter response~\cite{ATL-PHYS-PUB-2010-013} instead of the full \GEANT-based simulation.
 
For the background estimation, the normalisation of all Higgs boson samples also accounts for the decay branching fraction calculated with HDECAY~\cite{Djouadi:1997yw,Spira:1997dg,Djouadi:2006bz} and \PROPHECY~\cite{Bredenstein:2006ha,Bredenstein:2006rh,Bredenstein:2006nk}.
% End of text imported from the .//sections/samples.tex input file

\section{Object and event selection}
% The next lines are included from the .//sections/selection.tex input file
\subsection{Object definitions and event cleaning}
\label{sec:objects}
 
Each event is required to have a primary vertex, which is the vertex with the highest sum of squared transverse momenta from at least two associated tracks with $\pT>0.5\,\GeV{}$~\cite{ATL-PHYS-PUB-2015-026}.

Electrons are reconstructed by matching clustered energy deposits in the EM calorimeter to a track in the ID~\cite{EGAM-2018-01}, including the transition regions between barrel and endcap EM calorimeters at $1.37{\,<\,}|\eta|{\,<\,}1.52$.
Electron candidates must satisfy $\pT{\,>\,}4.5\,\GeV$, $|\eta|{\,<\,}2.47$, and a `loose' identification criterion.
Depending on the \pT and $|\eta|$ range, muons are reconstructed as either an ID track matched to an MS track or track segment(s), an ID track matched to a calorimeter energy deposit compatible with a minimum-ionising particle, or an MS track passing a requirement on compatibility with originating from the IP~\cite{MUON-2018-03}.
Muon candidates must satisfy $\pT{\,>\,}4\,\GeV$, $|\eta|{\,<\,}2.7$, and a `very loose' identification criterion.
No isolation requirements are placed on electron and muon candidates.
 
To be considered as a `signal lepton' stemming from a leptonic decay of a vector boson or a $\tau$-lepton, the leading lepton has to have $\pT{\,>\,}30\,\GeV{}$ and fulfil a `loose' isolation criterion.
For events with one lepton, i.e.\ for the control samples used to constrain the background from \Wlnjets processes, the lepton identification uses a `tight' (`medium') criterion for electrons (muons)~\cite{EGAM-2018-01,MUON-2018-03}.
If two leptons are required, as in the $Z\to{\ell\ell}$ control region, the identification uses `loose' criteria for electrons and muons.
Furthermore, the lepton has to be compatible with originating from the primary vertex.
 
Photon candidates are reconstructed from clustered energy deposits in the EM calorimeter~\cite{EGAM-2018-01}.
To be considered by the analysis, photons are required to have $\pT{\,>\,}20\,\GeV$, to fulfil `tight' identification and isolation criteria and to be within $|\eta|{\,<\,}2.37$, excluding the transition regions between barrel and endcap EM calorimeters.
Events that contain an isolated photon are rejected to ensure orthogonality to other ATLAS analyses~\cite{EXOT-2021-17}.
 
Particle-flow (PFlow) jets are constructed using the \antikt algorithm~\cite{Cacciari:2008gp,Fastjet} with a radius parameter of \(R=0.4\), using charged constituents associated with the primary vertex and neutral PFlow constituents as inputs~\cite{PERF-2015-09}.
Jets are calibrated to the particle scale using a sequence of corrections, including pile-up subtraction and in situ calibration~\cite{JETM-2018-05}.
 
Jets are required to have $\pt{\,>\,}20\,\GeV$ and $|\eta|{\,<\,}4.5$.
For jets with $\pt{\,<\,}60\,\GeV$ and $|\eta|{\,<\,}2.5$, the jet vertex tagger (JVT) discriminant is used to identify jets originating from the hard-scatter interaction through the use of tracking and vertexing information~\cite{ATLAS-CONF-2014-018}.
The chosen selection criterion accepts hard-scatter jets with an efficiency of about 97\%.
If either of the two leading jets has $|\eta|{\,>\,}2.5$ and $\pT{\,<\,}120\,\GeV$, it must satisfy a threshold for the forward jet vertex tagger (fJVT) discriminant~\cite{ATL-PHYS-PUB-2015-034,ATL-PHYS-PUB-2019-026} that has an efficiency for hard-scatter jets of about 93\% and rejects about 58\% of pile-up jets with $\pT{\,>\,}50\,\GeV$.
To further suppress the multijet background for selections with $\met{\,<\,}200\,\GeV{}$ a tighter fJVT criterion has to be satisfied, which has an efficiency  of 70\%--90\% for hard-scatter jets and rejects more than 80\% of pile-up jets depending on the $\pT$ and $\eta$ for jets with $\pT{\,>\,}50\,\GeV$.
 
Jets containing $b$-hadrons ($b$-jets) are identified using a multivariate discriminant (MV2c10)~\cite{PERF-2016-05}.
The chosen criterion is estimated to have 77\% efficiency for selecting $b$-jets in an inclusive \ttbar{} sample, with associated rejection factors of 5 and 134 for jets with charm and without heavy-flavour content, respectively.
 
To avoid double counting of energy deposits, the reconstructed objects are required to be separated in $(y,\phi)$-space according to the procedure detailed in Table~\ref{tab:overlap_removal}.
For leptons in the vicinity of jets, the $\Delta R$ threshold depends on the transverse momentum of the lepton to account for the collimation of boosted objects.
 
\begin{table}[]
\centering
\caption{Overview of the overlap removal procedure applied to pairs of objects and the corresponding matching criteria, listed according to priority.
}\label{tab:overlap_removal}
\begin{tabular}{ccc}
\toprule
Remove & Keep & Matching criteria \\
\midrule
electron	& electron	& shared inner-detector track, electron with lower \pt{} removed \\
muon		& electron	& muon with calorimeter energy deposits and shared inner-detector track \\
electron	& muon		& shared inner-detector track \\
photon	& electron	& $\Delta R<$ 0.4 \\
photon	& muon		& $\Delta R<$ 0.4 \\
jet			& electron	& $\Delta R<$ 0.2 \\
electron	& jet				& $\Delta R<\min(0.4, 0.04 + 10\,\GeV/p_\text{T}^e)$ \\
jet			& muon		& number of tracks $<$ 3 and $\Delta R<$ 0.2 \\
muon		& jet				& $\Delta R<\min(0.4, 0.04 + 10\,\GeV/p_\text{T}^{\mu})$\\
jet			& photon				& $\Delta R<$ 0.4 \\
\bottomrule
\end{tabular}
\end{table}
 
The missing transverse momentum, with magnitude $\MET$, is defined as the negative vectorial sum of the transverse momenta of all selected electrons, muons, photons, and jets, as well as tracks compatible with the primary vertex but not matched to any of those objects; this last contribution is called the `soft term' in ATLAS analyses~\cite{PERF-2016-07,ATLAS-CONF-2018-023}.
In the control samples with one or two signal leptons, the $\MET$ is adjusted such that these leptons are treated as invisible particles and referred to as `$E_\text{T}^\text{miss}$ (without leptons)'.
For the $W_{e\nu}$ control regions (see Section~\ref{sec:WZbg}), an $\MET$ significance variable is used to enrich the sample with multijet events containing a fake electron, i.e.\ a jet that is misidentified as an electron: 
\begin{equation}\nonumber
S_\text{MET}=\frac{\MET}{\sqrt{\pT^\text{j1}+\pT^\text{j2}+\pT^e}},
\end{equation}
where the upper indices specify the two highest-\pT jets (j1 and j2) as well as the electron ($e$).
To suppress the background from multijet events for which the $\MET$ stems from unidentified pile-up jets, or jets mistagged as pile-up, the variable $\MHT$ is defined as the magnitude of the vectorial sum of the transverse momenta of all jets with $\pT{\,>\,}20\,\GeV{}$, including jets tagged as pile-up jets.
 
Several requirements are applied to suppress non-collision background events~\cite{ATLAS-CONF-2015-029} with fake jets that do not stem from collisions at the IP.
Such fake jets can be caused by electronic noise, and jets from collisions are identified by requiring a good fit to the expected pulse shape for each constituent calorimeter cell.
Another source of fake jets are beam-halo interactions with the LHC collimators.
Those jets are identified by requirements on their energy distribution in the calorimeters and the fraction of their constituent tracks that originate from the primary vertex.
The event is rejected if any selected jet is identified as a fake jet.
 
\subsection{Signature from vector-boson fusion}
 
The VBF process exhibits characteristic features which are exploited in this analysis.
The two leading jets are typically in opposite hemispheres of the detector and are more forward than jets from non-VBF processes at a comparable parton interaction scale $\sqrt{\hat{s}}$.
This leads to large values of the pseudorapidity separation $\detajj$ and dijet invariant mass \mjj.
In QCD multijet events, the two leading jets exhibit a characteristic back-to-back topology ($\dphijj\sim\pi$) in the transverse plane, while for the VBF signal, the Higgs boson tends to have significant \pt{}, which results in smaller values of the azimuthal separation $\dphijj$.
Hence, a selection based on $\dphijj$ is an effective way to suppress the multijet background.
 
Another distinct feature of the VBF process is the reduced hadronic activity within the rapidity gap between the two leading jets, caused by the absence of colour connection between the two quarks.
Therefore, strong \Vjets production can be significantly suppressed by requiring that no further jet with $\pT{\,>\,}25\,\GeV{}$ is present.
However, initial- and final-state radiation in VBF processes can result in additional jets with $\pT{\,>\,}25\,\GeV{}$.
Therefore, to enhance the sensitivity of the analysis, events with three or four jets are considered as well, but with additional requirements.
The centrality $C_{i}$ of a third- and fourth-highest-\pT{} jet ($i{\,=\,}\text{j3}$ and $i{\,=\,}\text{j4}$)~\cite{HIGG-2013-13} is defined as
\begin{equation}\nonumber
C_{i}=\exp \left( -\frac{4}{\left(\eta^\text{j1}-\eta^\text{j2}\right)^2}\left(\eta^i-\frac{\eta^\text{j1}+\eta^\text{j2}}{2}\right)^2\right).
\end{equation}
QCD processes tend to have sub-leading jets emitted close to the midpoint of the rapidity gap, which gives $C_i \approx 1$, while sub-leading jets from final-state radiation in a VBF process tend to have smaller values of $C_i$.
Another observable that tests whether a sub-leading jet is compatible with having been radiated from one of the two leading jets in the VBF process is the smaller of the values of the invariant mass of this sub-leading jet and one of the two leading jets, relative to \mjj{}:
\begin{equation}\nonumber
m^\text{rel}_{i}=\frac{\text{min}\{m_{\text{j1,}i},m_{\text{j2,}i}\}}{\mjj{}}.
\end{equation}
Small values of $m^\text{rel}_{i} $ indicate that the additional jet is compatible with final-state radiation.
 
\subsection{Event selection}
\label{sec:defSR}
 
To enter the signal region (SR), the following event requirements must be satisfied:
\begin{itemize}
\item The event contains no lepton candidate, nor a photon.
\item The event has two, three or four jets with $\pt{\,>\,}25\,\GeV$ that fulfil the JVT requirement if they have $|\eta|{\,<\,}2.5$.
\item The leading and sub-leading jets have $\pT{\,>\,}80\,\GeV$ and $50\,\GeV$, respectively. Both jets must fulfil fJVT requirements if they have $|\eta|{\,>\,}2.5$ and $\pT{\,<\,}120\,\GeV$. 
\item To ensure orthogonality to other searches, it is required that not more than one of the jets are $b$-tagged. The fraction of events removed by this requirement is less than 0.02\% because most of the jets are outside the acceptance of the tracking detector.
\item Any existing third ($i=\text{j3}$) or fourth ($i=\text{j4}$) jet must have $C_i{\,<\,}0.6$ and $m^\text{rel}_{i}{\,<\,}0.05$, to be compatible with VBF final-state radiation.
\item The event has $\MET{\,>\,}160\,\GeV$, which strongly suppresses the multijet background.
\item The soft term of the \met{} is smaller than $20\,\GeV$. This requirement removes events with $W\to\mu\nu$ decays where the muon is not identified but is reconstructed as a track.
\item Further suppression of the multijet background is achieved by requiring $\MHT{\,>\,}140\,\GeV$.
\item The two leading jets are not back-to-back: $\dphijj{\,<\,}2$.
\item The two leading jets must fulfil the VBF topology requirements of opposite longitudinal hemispheres ($\eta^\text{j1}\cdot\eta^\text{j2}{\,<\,}0$), large pseudorapdidity separation ($\detajj{\,>\,}3.8$), and large invariant mass ($\mjj{\,>\,}0.8\,\TeV$).
\end{itemize}
 
For the VBF process, which contributes approximately 90\% of the signal events in the SR, the selection efficiency for an invisibly decaying Higgs boson is about 1.1\%.
For the ggF process, the selection efficiency is two orders of magnitude smaller.
The selected events in the SR are separated into 16 bins of different signal purity. 
Events with exactly two jets and $\MET{\,>\,}200\,\GeV$ are separated into five \mjj bins ($0.8$--$1.0\,\TeV{}$, $1.0$--$1.5\,\TeV{}$, $1.5$--$2.0\,\TeV{}$, $2.0$--$3.5\,\TeV{}$, ${>\,}3.5\,\TeV{}$) and are further separated into two bins with $\dphijj{\,<\,}1$ and $1{\,\leq\,}\dphijj{\,<\,}2$.
This results in ten SR bins for two-jet events.
The highest signal-to-background ratio is obtained for the largest values of \mjj, while the lower range in \dphijj also shows better background suppression.
Events with three or four jets and $\MET{\,>\,}200\,\GeV$ form another three bins defined by three \mjj ranges: $1.5$--$2.0\,\TeV{}$, $2.0$--$3.5\,\TeV{}$, and ${>\,}3.5\,\TeV{}$.
Finally, the same three \mjj intervals are used to define three more bins with two jets but $160\,\GeV{\,<\,}\MET{\,\le\,} 200\,\GeV$.
A schematic view of the binning strategy is shown in Figure~\ref{fig:SRbinning}.
 
\begin{figure}
\centering
\includegraphics[width=1.0\linewidth]{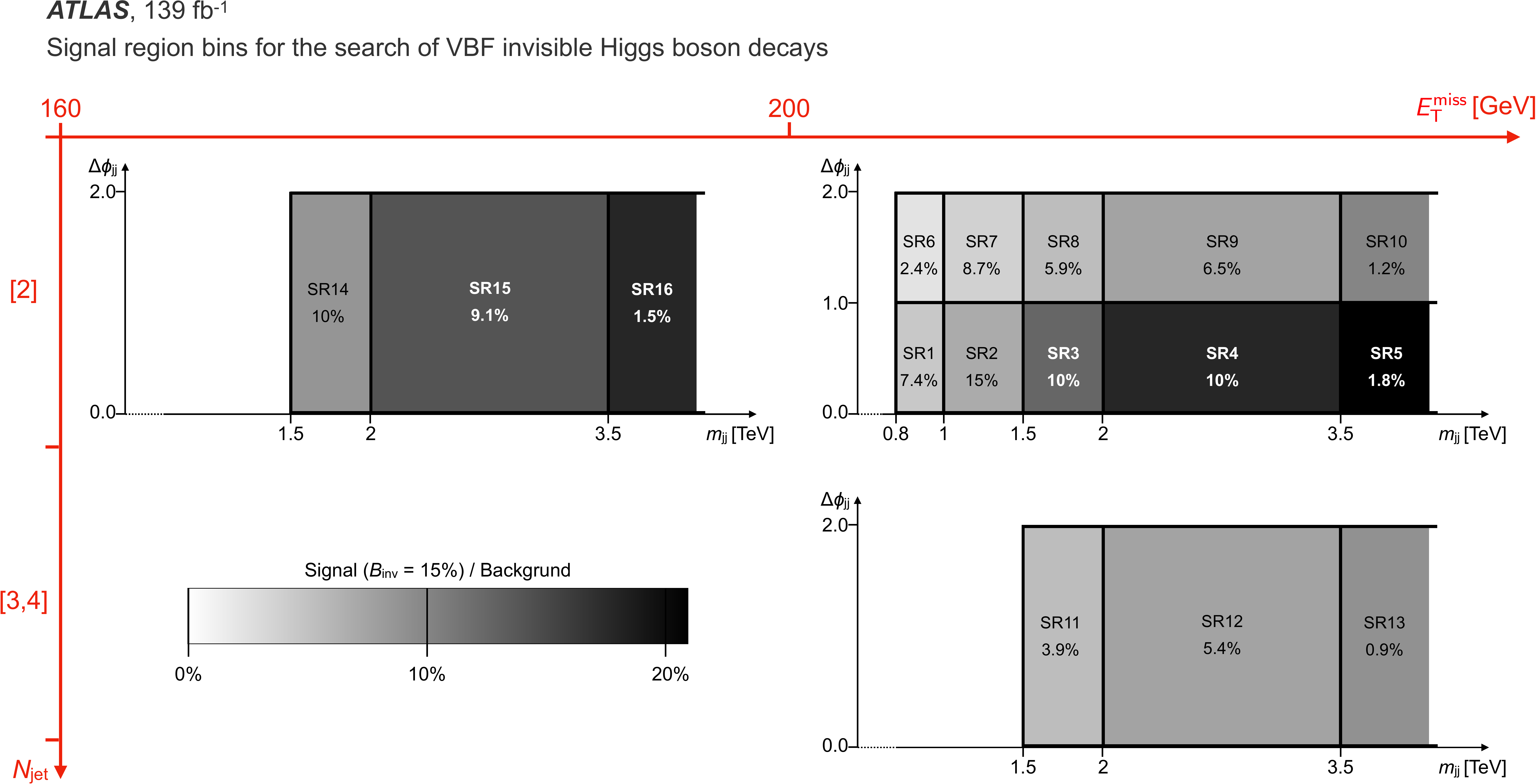}
\caption{Schematic view of the 16 bins in the signal region.
The shading indicates the signal-to-background ratio, assuming a branching fraction of the Higgs boson to invisible particles of $\mathcal{B}_\text{inv}{\,=\,}0.15$.
The percentage gives the approximate contribution of signal from invisibly decaying Higgs bosons to each of the bins relative to the total signal yield in the inclusive SR.}
\label{fig:SRbinning}
\end{figure}

% End of text imported from the .//sections/selection.tex input file
 
\section{Background estimation}
% The next lines are included from the .//sections/backgrounds.tex input file
Events from \Vjets processes comprise about 95\% of the background that enters the SR and their contributions are estimated using dedicated control regions (CRs).
A smaller contribution arises from multijet (MJ) processes and is estimated from data.
Minor backgrounds from $t\bar{t}$, $VV$, $VVV$, and VBF $H$ production are estimated by MC simulation.
 
\subsection{\Vjets background}
\label{sec:WZbg}
 
The backgrounds in the SR from \Vjets processes are challenging to model, and hence have large associated theoretical uncertainties.
To improve the estimation of these backgrounds, a data-driven technique is implemented that uses CRs defined in kinematic regions analogous to the SR but containing selected \Zlljets and \Wlnjets events.
The CRs are divided into 16 bins following the same selection in \mjj{}, \dphijj{}, \met{}, and $N_\text{jet}$ as for the SR, but using `$E_\text{T}^\text{miss}$ (without leptons)' instead of \met{}.
By construction, the strong and electroweak \Vjets processes are probed simultaneously in those CR bins.
 
The $Z_{\ell\ell}$ CR is based on the same selection criteria as the SR, but the lepton veto is replaced by a requirement of exactly two same-flavour opposite-sign `signal leptons' ($\ell$) with $|m_{\ell\ell}-m_{Z}|{\,<\,}25\,\GeV$.
The $\pT$ threshold for the leading lepton is $30\,\GeV{}$, while for the sub-leading lepton no further requirement on $\pT$ is applied.
To suppress contamination from $VV$ and $VVV$ production, as well as contributions from VBF $H\to W^\ast W$ or $H\to\tau^+ \tau^-$, the events are required to have $\MET{\,<\,}70\,\GeV$, where the leptons are included in the \MET calculation.
This exploits the fact that events from those processes are expected to have larger amounts of \MET, due to the neutrinos produced by leptonically decaying $W$ bosons or $\tau$-leptons, than events from \Zlljets processes, for which the \MET arises mainly from jet mismeasurements and peaks at values of $\mathcal{O}(10\,\GeV)$.
These selection criteria ensure a purity in \Zjets events greater than $90\%$ for all bins.
A comparison of the observed yields in the $Z_{\ell\ell}$ CR and the expectation from simulation is shown for selected distributions in Figure~\ref{fig:zcr_zll_prefit139}.
 
\begin{figure}
\centering
\subfloat[]{
\parbox{0.49\textwidth}{
\centering
\includegraphics[width=0.49\textwidth]{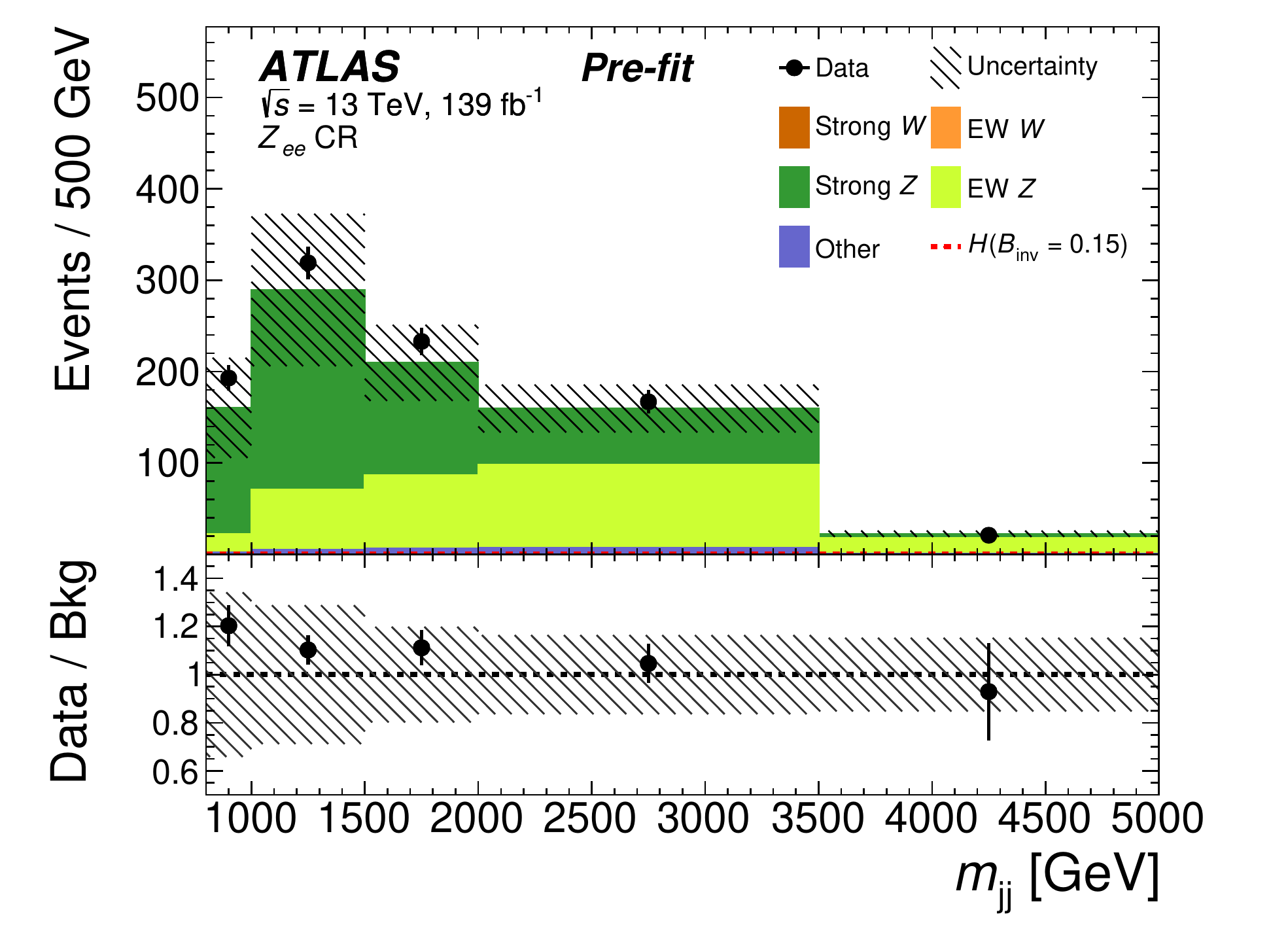}
}
\label{fig:zcr_zll_prefit139_a}
}
\subfloat[]{
\parbox{0.49\textwidth}{
\centering
\includegraphics[width=0.49\textwidth]{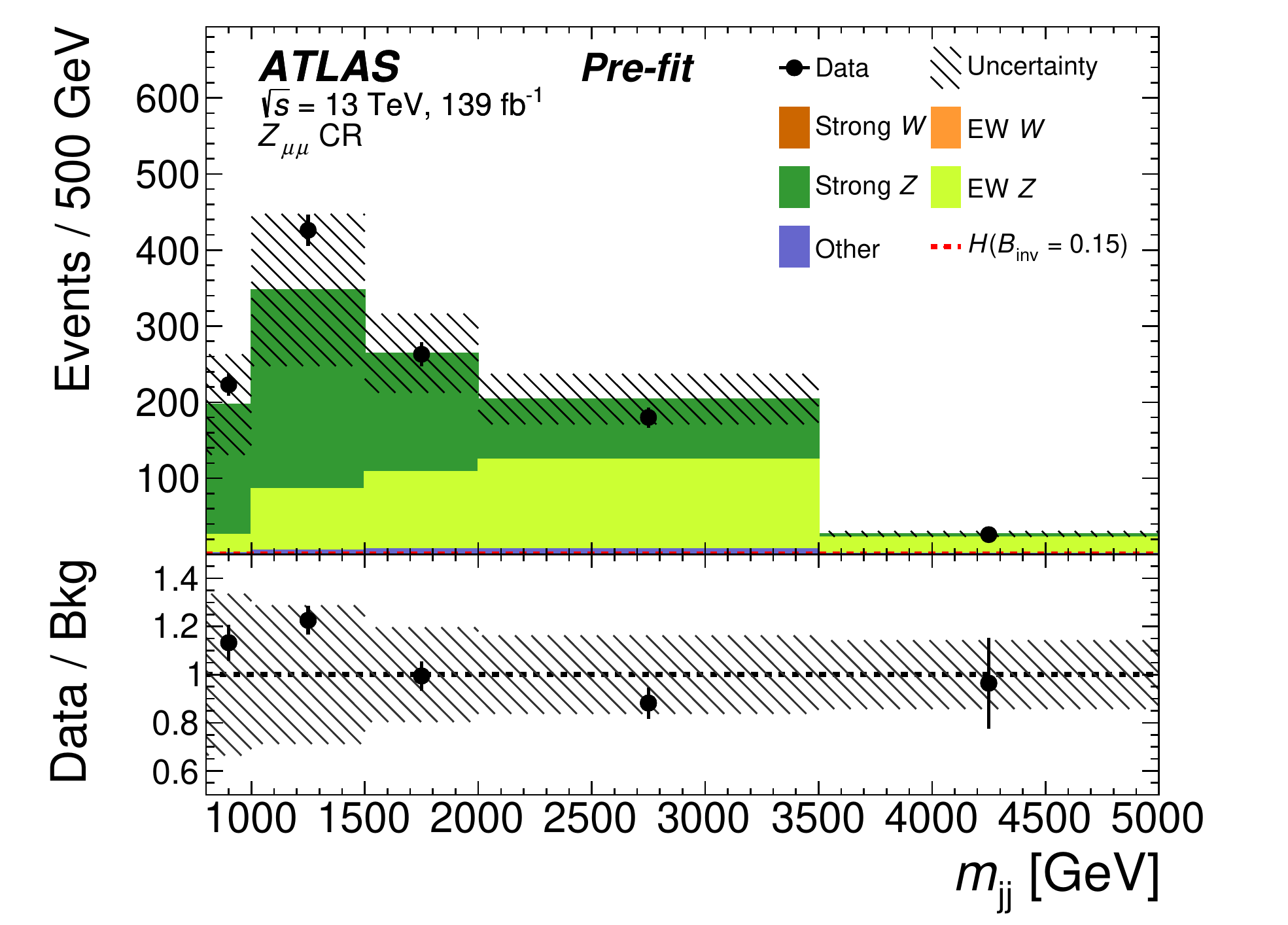}
}
\label{fig:zcr_zll_prefit139_b}
}
 
\subfloat[]{
\parbox{0.49\textwidth}{
\centering
\includegraphics[width=0.49\textwidth]{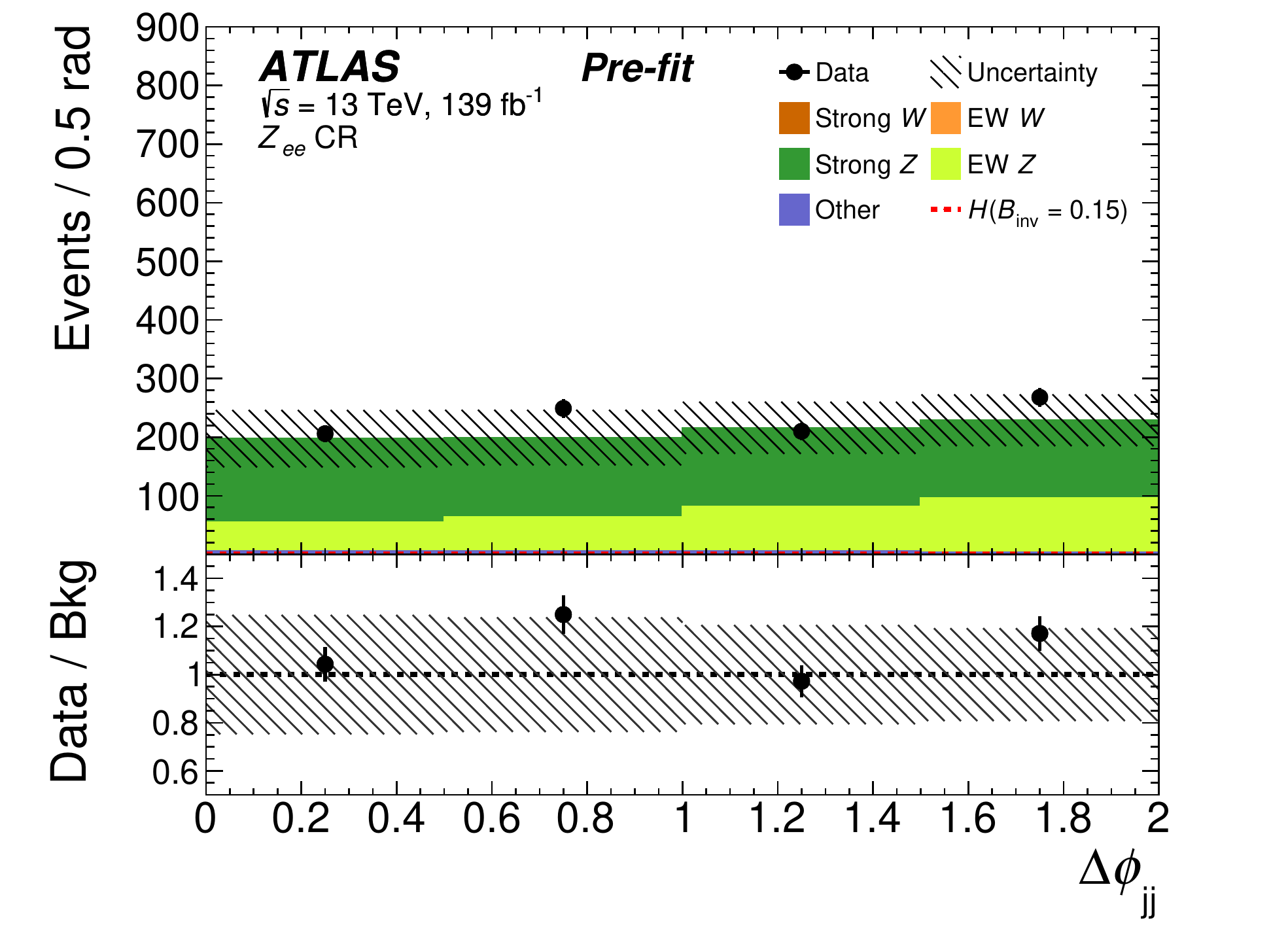}
}
\label{fig:zcr_zll_prefit139_c}
}
\subfloat[]{
\parbox{0.49\textwidth}{
\centering
\includegraphics[width=0.49\textwidth]{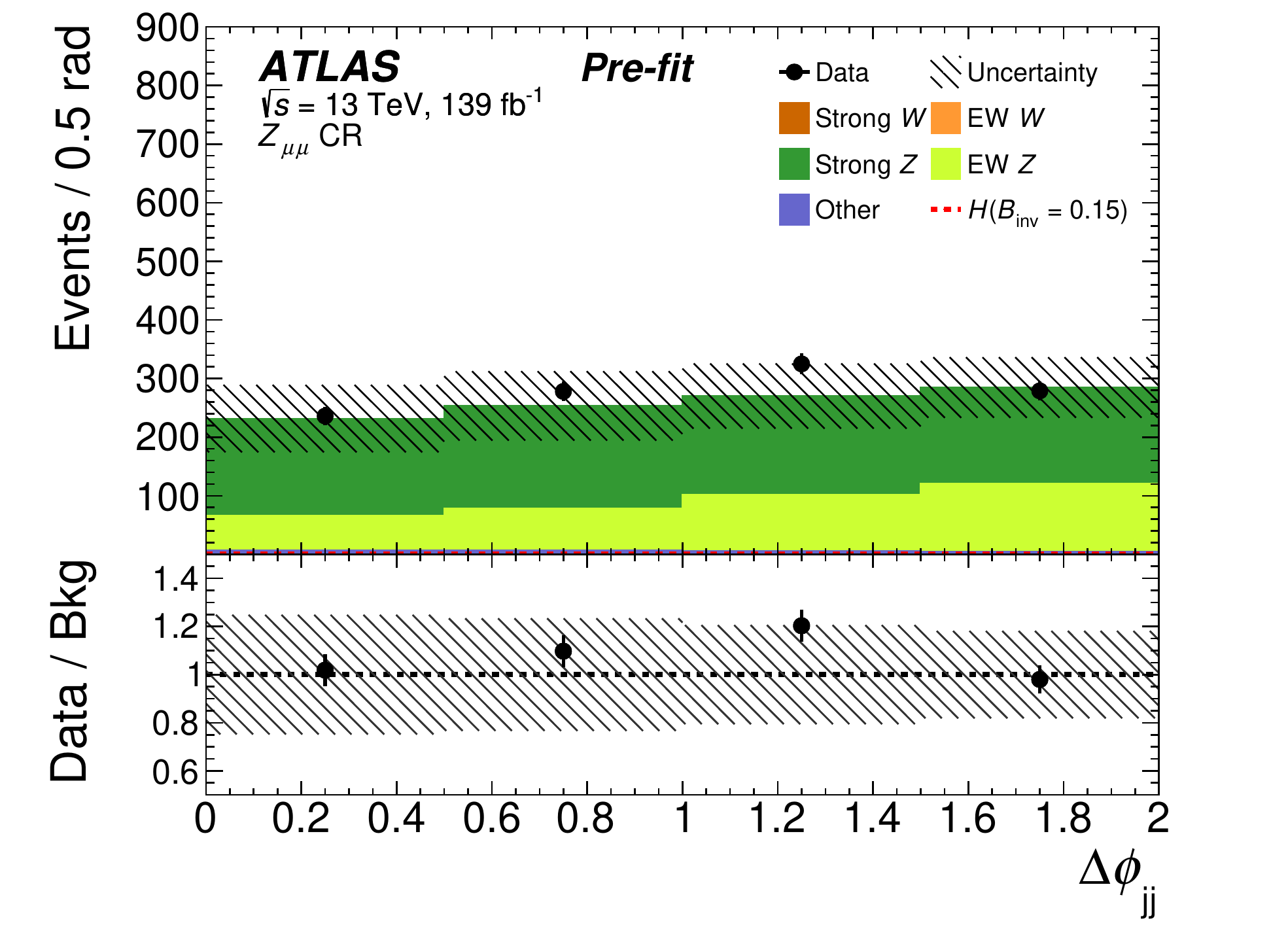}
}
\label{fig:zcr_zll_prefit139_d}
}
 
\subfloat[]{
\parbox{0.49\textwidth}{
\centering
\includegraphics[width=0.49\textwidth]{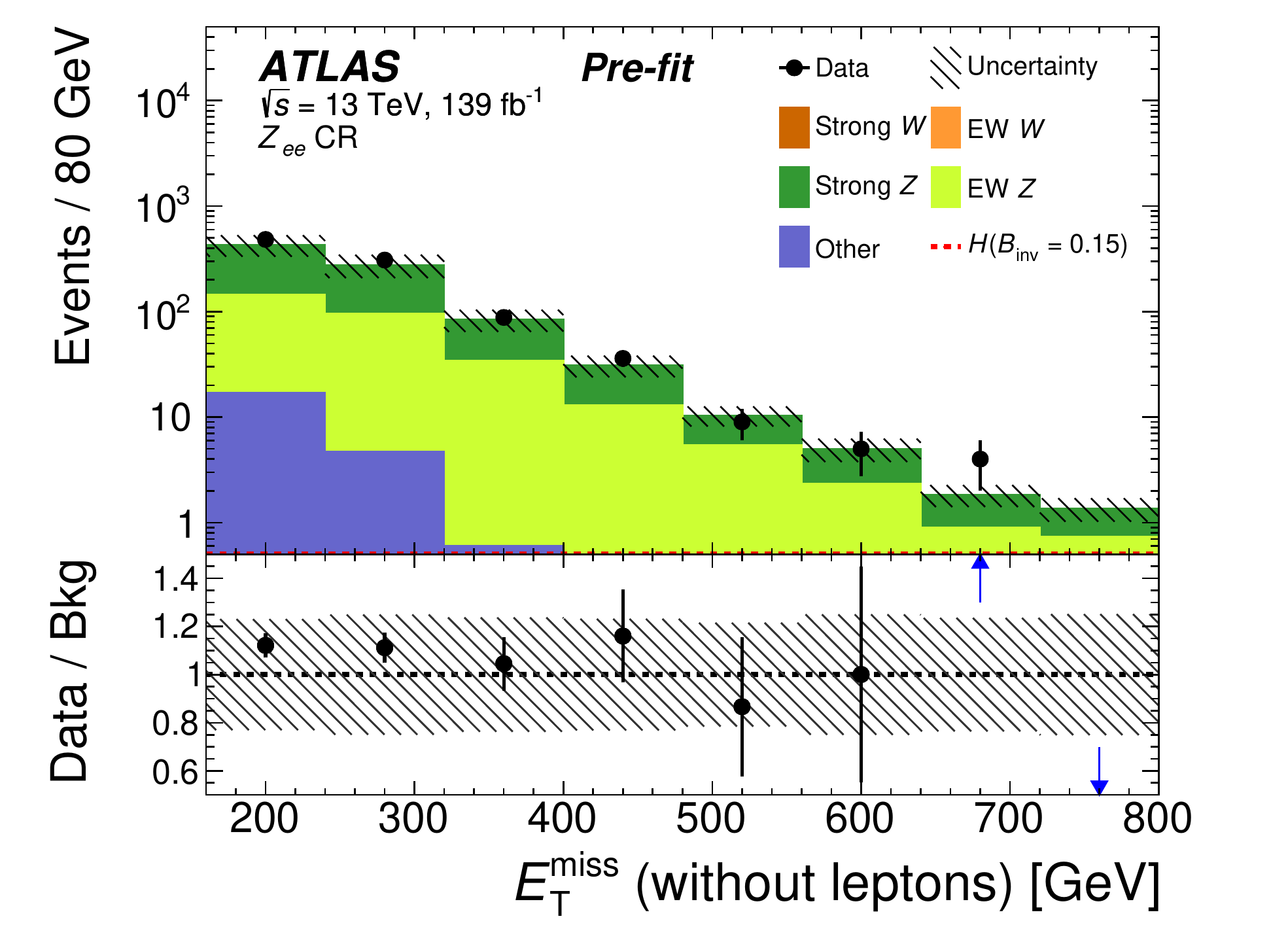}
}
\label{fig:zcr_zll_prefit139_e}
}
\subfloat[]{
\parbox{0.49\textwidth}{
\centering
\includegraphics[width=0.49\textwidth]{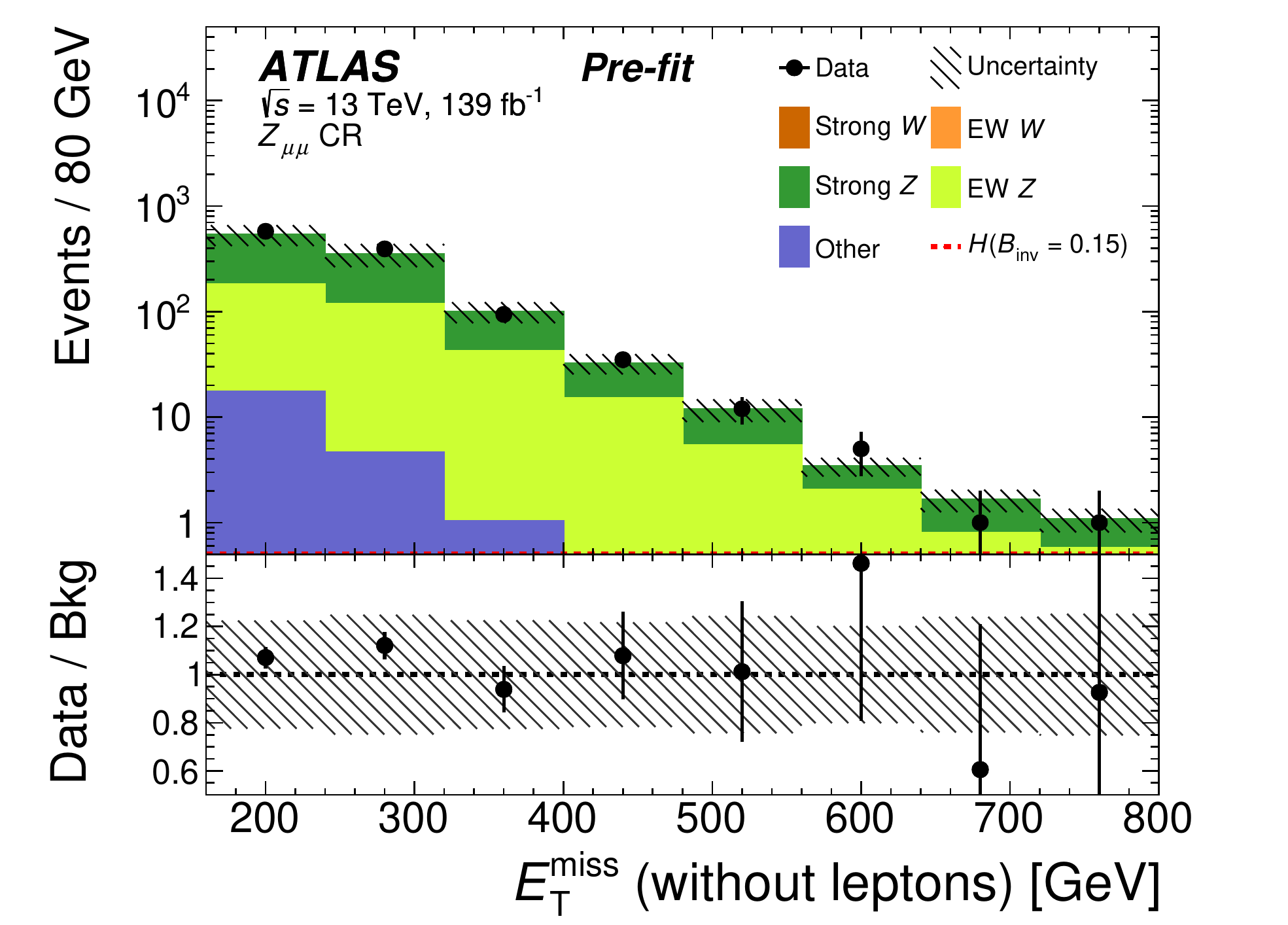}
}
\label{fig:zcr_zll_prefit139_f}
}
\caption{Observed and predicted distributions of $m_\text{jj}$, $\Delta\phi_{\text{jj}}$, and $E_\text{T}^\text{miss}$ (without leptons) in the $Z_{\ell\ell}$ control regions, separately for $Z\to e^+e^-$ (\protect\subref{fig:zcr_zll_prefit139_a}, \protect\subref{fig:zcr_zll_prefit139_c}, and \protect\subref{fig:zcr_zll_prefit139_e}) and $Z\to \mu^+\mu^-$ (\protect\subref{fig:zcr_zll_prefit139_b}, \protect\subref{fig:zcr_zll_prefit139_d}, and \protect\subref{fig:zcr_zll_prefit139_f}).
The MC predictions are normalised to cross sections times luminosity and are shown prior to the likelihood fit.
Minor contributions from $t\bar{t}$, $VV$, $VVV$, and VBF $H$ with $H\to\tau^+\tau^-$ or $H\to W^\ast W$ are combined and labelled `other'.
The tiny contributions from $W\text{+\,jets}$ processes are not visible for the given value range.
The hatched band indicates statistical and experimental systematic uncertainties.
The dielectron and dimuon channels are combined to a dilepton CR in the statistical analysis.
}
\label{fig:zcr_zll_prefit139}
\end{figure}
 
The $W$ CRs require exactly one `signal lepton' with $\pT{\,>\,}30\,\GeV$.
The remaining selection criteria are identical to those of the signal region.
Events that pass these requirements are separated according to the lepton flavour.
The single-electron sample is expected to have a contribution of events with fake electrons originating from multijet processes in which a jet is misidentified as an isolated electron.
To suppress this contribution, the events in the single-electron control region ($W_{e\nu}$ CR) are also required to satisfy $S_\text{MET}{\,>\,}4\sqrt{\GeV}$.
The single-electron events that fail this $S_\text{MET}$ selection define the `fake-$e$ CR', which is enriched with fake electrons since multijet events have no or very little \MET and hence are expected to have low $S_\text{MET}$ values, as checked in simulation.
In this CR, the fake-electron contribution can be extracted in all 16 bins by subtracting from the data the MC predictions for processes with real electrons.
To evaluate the fake-electron background in the $W_{e\nu}$ CR, another control region ($W_{e\nu}$ anti-ID) enriched with fake electrons is defined with the electron passing a loose identification requirement but not the tight one for `signal leptons' (Figure~\ref{fig:fake_lep_a}).
For this selection the ratio $R_S$ of the number of data events with high $S_\text{MET}$ to the number with low $S_\text{MET}$ is calculated after the $W$, $Z$, $VV$, $VVV$ and $t\bar{t}$ backgrounds are subtracted.
The average $R_S$ for all 16 bins is $0.12{\,\pm \,}0.01$ and its value is evaluated for each \mjj{} range for $160\,\GeV{}{\,<\,}\MET{\,<\,}200\,\GeV{}$ and $\MET{\,>\,}200\,\GeV{}$ separately.
The fake-electron contribution entering bin $i$ of the $W_{e\nu}$ CR is evaluated as $R_{S,i}\cdot n_{\text{fake-$e$},i}$, where $n_{\text{fake-$e$},i}$ is the excess of data events over the expected prompt background in the corresponding bin of the `fake-$e$ CR'.
This procedure assumes that the identification efficiencies of electrons and $S_\text{MET}$ are uncorrelated.
 
\begin{figure}
\centering
\subfloat[]{
\parbox{0.49\textwidth}{
\centering
\includegraphics[width=0.49\textwidth]{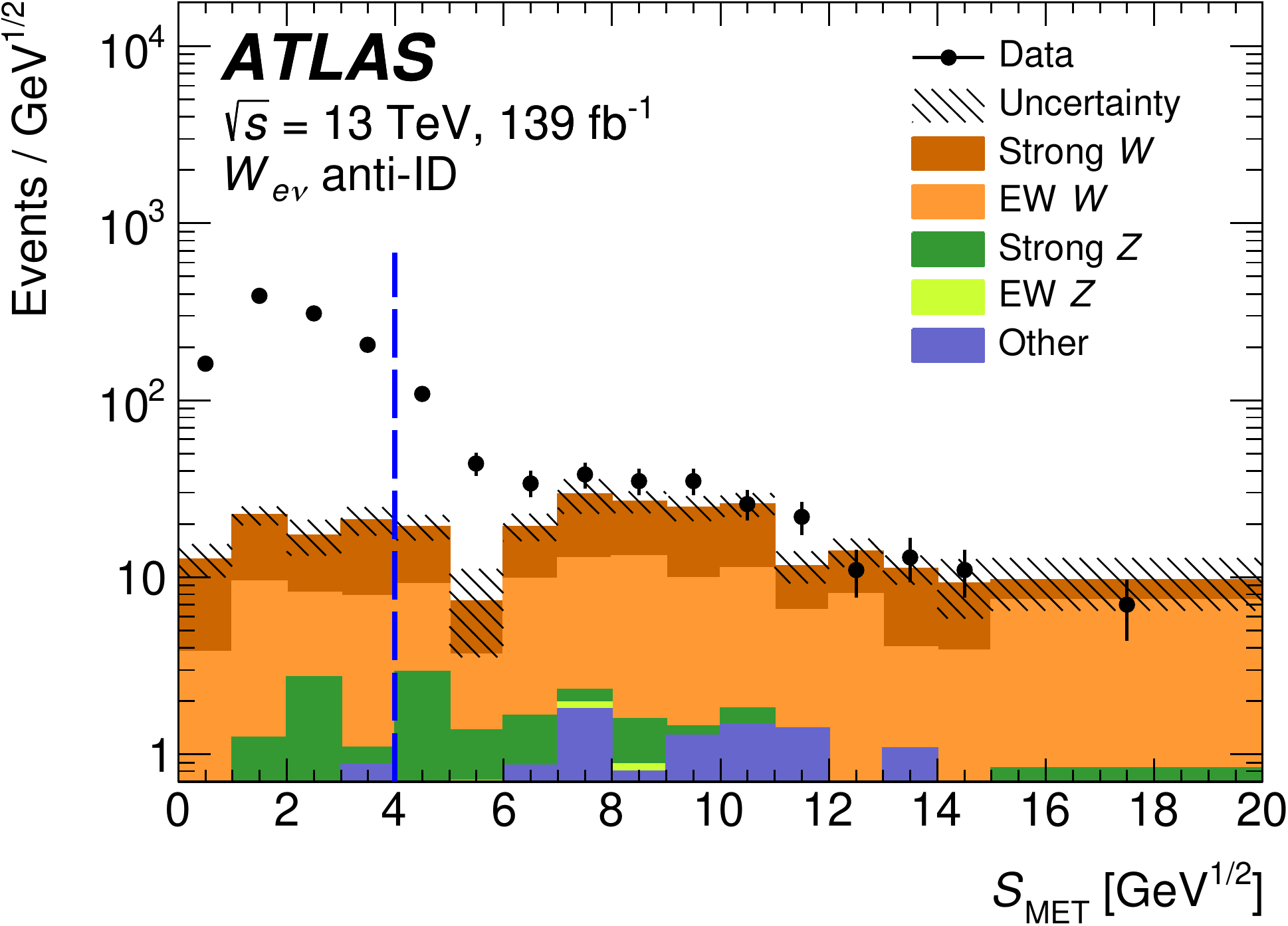}
}
\label{fig:fake_lep_a}
}
\subfloat[]{
\parbox{0.49\textwidth}{
\centering
\includegraphics[width=0.49\textwidth]{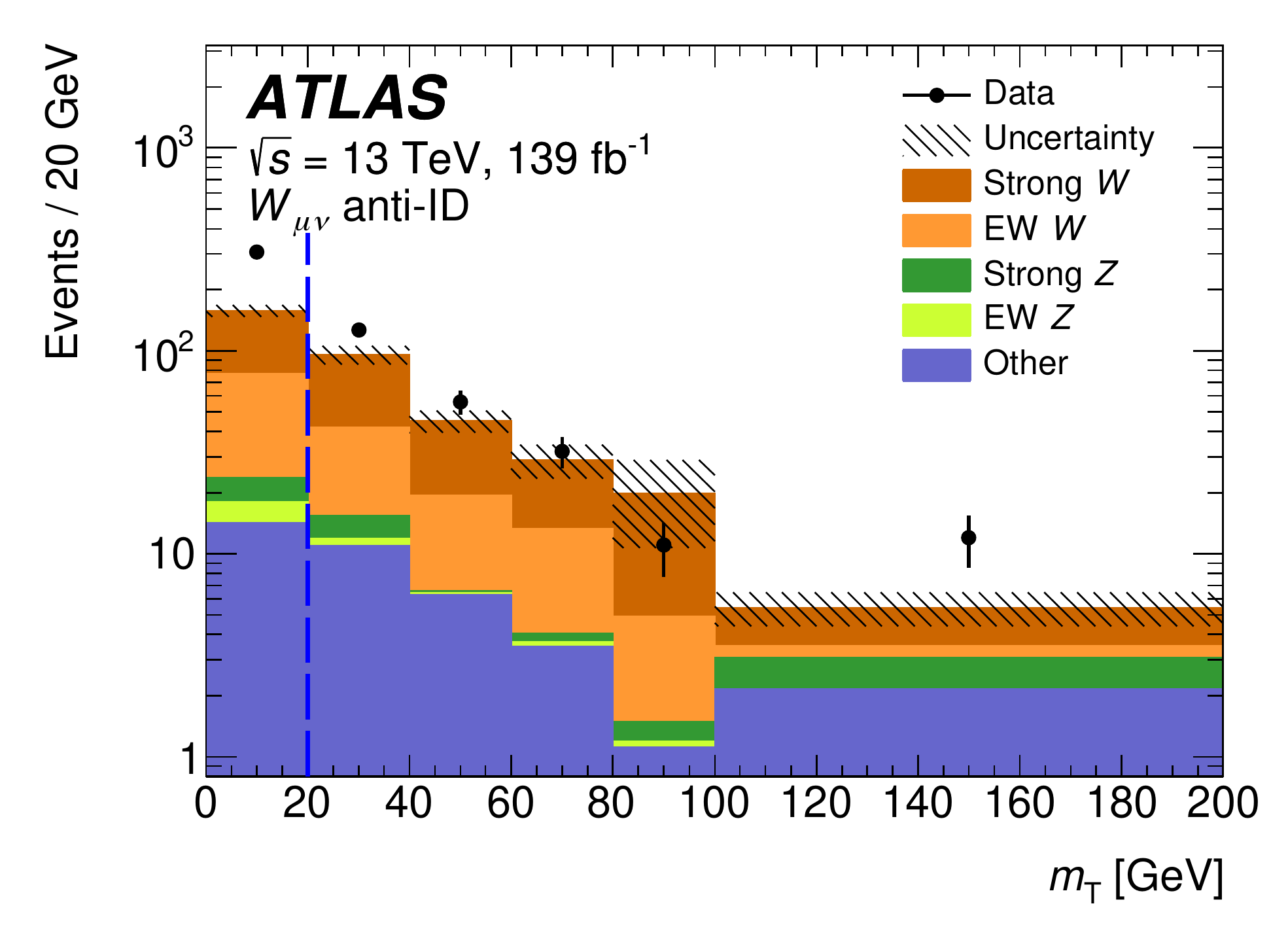}
}
\label{fig:fake_lep_b}
}
\caption{The figure labelled \protect\subref{fig:fake_lep_a} shows a comparison of the $S_\text{MET}$ distribution in data and simulation for the $W_{e\nu}$ anti-ID control region, for which the electron passes a loose identification criterion but not the tight one.
The figure labelled \protect\subref{fig:fake_lep_b} shows a comparison of the transverse mass $m_\text{T}$ distribution in data and simulation for the $W_{\mu\nu}$ anti-ID control region, for which the muon candidate does not pass the loose identification criterion.
Minor contributions from $t\bar{t}$, $VV$, $VVV$, and VBF $H$ with $H\to\tau^+\tau^-$ or $H\to WW^\ast$ are combined and labelled `other'.
The excesses in data, observed at low $S_\text{MET}$ and low $m_\text{T}$ values, are caused by multijet events containing a fake lepton.
The hashed bands display the combined uncertainty from MC sample sizes and experimental sources.
}
\end{figure}
 
A small contribution of events with fake muons from misidentified jets is expected in the $W_{\mu\nu}$ control region.
This contribution is estimated similarly to the fake-electron contribution in the $W_{e\nu}$ CR discussed above.
Since events with fake muons tend to contain a significant amount of \MET, the variable $S_\text{MET}$ is not well suited to increasing the contribution of events with $W$ boson decays.
Instead, events in the single-muon control region ($W_{\mu\nu}$ CR) must have a large transverse mass $\mT=\sqrt{2E^\ell_\text{T}\MET (1-\cos\phi(\ell,\MET))}>20\,\GeV$ for the muon candidate.
Events that fail this additional requirement form the `fake-$\mu$' control region.
The requirement on \mT{} suppresses fake muons, as those are typically aligned with the direction of the missing transverse momentum and consequently exhibit smaller values of \mT.
The contribution of events with fake muons to each bin $i$ of the $W_{\mu\nu}$ CR is given by $R_{M}\cdot n_{\text{fake-$\mu$},i}$.
Here, $n_{\text{fake-$\mu$},i}$ is the excess of data events over the expected prompt background in the corresponding bin of the `fake-$\mu$' CR and $R_M$ is the ratio of the number of events with $\mT{\,>\,}20\,\GeV$ to the number with $\mT{\,<\,}20\,\GeV$.
To determine $R_M$ from the data, another control region ($W_{\mu\nu}$ anti-ID) enriched with fake muons is defined with a muon candidate being present, but not passing the loose identification requirement for `signal muons' (Figure~\ref{fig:fake_lep_b}).
For this method, we assume no correlation of the identification efficiencies of muons and \mT.
As before, the $W$, $Z$, $VV$, and $t\bar{t}$ backgrounds are subtracted from the observed yields in the $W_{\mu\nu}$ anti-ID control region.
No statistically significant dependence of $R_M$ on \mjj{} is observed and its averaged value is $0.29{\,\pm\,}0.15$, where the uncertainty is from the limited number of events in the anti-ID control region.
A comparison of the observed yield in the $W$ CRs and the expectation from simulation is shown in Figure~\ref{fig:wcr_wen_prefit139}.
 
\begin{figure}
\centering
\subfloat[]{
\parbox{0.49\textwidth}{
\centering
\includegraphics[width=0.49\textwidth]{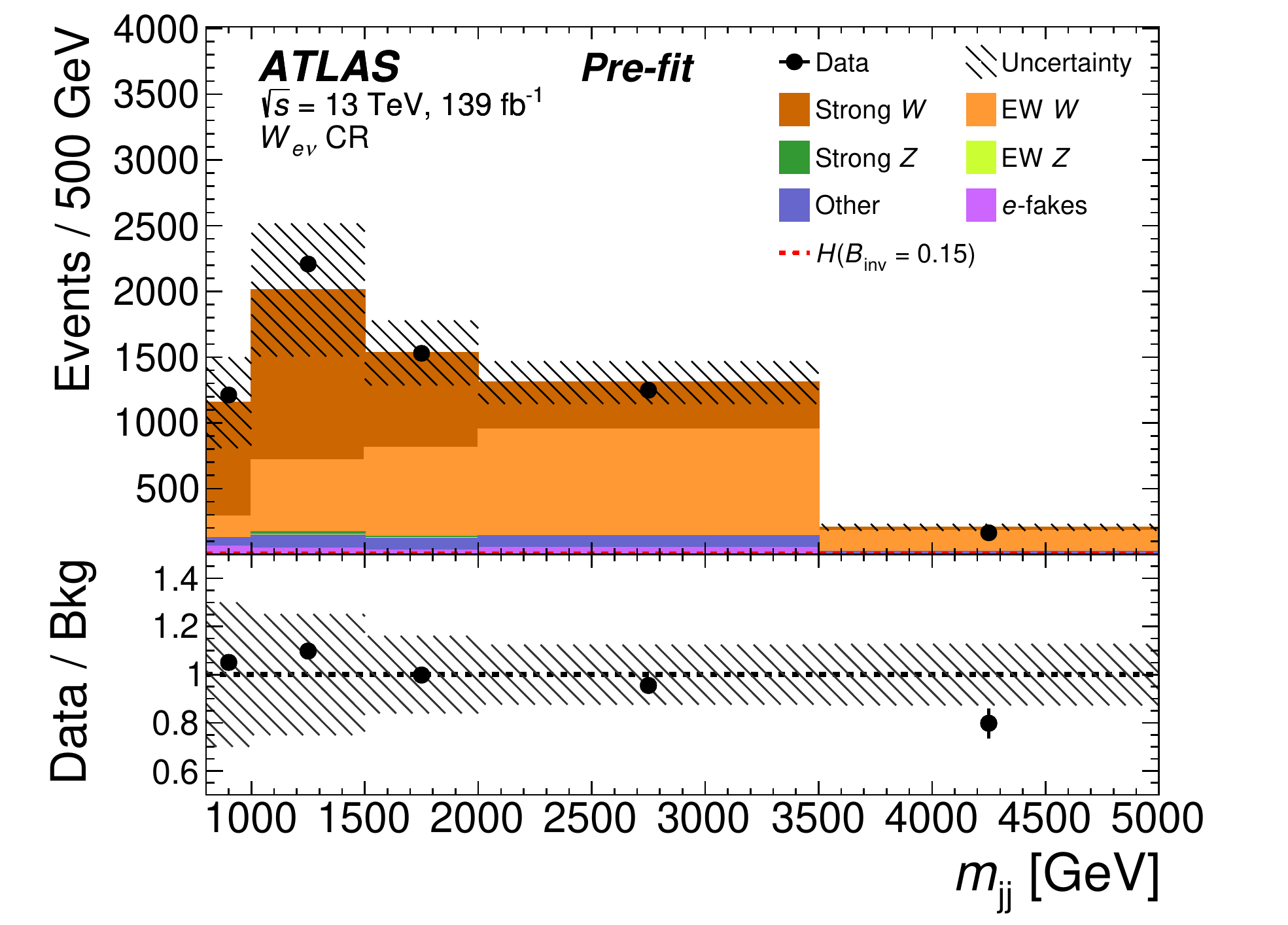}
}
\label{fig:wcr_wen_prefit139_a}
}
\subfloat[]{
\parbox{0.49\textwidth}{
\centering
\includegraphics[width=0.49\textwidth]{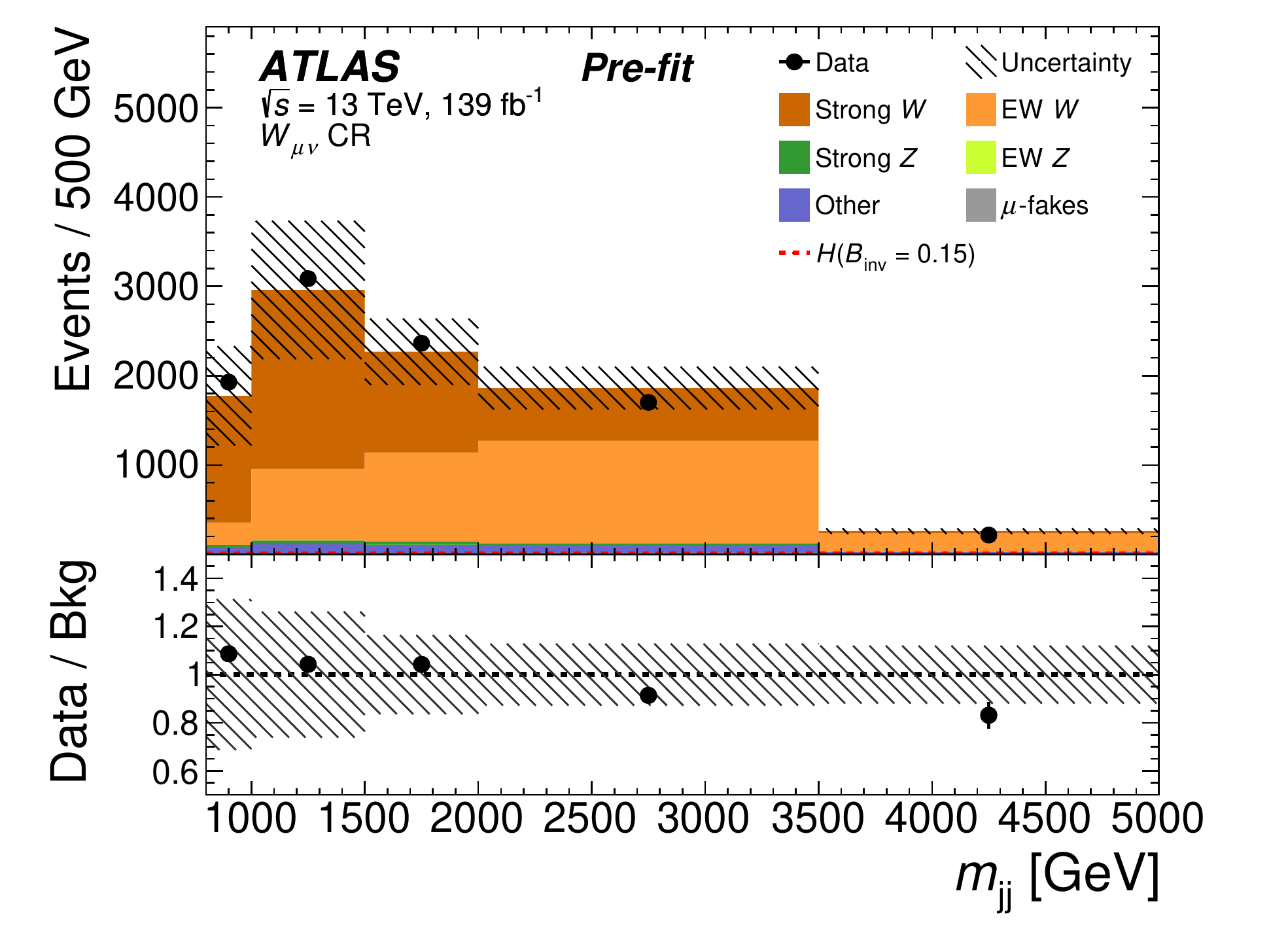}
}
\label{fig:wcr_wen_prefit139_b}
}
 
\subfloat[]{
\parbox{0.49\textwidth}{
\centering
\includegraphics[width=0.49\textwidth]{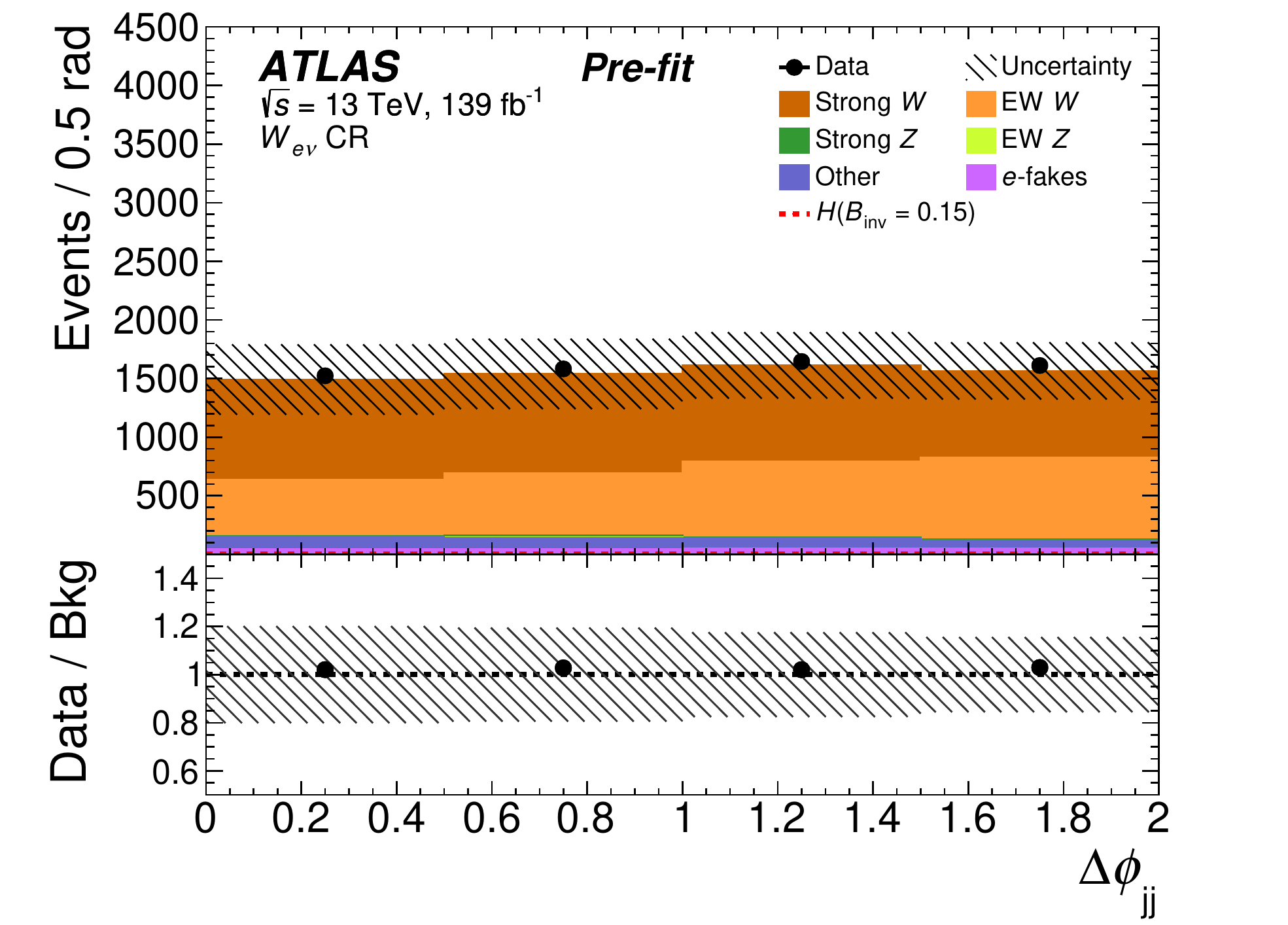}
}
\label{fig:wcr_wen_prefit139_c}
}
\subfloat[]{
\parbox{0.49\textwidth}{
\centering
\includegraphics[width=0.49\textwidth]{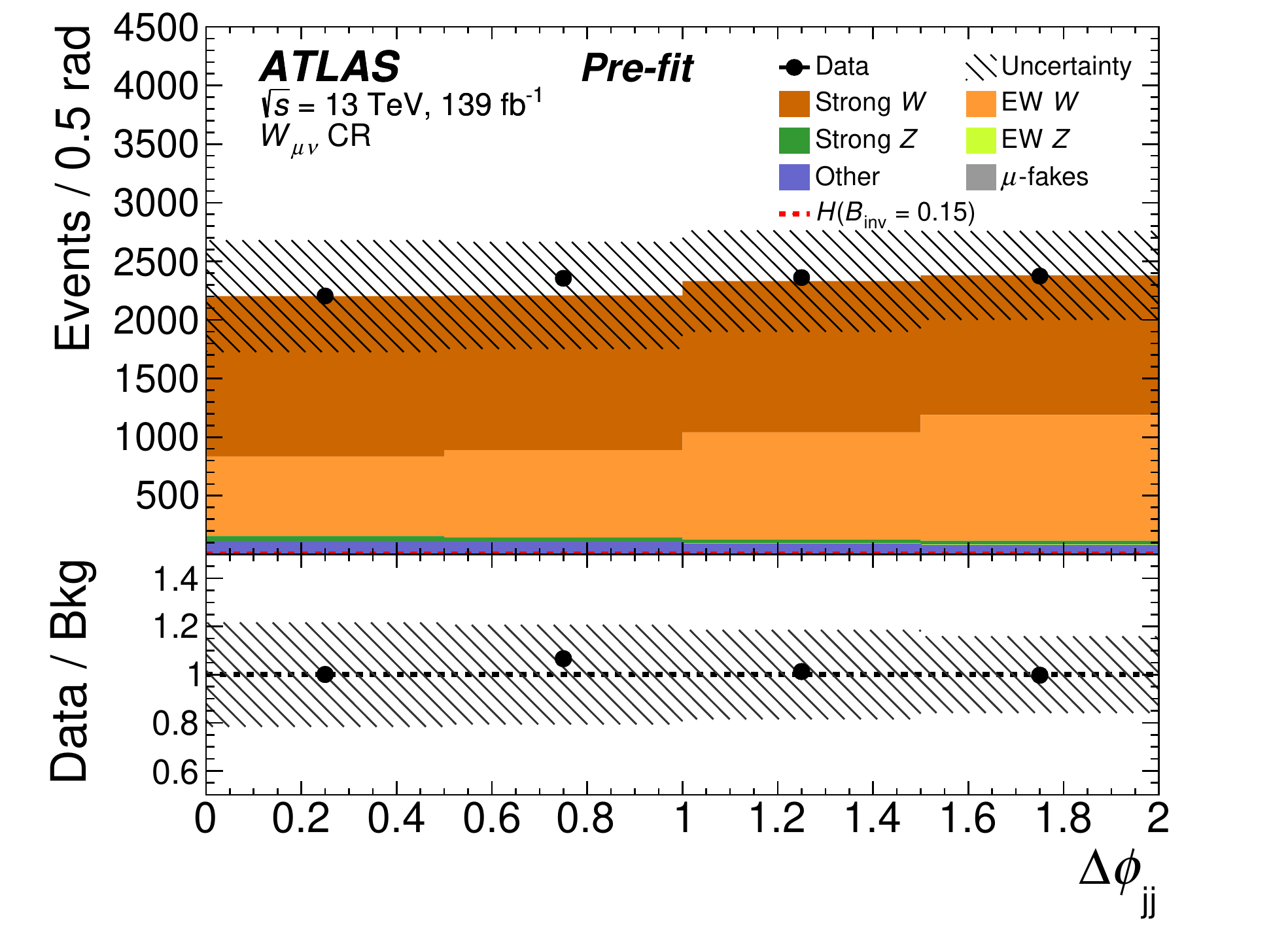}
}
\label{fig:wcr_wen_prefit139_d}
}
 
\subfloat[]{
\parbox{0.49\textwidth}{
\centering
\includegraphics[width=0.49\textwidth]{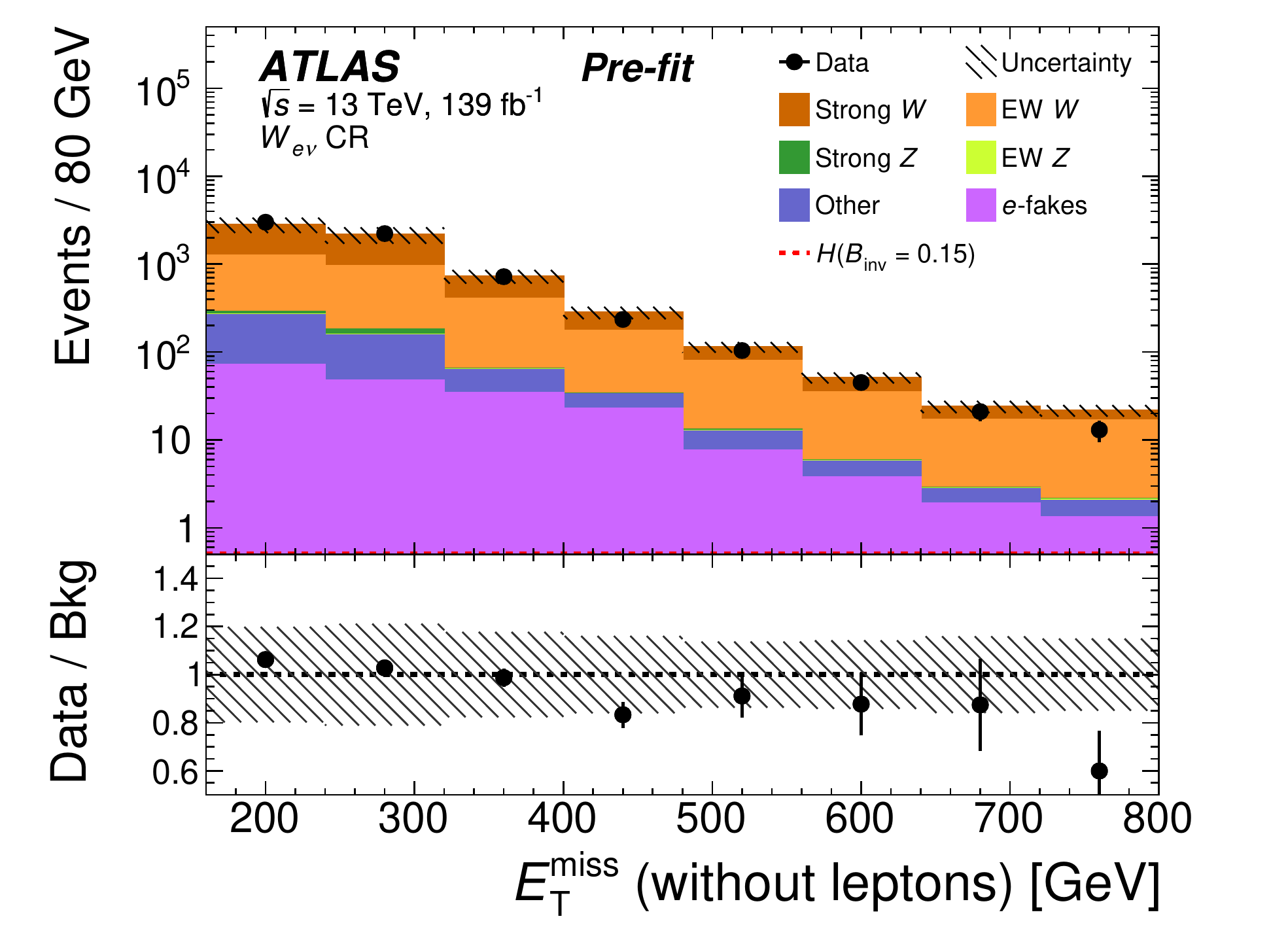}
}
\label{fig:wcr_wen_prefit139_e}
}
\subfloat[]{
\parbox{0.49\textwidth}{
\centering
\includegraphics[width=0.49\textwidth]{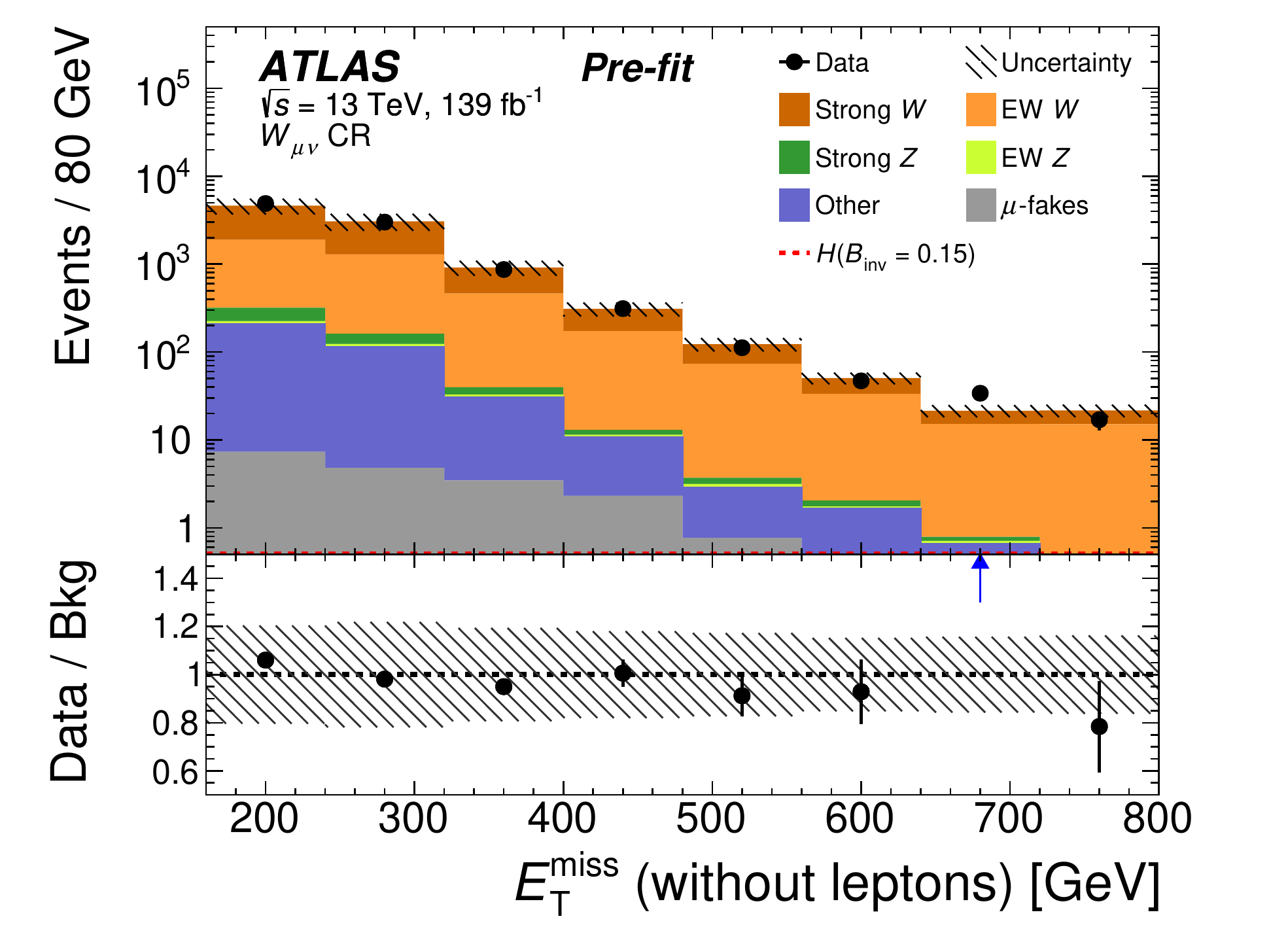}
}
\label{fig:wcr_wen_prefit139_f}
}
\caption{Observed and predicted distributions of $m_\text{jj}$, $\Delta\phi_{\text{jj}}$, and $E_\text{T}^\text{miss}$ (without leptons) in the $W_{e\nu}$ (\protect\subref{fig:wcr_wen_prefit139_a}, \protect\subref{fig:wcr_wen_prefit139_c}, and \protect\subref{fig:wcr_wen_prefit139_e}) and $W_{\mu\nu}$ (\protect\subref{fig:wcr_wen_prefit139_b}, \protect\subref{fig:wcr_wen_prefit139_d}, and \protect\subref{fig:wcr_wen_prefit139_f}) control regions.
The MC predictions are normalised to cross sections times luminosity and are shown prior to the likelihood fit.
The small contribution from fake electrons is the expected contribution after the combined likelihood fit.
Minor contributions from $t\bar{t}$, $VV$, $VVV$, and VBF $H$ with $H\to\tau^+\tau^-$ or $H\to W^\ast W$ are combined and labelled `other'.
The hatched band indicates statistical and experimental systematic uncertainties.
}
\label{fig:wcr_wen_prefit139}
\end{figure}
 
The event yield in the $Z_{\ell\ell}$ CR is significantly lower than that of the \Znn{} background in the SR because of the small branching fraction of \Zll{} and the need to identify two leptons within the detector acceptance.
Therefore, using only the $Z_{\ell\ell}$ CR to constrain the $Z$ background in the SR results in sizeable statistical uncertainties that limit the sensitivity of the analysis.
To improve the \Zjets background estimation, the $W_{\ell\nu}$ CRs are used to determine not only the \Wjets contribution to the SR, but also the \Zjets background.
This is motivated by the similarity of the \Wjets and \Zjets processes: they have identical QCD couplings for the hadronic part of the process, with only small differences in the mass, spin- and flavour structure.
This approach requires an accurate estimate of the ratio \RZW{\mathrm{TH}} of the \Zjets and \Wjets cross sections, which is provided by a dedicated NLO QCD and NLO electroweak calculation in the phase space of a VBF-like signature separately for the electroweak and strong processes~\cite{Lindert:2021xxx}.
The ratio \RZW{\mathrm{TH}} is evaluated as a function of generator-level \mjj{}, and the expectations for \Zjets processes in all CRs and SRs are reweighted by $\RZW{\mathrm{TH}}\,\big/\,\RZW{\textsc{MC}}$ where \RZW{\textsc{MC}} is the $Z$ to $W$ cross-section ratio in the same VBF-like region as is used for the numerator, with `MC' being \SHERPA for strong \Vjets samples and \HERWIG for electroweak \Vjets samples.
The correction to the strong \Zjets prediction is below 5\% across the whole \mjj range (Figure \ref{fig:wz_weight_a}); however, for the electroweak
\Zjets prediction it varies between 20\% at low \mjj values and 5\% at large \mjj values (Figure \ref{fig:wz_weight_b}).
The main reasons for these significant corrections at low \mjj are missing NLO $s$-channel contributions~\cite{Baglio:2011juf} and missing interference between $s$- and $t$-channel processes in the simulated event samples.
 
\begin{figure}
\centering
\subfloat[]{
\parbox{0.49\textwidth}{
\centering
\includegraphics[width=0.49\textwidth]{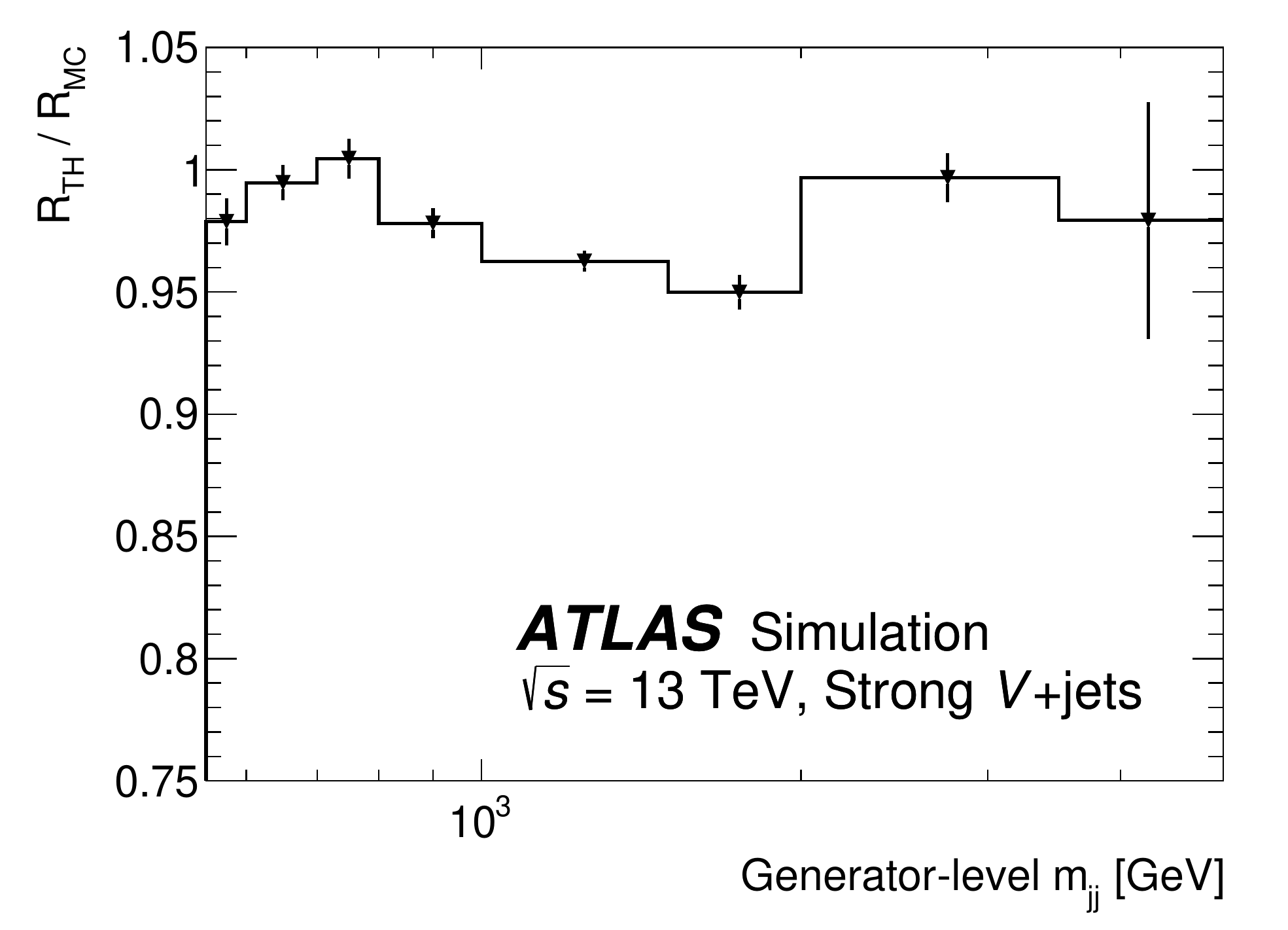}
}
\label{fig:wz_weight_a}
}
\subfloat[]{
\parbox{0.49\textwidth}{
\centering
\includegraphics[width=0.49\textwidth]{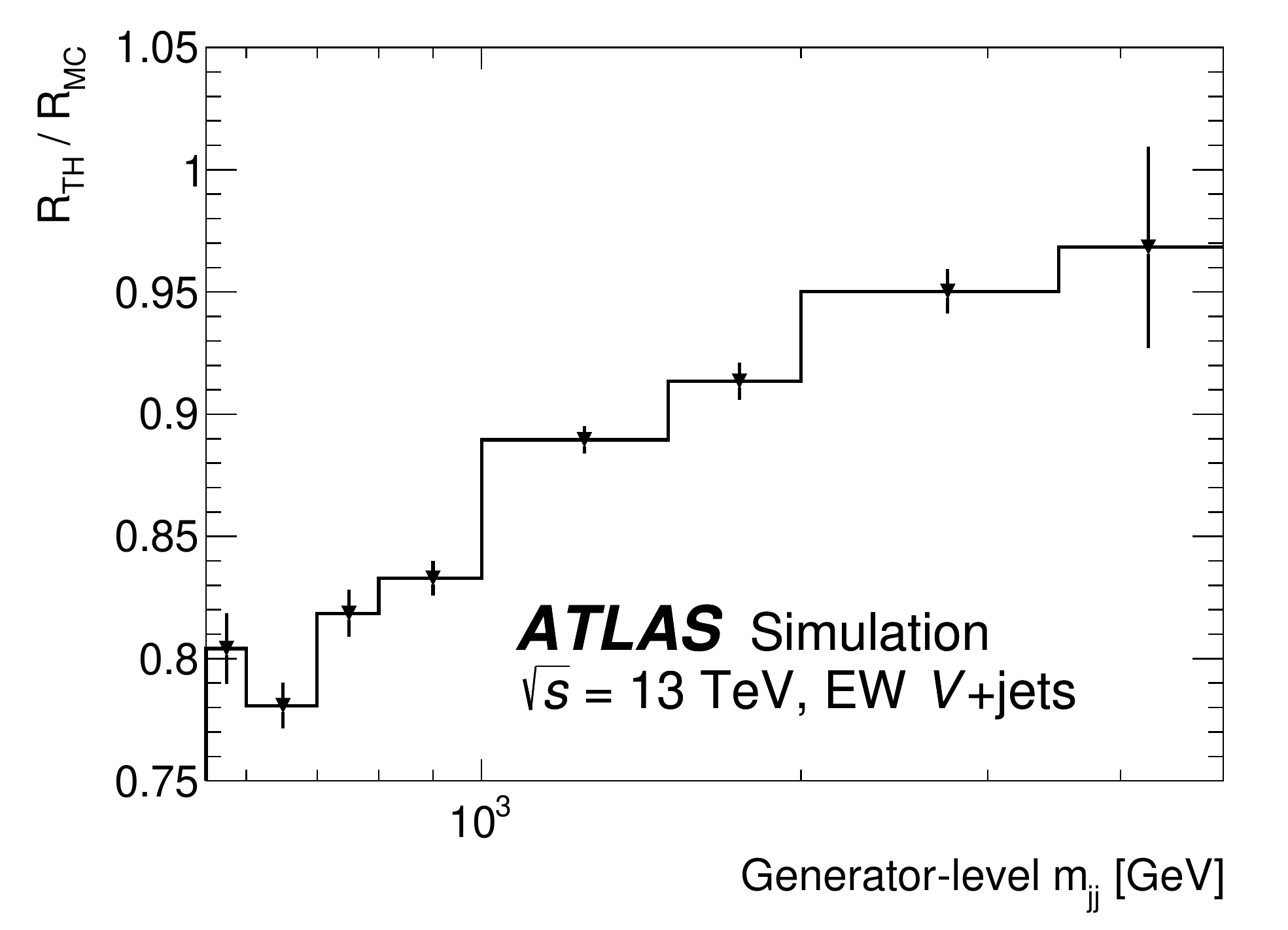}
}
\label{fig:wz_weight_b}
}
\caption{The double ratio $\mathcal{R}^{Z/W}_{\text{TH}}{\,\big/\,}\mathcal{R}^{Z/W}_{\text{MC}}$ as a function of generator-level $m_\text{jj}$, which is used to correct the \protect\subref{fig:wz_weight_a} strong and \protect\subref{fig:wz_weight_b} electroweak $V\text{+\,jets}$ simulation to achieve state-of-the art precision for the ratio of $Z\text{+\,jets}$ to $W\text{+\,jets}$ cross sections in the phase-space of a VBF-like signature as detailed in Reference ~\cite{Lindert:2021xxx}, which is similar to the kinematic region targeted by the analysis.
The error bars show the statistical uncertainty of the simulation.}
\end{figure}

\subsection{Multijet background}
\label{sec:MJ}
 
Events from multijet processes have momentum balance in the transverse plane at generator level, but can have some \MET due to neutrinos produced in subsequent decays of $c$- and $b$-hadrons.
At reconstruction level, significant \MET can occasionally be produced by experimental effects, primarily fluctuations in the detector's response to jets, quantified by the jet energy resolution (JER), but also effects such as very energetic jets not having their energy fully deposited within the calorimeters (`punch-through')
and misclassification of jets as originating from the hard-scatter interaction or a pile-up interaction.
The signal selection is designed to heavily suppress such events through the requirement of large \MET and small \dphijj.
However, in the presence of high pile-up activity a combination of jet mismeasurement and mistagging of pile-up jets can produce signal-like topologies.
In such cases, it is possible that the two highest-\pt{} jets in the event originate from different scattering processes and thus are uncorrelated and can have a small \dphijj separation. Furthermore, each of the two jets would have its own hadronic recoil system.
In most of these cases, the hadronic recoil to the selected jets will result in additional reconstructed jets, and the event will not pass the SR selection.
However, some or all of these additional jets might be lost, either because their reconstructed \pt{} is below the jet \pt{} requirement or because they are misclassified by the pile-up tagging.
The latter effect is a source of \MET that is of particular interest for the investigated signature since one or both of the leading jets are often located outside the tracker acceptance, where the pile-up tagging is less efficient.

The multijet background is estimated using two independent methods:
a modified version of the `Rebalance and Smear' (R+S) technique used in previous analyses~\cite{CMS-SUS-10-005, EXOT-2016-37} and a novel data-driven approach, called the `pile-up-CR method', using multijet-enriched control regions.
These methods use different approaches and independent data sets to obtain predictions of the multijet background in each SR bin with associated uncertainties.
As further detailed below, the final multijet estimate is obtained using either the more precise method, with the other prediction providing validation, or in some parts of the SR a combination of the estimates from the two methods.
An overview of all control regions used for the estimation and validation of the multijet background is presented in Table~\ref{tab:MJCRs}.
 
\begin{table}
\centering
\caption{Control regions used for the multijet background estimation. Only changes relative to the SR selection are given. The third column shows if a the control region is binned in $\Delta\phi_{\text{jj}}$ and $m_\text{jj}$ like the signal region. Except for the MJ CR, which uses single-jet-triggered data, all control regions use $E_\text{T}^\text{miss}$-triggered data.}
\label{tab:MJCRs}
\resizebox{\textwidth}{!}{
\begin{tabular}{cccl}
\toprule
Region & Definition & Binned & Purpose \\
\midrule
MJ CR & $100\,\GeV<\MET<200\,\GeV$ and $\mjj>400\,\GeV$ & no & Determine HS-only and HS+PU \\
& $N_\text{jet}=2$, $\detajj>2.5$ & & composition\\
& no requirement on \MHT & & \\
\midrule
low-\mjj CR & $150\,\GeV<\MET<200\,\GeV$  & no & Multijet (R+S) normalisation \\
& $2\leq N_\text{jet} \leq 4$, $200\,\GeV<\mjj<800\,\GeV$ & & Estimate non-closure uncertainty \\
& $\detajj>2.5$, ($|\eta^\text{j1}|>2.4$ or $|\eta^\text{j2}|>2.4$)& \\
& $\dphijj<1$, no requirement on \MHT & & \\
\midrule
mid-\mjj CR & $160\,\GeV<\MET<200\,\GeV$ & no & Same as low-\mjj CR\\
& $N_\text{jet}=2$, $800\,\GeV<\mjj<1500\,\GeV$ & & \\
& $\dphijj<1$, no requirement on \MHT & & \\
\midrule
pile-up CR & |$\eta^{\text{j}1}|>2.4$ and $\text{j}1$ is PU tagged  & yes & Scaled by $R_\text{MJ}$ to \\
& no pile-up-tagging requirement for $\text{j}2$ & & obtain multijet yield in SR \\
& $2\leq N_\text{jet} \leq 4$ & &  \\
& no $m^\text{rel}_i$ requirement for $1<\dphijj<2$  & & \\
\midrule
low-MET CR & $100\,\GeV<\MET<160\,\GeV$ & yes & Numerator of $R_\text{MJ}$ \\
\midrule
low-MET PU-CR & same as pile-up CR, but & yes & Denominator of $R_\text{MJ}$ \\
& $100\,\GeV<\MET<160\,\GeV$ & & \\
\midrule
low-MET VR & $110\,\GeV<\MET<150\,\GeV$ & no & Performance check of \\
& loose fJVT requirement & & pile-up CR method \\
& no requirement on \MHT & & \\
\bottomrule
\end{tabular}
}
\end{table}
 
The R+S method has two distinct steps.
First, a multijet event sample is constructed in which each event exhibits momentum balance in the transverse plane and hence approximates the particle level.
Next, the detector response and event reconstruction is emulated multiple times per event, which will occasionally produce events that fulfil the SR requirements.
The R+S method is performed twice using different input events, namely single-jet-triggered data events and simulated multijet events.
These approaches make different assumptions and result in two separate predictions.
Each jet is classified as either a hard-scatter (HS) jet or a pile-up (PU) jet.
When simulated events are used, a jet is identified as a HS jet if it is matched with $\DeltaR{\,<\,}0.1$ to a generator-level jet from the hard-scatter process, while jets that are not matched are classified as PU jets.
Events are only considered further if the transverse momentum imbalance calculated from all matched generator-level jets is less that $35\,\GeV{}$.
This ensures that all significant HS jets have been identified.
The subset of events with two or more HS jets but no PU jet with $\pT{\,>\,}50\,\GeV$ defines a HS-only sample, while the HS+PU sample contains events with two or more HS jets and at least one PU jet with $\pT{\,>\,}50\,\GeV$.
When data events are used, HS jets cannot be identified by a $\Delta R$ matching to generator-level jets.
Instead, information from the pile-up tagging is used.
Those with no pile-up-tagged jets with $\pT{\,>\,}25\,\GeV$ define the HS-only sample.
PU jets stemming from the same scattering process are identified by requiring two pile-up-tagged jets with $\pT{\,>\,}25\,\GeV$, which also fulfil the requirements that their average \pT is above $40~\GeV{}$ and that they are back-to-back ($\dphijj{\,>\,}2.7$).
Events with exactly two such jets define the HS+PU sample.
If simulated events are used as input, the advantage is a clean separation of the HS and PU jets by using available generator-level information.
If data events are used as input, the method relies on pile-up tagging to identify the PU jets, which is less efficient.
But in the latter case, the method takes the pile-up contribution from data, which is expected to be more accurate than if it were taken from simulation.
 
The HS jets are `rebalanced', i.e.\ the momenta of the jets are adjusted within their experimental resolution such that no transverse momentum imbalance is present.
If simulated seed events are used, PU jets remain unchanged in this procedure, but if seed events from data are used the two back-to-back PU jets are rebalanced as well.
The rebalanced jets are then `smeared' according to the expected jet response function, which is evaluated using simulated events and parameterised as a function of jet energy and $\eta$ separately for light- and heavy-flavour jets.
Following the smearing step, the pile-up jet tagging (JVT and fJVT selection) is re-evaluated since the selection efficiency depends on the jet momentum, which is altered during the smearing.
To account for large event weights stemming from huge cross sections for some simulated processes or large prescale factors for single-jet triggers, the events are smeared up to $8000$ times with accordingly reduced event weights.
Since significant fluctuations of the jet response are the primary source of \MET in multijet events, the tails of the jet response function are sampled with increased probability for the three leading jets.
To further increase the size of the HS+PU sample, the HS and PU components of each event are rotated relative to each other by a random azimuthal angle multiple times.
Following this procedure, the event selection is applied to populate the SR and CR bins.
 
Independently of how the HS-only and the HS+PU samples are defined, the two multijet background components are normalised by a fit to the \dphijj distribution in a loose multijet control region using single-jet-triggered data (MJ CR, defined in Table \ref{tab:MJCRs}).
The \dphijj distribution is well suited for this, since HS-only topologies are expected to accumulate at large \dphijj, while at low \dphijj the HS+PU topology should be dominant.
The shapes of the \mjj and \dphijj distributions after the normalisation are shown in Figure~\ref{fig:RnS_preselTJV} for the case of the HS-only and HS+PU samples being derived from simulated events.
 
\begin{figure}
\centering
\subfloat[]{
\parbox{0.49\textwidth}{
\centering
\includegraphics[width=0.49\textwidth]{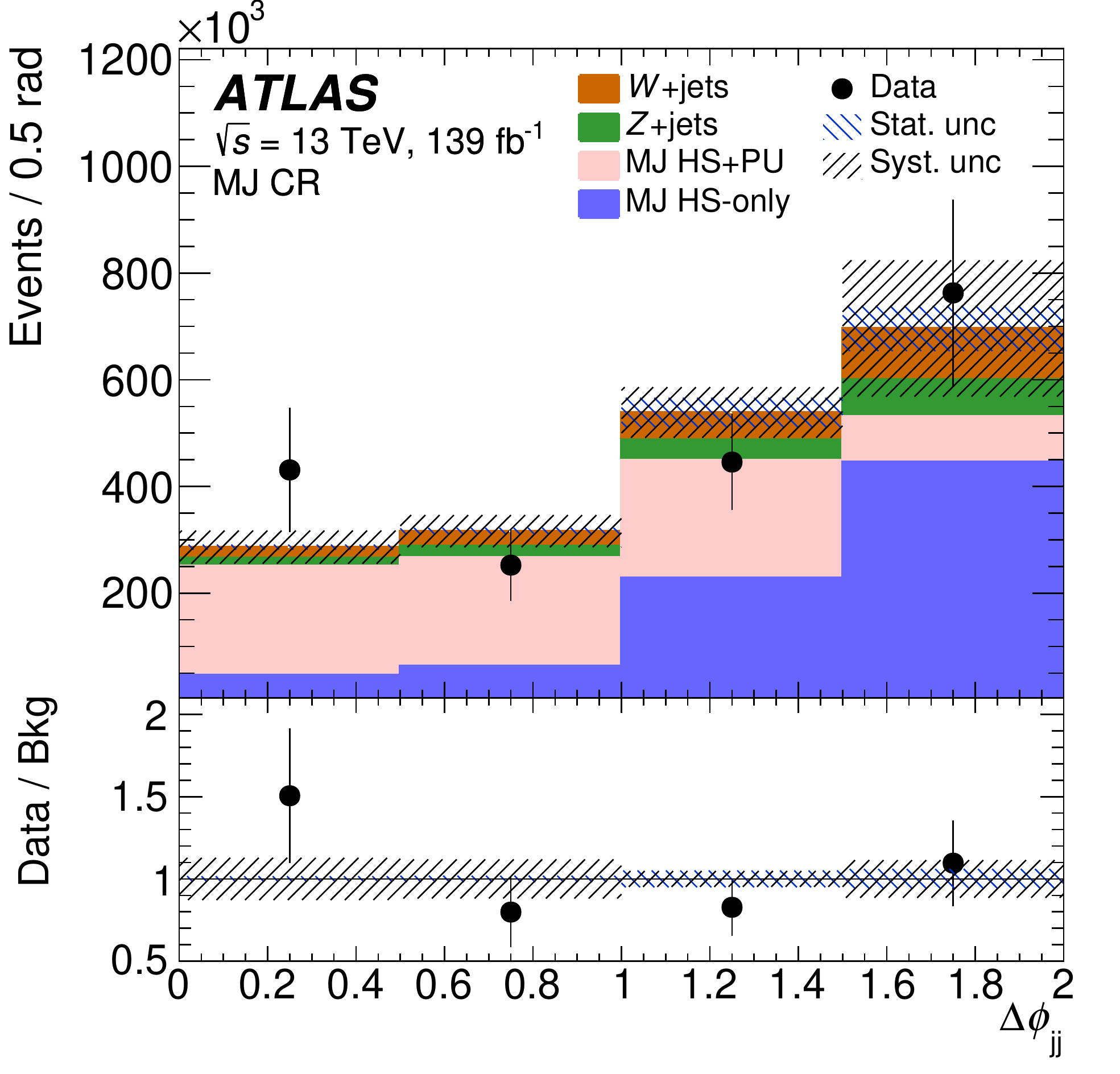}
}
\label{fig:RnS_preselTJV_a}
}
\subfloat[]{
\parbox{0.49\textwidth}{
\centering
\includegraphics[width=0.49\textwidth]{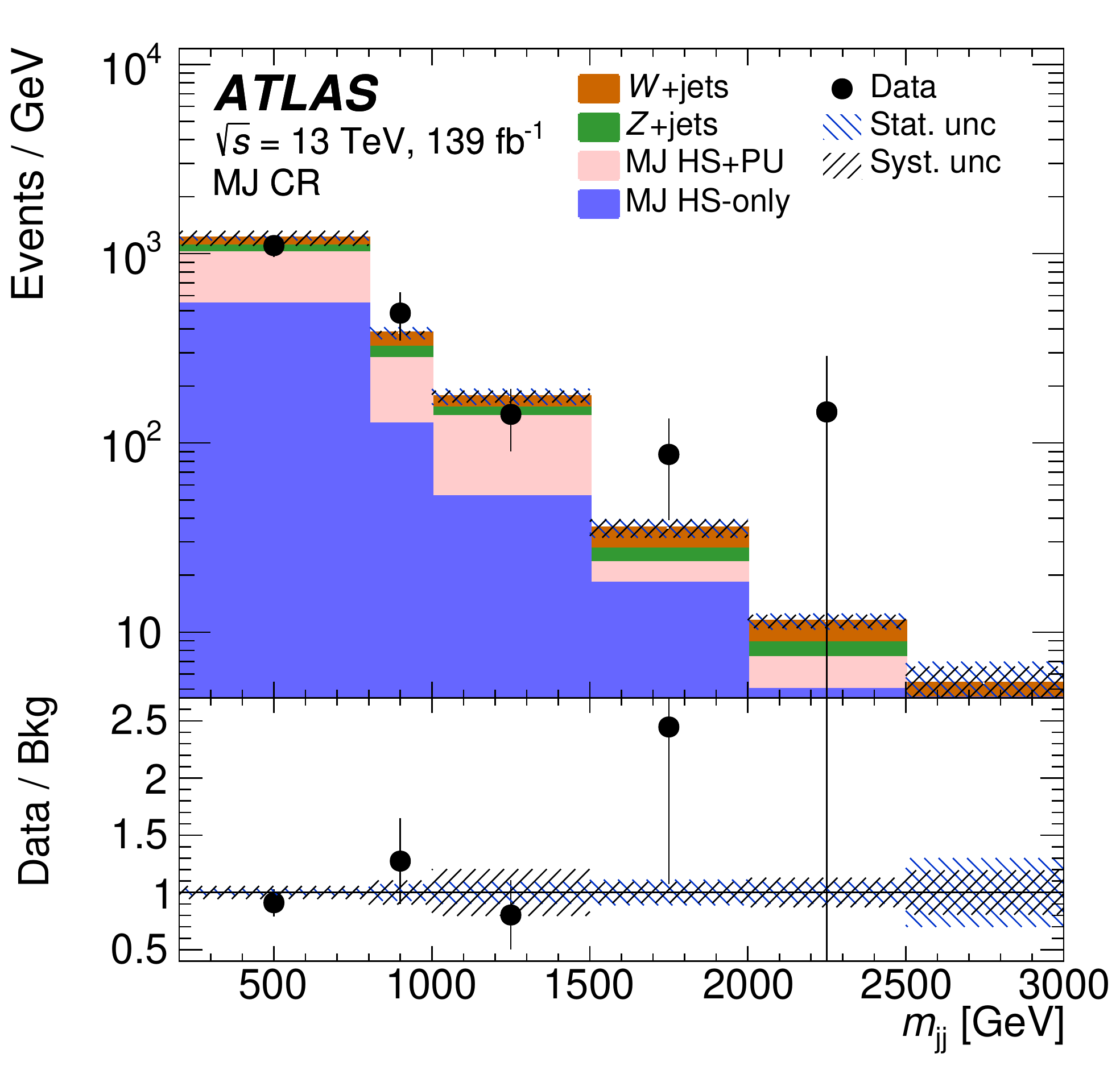}
}
\label{fig:RnS_preselTJV_b}
}
\caption{Distributions of \protect\subref{fig:RnS_preselTJV_a} $\Delta\phi_{\text{jj}}$ and \protect\subref{fig:RnS_preselTJV_b} $m_\text{jj}$ are shown in the MJ CR.
The shapes of the multijet background components (HS-only and PU+HS) are estimated by the R+S technique using simulated events as input.
The components are normalised using a fit to the observed $\Delta\phi_{\text{jj}}$ distribution \protect\subref{fig:RnS_preselTJV_a}.
The hatched bands indicate the statistical and the systematic uncertainties separately.}
\label{fig:RnS_preselTJV}
\end{figure}
 
The final step in the multijet background estimation is to apply a correction to account for the inefficiency of the \MET triggers used to collect the data in the SR bins.
For \MET values above 160\,\GeV, the \MET triggers are almost fully efficient and the corrections are small compared with the other uncertainties, which are summarised later.
The multijet background is estimated separately for three data-taking periods (2015--2016, 2017, and 2018) to account for different pile-up, beam, and trigger conditions.
 
The multijet prediction from the R+S method is tested in two dedicated control regions.
The first (low-\mjj CR) is defined at low \mjj{} ($200\,\GeV{\,<\,}\mjj{\,<\,}800\,\GeV$) for \MET values between 150 and 200~\GeV{}.
The selection criterion for \detajj{} is relaxed to $\detajj{\,>\,}2.5$ but one of the two leading jets is required to be forward with $|\eta|{\,>\,}2.4$.
Another control region (mid-\mjj CR) is defined at intermediate \mjj ($800\,\GeV{\,<\,}\mjj{\,<\,}1500\,\GeV$) for \MET values between 160 and 200~\GeV{} and $N_\text{jet}{\,=\,}2$.
Both control regions have no requirement on \MHT and furthermore the events have to fulfil $\dphijj{\,<\,}1$ to increase each sample's purity in multijet events.
The remaining selection requirements are the same as for the SR, i.e.\ the used data are collected by \MET triggers in contrast to the MJ CR, which uses single-jet triggers.
For each of the two control regions, the multijet prediction is normalised to the observed data.
The average normalisation factor from the two multijet control regions is used for the nominal multijet prediction, while half of the difference between the two normalisation factors is assigned as a `non-closure' uncertainty of the method.
 
Further systematic uncertainties in the R+S multijet prediction arise from three sources.
The smearing is repeated with systematically varied jet energy response distributions to account for potential differences between data and simulation.
Two uncertainty variations of the jet energy response are considered: a 15\% widening of its `core' part, which is modelled by a Gaussian distribution, and a 50\% increase of the non-Gaussian tails.
These uncertainties are motivated by finely binned data--MC comparisons of the widths and the non-Gaussian tails of the dijet \pT asymmetry, which is closely related to the calorimeter's response to jets~\cite{JETM-2018-05}.
The differences from the nominal prediction are used as an additional systematic uncertainty.
The prediction of the multijet background is made with seed events from simulation and from data, and the resulting difference is used as a further systematic uncertainty.
Table~\ref{tab:MJSummary} summarises the inclusive multijet predictions and the assigned systematic uncertainties.
 
An independent and novel approach used to predict the multijet background is the pile-up-CR method.
A control region enriched with multijet events (PU-CR) is defined by inverting the fJVT criterion for the leading jet, i.e.\ by requiring that the leading jet in $\pT$ is identified as a forward pile-up jet.
No requirement is made on the fJVT value of the sub-leading jet in \pT.
This ensures that the PU-CR contains events with both leading jets originating from different interactions, as well as events in which they do originate from the same scattering process.
This control region's purity in multijet events is around 95\% for \MET larger than 200\,\GeV\ and greater than 92\% for \MET{} between 160 and 200\,\GeV.
The pile-up CR is binned in \dphijj, \mjj, and \MET{} bins corresponding to the signal region bins.
The event yield $N_{i}^\text{PU-CR}$ in the pile-up CR bin $i$ is scaled to the expected multijet background in the corresponding SR bin by a normalisation factor $R_{\text{MJ},i}$, which is measured at lower \MET{} where the multijet background is a significant contribution.
Because forward pile-up jets are identified only up to \pT values of $120\,\GeV{}$, the yields in the pile-up CRs are biased to lower values in \MET{} compared with the SR if only events with two jets are used.
To ensure that the normalisation factor is valid over the considered range in \MET{} the event selection for the pile-up CR allows up to four jets, following the jet requirements on $C_i$ and $m^\text{rel}_{i}$ for $i{\,=\,}3$ or 4 as defined in Section~\ref{sec:defSR}.
Because not many events contribute to the pile-up CR bins for $1{\,<\,}\dphijj{\,<\,}2$ the requirement on $m^\text{rel}_{i}$ for a possible third or fourth jet is dropped.
The normalisation factor $R_{\text{MJ},i}$ is calculated from the event yields $N_{i}^{\text{low-MET}}$ and $N_{i}^\text{PU-CR,low-MET}$ at low \MET between 100\,\GeV{} and 160\,\GeV{} with the remaining selection criteria being identical to the SR and pile-up CR bin definitions, respectively:
\begin{equation}\nonumber
R_{\text{MJ},i}=\frac{ N_{i}^{\text{low-MET}} - B_{\text{non-MJ},i}^\text{low-MET}}{ N_{i}^\text{PU-CR,low-MET} - B_{\text{non-MJ},i}^\text{PU-CR,low-MET} }.
\end{equation}
Here, $B_{\text{non-MJ},i}^\text{low-MET}$ and $B_{\text{non-MJ},i}^\text{PU-CR,low-MET}$ are the contributions from non-multijet processes to the corresponding bins.
The normalisation factor is then applied to the event yield in the pile-up CR after subtraction of the non-multijet backgrounds to obtain a prediction of the multijet yield in the SR:
\begin{equation}\nonumber
B^\text{SR}_{\text{MJ},i}=R_{\text{MJ},i} \cdot \left( N_{i}^\text{PU-CR} - B_{\text{non-MJ},i}^\text{PU-CR} \right),
\end{equation}
with the non-multijet backgrounds $B_{\text{non-MJ},i}^\text{PU-CR}$ taken from simulation but normalised to the data in appropriate control regions that have no requirement on the fJVT value of the leading jet.
For SR bins 14 to 16 ($160\,\GeV{\,<\,}\MET{}{\,<\,}200\,\GeV{}$), the ratio $R_{\text{MJ},i}$ and the pile-up CRs use the tighter fJVT criterion than for the remaining SR bins as mentioned in Section~\ref{sec:objects}.
The stability of $R_{\text{MJ},i}$ as a function of \MET was checked in bins of \MET with a width of $10\,\GeV$ between values of $100\,\GeV$ and $160\,\GeV$ for the SR bins with $160\,\GeV{\,<\,}\MET{\,<\,}200\,\GeV$ and between values of $100\,\GeV$ and $200\,\GeV$ for SR bins with $\MET{\,>\,}200\,\GeV$.
To increase the statistical precision of this test, the bins 1--5, 6--10 and 14--16 are merged.
No statistically significant dependence on \MET is observed and an uncertainty of 20\% is assigned for this effect, which corresponds to the statistical precision of the test.
For the individual bins, the statistical uncertainty of the measured $R_{\text{MJ},i}$ is added in quadrature to this value, resulting in an uncertainty of
20\% to 27\% for the normalisation factor.
The method is validated by using a selection enriched in multijet events (low-MET VR) with a reduced \MET requirement ($110\,\GeV{\,<\,}\MET{\,<\,}150\,\GeV$), no requirement on \MHT and the same loose fJVT requirement as used for the SR bins with $\MET{\,>\,}200\,\GeV$ (Figure \ref{fig:FJVT_DataClosure}).
The contributions from \Vjets processes are normalised in appropriate CRs and the data agrees with the background prediction within the uncertainties.
 
\begin{figure}
\centering
\subfloat[]{
\parbox{0.49\textwidth}{
\centering
\includegraphics[width=0.49\textwidth]{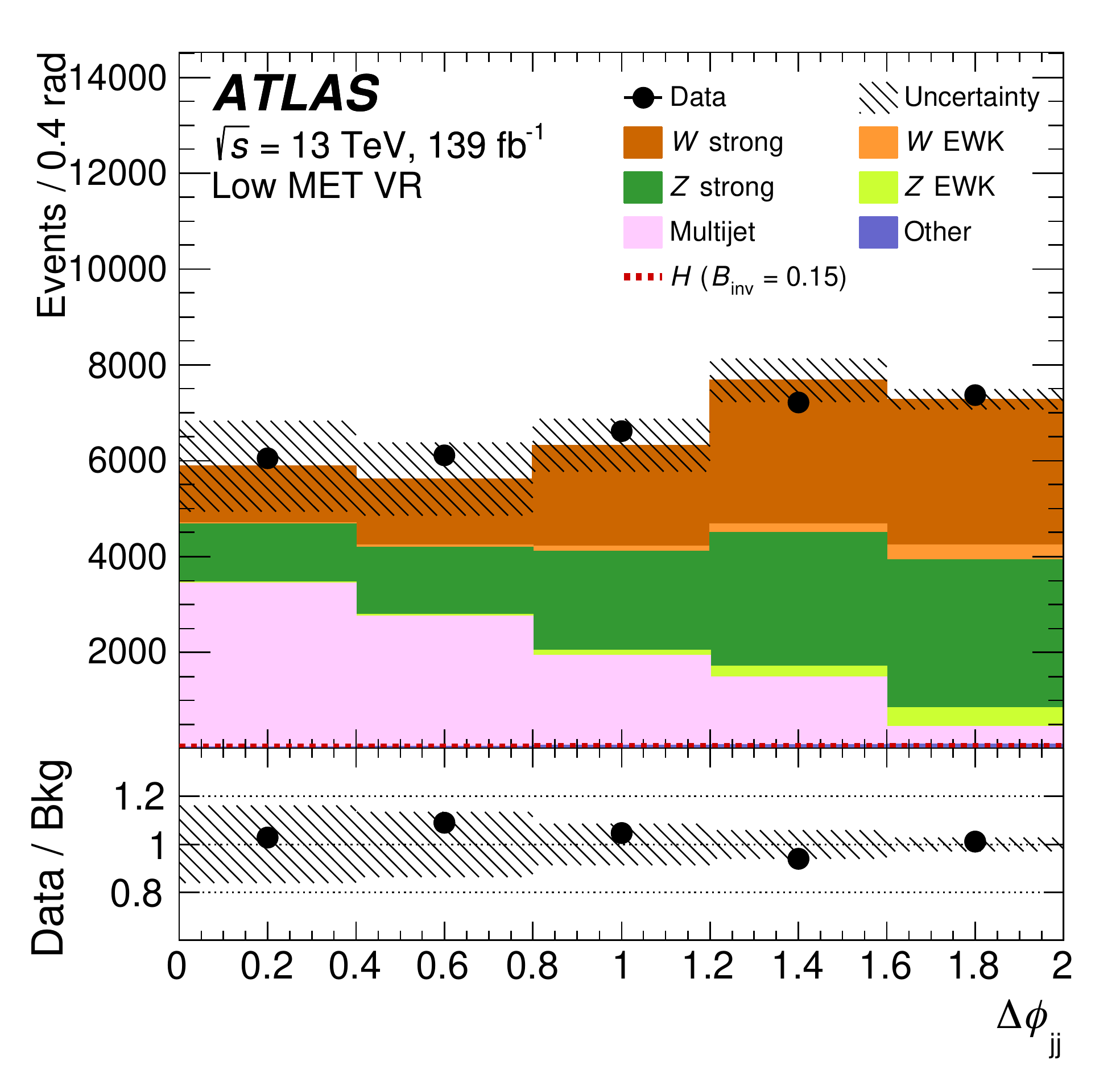}
}
\label{fig:FJVT_DataClosure_a}
}
\subfloat[]{
\parbox{0.49\textwidth}{
\centering
\includegraphics[width=0.49\textwidth]{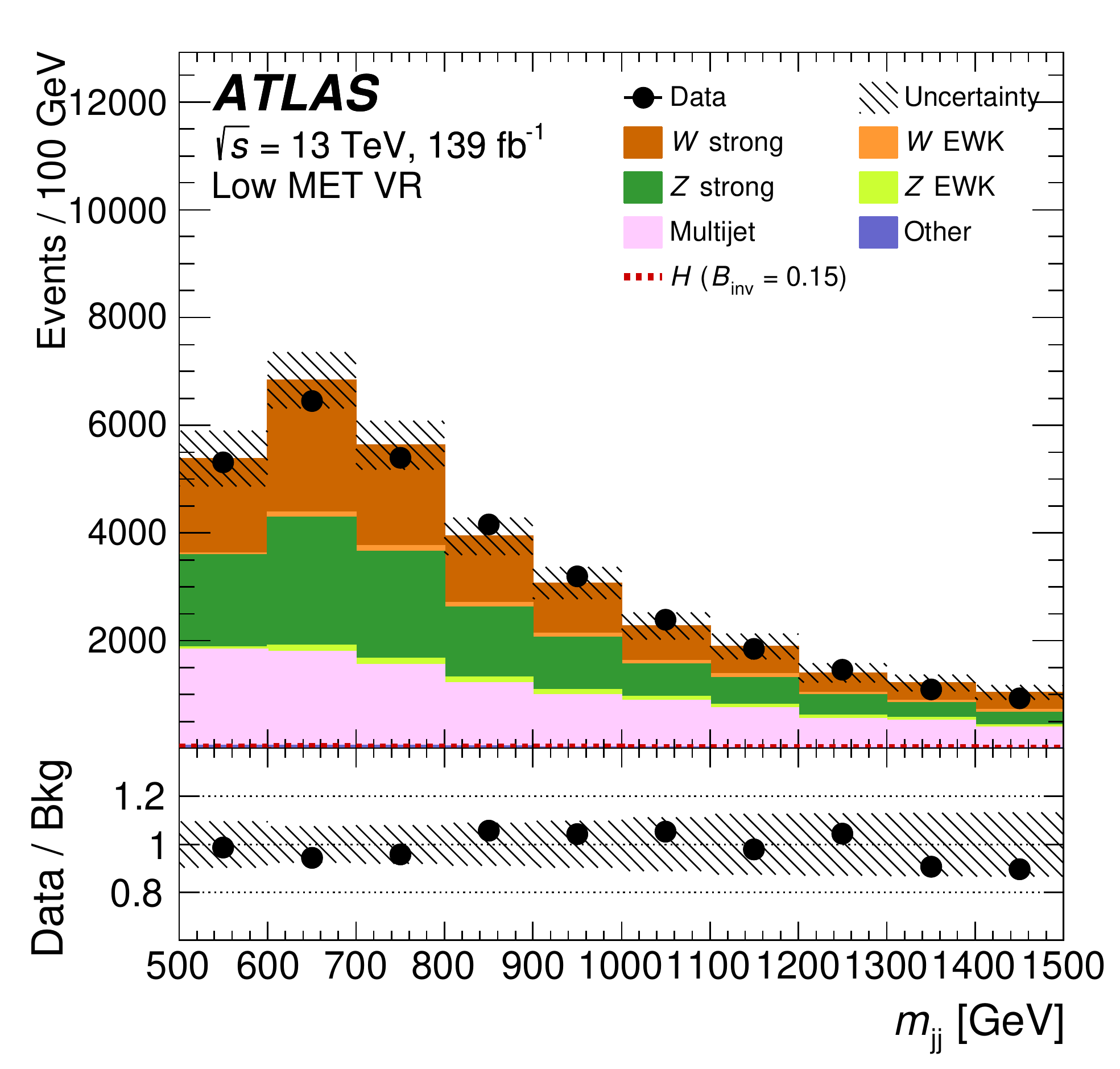}
}
\label{fig:FJVT_DataClosure_b}
}
\caption{The distribution of \protect\subref{fig:FJVT_DataClosure_a} $\Delta\phi_{\text{jj}}$ and \protect\subref{fig:FJVT_DataClosure_b} $m_\text{jj}$ in the low-$E_\text{T}^\text{miss}$ validation region.
The multijet background is predicted by the pile-up-CR method.
Minor contributions from $t\bar{t}$, $VV$, $VVV$, and VBF $H$ with $H\to\tau^+\tau^-$ or $H\to W^\ast W$ are combined and labelled `other'.
The hatched band indicates the statistical and systematic uncertainties of the backgrounds.
}
\label{fig:FJVT_DataClosure}
\end{figure}
 
\begin{table}
\centering
\caption{Predicted multijet background, before the likelihood fit, entering the signal region bins with $N_\text{jet}{\,=\,}2$ with associated uncertainties from the Rebalance and Smear method and the pile-up-CR method. The quoted uncertainty in the inclusive prediction is estimated by summing uncertainties from all sources in quadrature over the SR bins. The individual sources of uncertainty are given for the inclusive SR but also as a range for the individual SR bins. The range for the statistical uncertainties of the pile-up-CR method are only shown for bins 1 to 5; for bin 6 to 10 no bin-by-bin uncertainties are evaluated since those are merged, and for bins 14 to 16 the statistical uncertainties are negligible.}
\label{tab:MJSummary}
\begin{tabular}{ccc|ccc}
\toprule
& \multicolumn{2}{c}{Rebalance and Smear} & & \multicolumn{2}{c}{Pile-up-CR method} \\
Multijet pred. (SR incl.) & \multicolumn{2}{c}{$840\pm 367$} & & \multicolumn{2}{c}{$761\pm 186$}\\
\midrule
\multicolumn{6}{c}{Uncertainties [\%]} \\
Source & inclusive & per bin & Source & inclusive & per bin\\
\midrule
Statistical & 9 & 6--46 & Stat. & ~~~8.6 & 16--49 (only bins 1--5)\\
Non-closure & 40~~ & 37--43~~ & Syst. & 20~~ & 20--27 \\
Jet energy response (core) & ~~~4.8 & 1--41 & & &\\
Jet energy response (tail) & ~~~5.7 & 1--53 & & &\\
Data vs simulation & ~~~4.9 & 2--30 & & &\\
\bottomrule
\end{tabular}
\end{table}
 
The predicted yields and their uncertainties are summarised in Table~\ref{tab:MJSummary} for SR bins with $N_\text{jet}{\,=\,}2$.
The SR bins with $3{\,\le\,}N_\text{jet}{\,\le\,}4$ are excluded from the comparison, because for those no predictions from the pile-up-CR method are available.
The multijet background predictions from the two methods for the inclusive signal region agree within their uncertainties.
Since for the SR bins with $\dphijj{}<1$ and the bins with $160\,\GeV{\,<\,}\MET{}{\,<\,}200\,\GeV{}$ the uncertainties in the expected yields from the pile-up-CR method are smaller than those from the R+S method, the former are used.
Because not many events contribute to the pile-up CR bins for $1{\,<\,}\dphijj{\,<\,}2$ associated with the SR bins 6--10, the prediction with the pile-up-CR method is made inclusively for these bins and the result is then distributed over the individual bins in accord with predictions from the R+S method.
For the SR bins with $3{\,\le\,}N_\text{jet}{\,\le\,}4$ the multijet prediction from the R+S approach is adopted in order to avoid using the same pile-up-CR bins as used for SR bins with $N_\text{jet}{\,=\,}2$.
% End of text imported from the .//sections/backgrounds.tex input file
 
\section{Systematic uncertainties}
\label{sec:systematics}
% The next lines are included from the .//sections/uncertainties.tex input file
\subsection{Theoretical uncertainties}
 
The theoretical uncertainties affect both the signal and background processes and originate from the limited order at which the matrix elements are calculated, the matching of those calculations to parton showers, and the uncertainty of the proton PDFs.
For the minor backgrounds from $t\bar{t}$, $VV$, $VVV$, and VBF $H$ production with $H\to\tau^+\tau^-$ or $H\to W^\ast W$ the theoretical uncertainties have been found to be negligible.
 
\subsubsection{\Wjets and \Zjets uncertainties}
\label{sec:WZjetTheorySyst}
 
The higher-order matrix-element effects and parton shower matching uncertainties are assessed by varying the four scale choices used in the event generation:
the renormalisation, factorisation, resummation, and CKKW matching scales~\cite{Lonnblad:2011xx}.
 
The factorisation and renormalisation scales are varied upwards and downwards by a factor of two using on-the-fly varied event weights in the \Sherpa MC simulation.
The use of event weights reduces the statistical uncertainty of the estimate compared with using independently generated samples.
The corresponding uncertainties are calculated by taking an envelope of the seven factorisation/renormalisation scale variations:
the central values, each scale varied separately, and both scales varied coherently.
For the strong \Vjets background, the effect of the seven-point scale variations on the expected yield in bins of \mjj are $^{+27\%}_{-18\%}$ at low \mjj, increasing to  $^{+43\%}_{-26\%}$ at high \mjj.
The corresponding values for the electroweak \Vjets background  are 2\%--14\% for \Zjets and 1\%--8\% for \Wjets.
 
The resummation scale is also varied by a factor of two, while the CKKW scale is decreased to $15\,\GeV$ and increased to $30\,\GeV$ from the nominal $20\,\GeV$ value.
These are applied to the samples using event-by-event reweighting factors calculated with \Sherpa[2.1], thereby reducing the overall CPU resources needed to generate the required samples.
The reweighting factors are binned in $\pT^{V}$, the transverse momentum of the produced vector boson, and the number of generator-level jets in the event.
The relative error, calculated as the average of the relative effects of the upwards and downwards variations, ranges from 4\% to 8\% for the resummation scale and from 4\% to 6\% for the CKKW scale.
 
The effects of the PDF uncertainties on the $W$ and $Z$ backgrounds are evaluated separately for each bin as the standard deviation of the yields obtained by using the different PDF replicas of the \NNPDF set.
This results in uncertainties ranging from 1\% to 2\%.
 
All theoretical uncertainties are treated as uncorrelated between strong and electroweak processes; however, uncertainty sources affecting \Zjets and \Wjets events from the same type of process are treated as correlated.
This means that an uncertainty variation that increases the strong \Zjets background in the SR also increases the strong \Wjets background in the SR and the corresponding backgrounds in the $Z$ and $W$ CRs.
This leads to a significant decrease in the uncertainty of the total background yield in each signal region bin.
To quantify how well the effects of these uncertainties cancel out, the variation of the ratio of each \Vjets background in the SR to that in the corresponding CR is calculated.
The uncertainty in the value of each ratio is 1\%--3\% and is dominated by the statistical uncertainties of the samples used.
 
The \Vjets prediction is estimated by the likelihood fit (see Section~\ref{sec:fitsetup}) with one normalisation factor per SR bin, so \Zjets and \Wjets backgrounds are assumed to be correlated.
Several uncertainties are then assessed by a dedicated NLO QCD and NLO electroweak calculation~\cite{Lindert:2021xxx} of the \RZW{} ratio, which is used to refine the MC prediction of the \Wjets and \Zjets backgrounds (see Section~\ref{sec:WZbg}).
The relative size of the impact of the NLO QCD corrections on the \RZW{} ratio is taken as the uncertainty, which amounts to ${\sim}2\%$ for both the strong and electroweak processes across \mjj bins.
The QCD and electroweak corrections are applied either multiplicatively (default) or additively, resulting in a difference of less than 1\% in all bins, which is assigned as an uncertainty.
For the strong process, parton shower uncertainties of 3\% to 5\% are calculated as the difference between \RZW{} values at NLO with and without parton shower simulation.
For the electroweak process, they are calculated as the difference between dipole and angular ordering in \HERWIG, resulting in uncertainties of less than 2\%.
The dedicated NLO QCD and NLO electroweak calculation does not use the same VBF selection as the analysis, nor does it give an \RZW{} value for each CR and SR bin.
Instead, for strong \Vjets production the  \RZW{} differences between the VBF-like region used to define \RZW{} and the individual CR and SR regions are modelled with \Sherpa.
The dominant effect is due to the jet-veto requirement, which is estimated from \Sherpa as the change in \RZW{} when applying the jet veto.
It ranges from $2\%$ at low \mjj to $13\%$ at high \mjj, where the contribution from strong \Vjets production is comparatively small.
For electroweak \Vjets production, the dominant effect is the difference between the \RZW{} values for the inclusive \dphijj{} range and the individual generator-level \dphijj{} ranges of 0--1, 1--2, and ${>}2$.
This is estimated from the dedicated NLO QCD and NLO electroweak calculation and ranges up to 11\%.
The difference between the requirement on \detajj for the SR definition ($\detajj{\,>\,}3.8$) and that for the NLO calculation ($\detajj{\,>\,}2.5$) is covered by the assigned uncertainties and has a negligible impact on the final results.
A summary of the theoretical uncertainties of \RZW{} is given in Figures~\ref{fig:wz_unc_a} and~\ref{fig:wz_unc_b} for the strong and electroweak \Vjets processes, respectively.

\begin{figure}
\centering
\subfloat[]{
\parbox{0.49\textwidth}{
\centering
\includegraphics[width=0.49\textwidth]{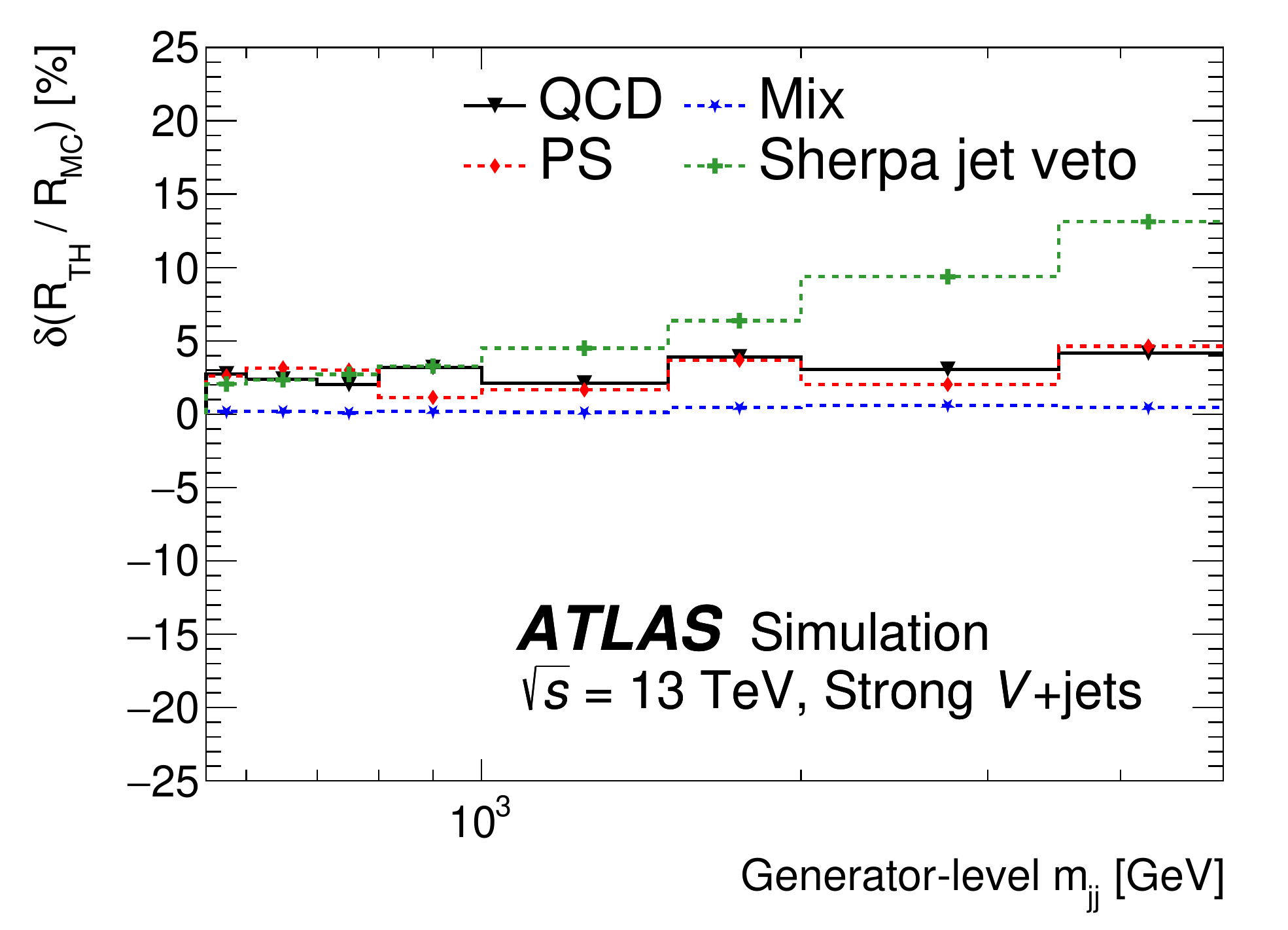}
}
\label{fig:wz_unc_a}
}
\subfloat[]{
\parbox{0.49\textwidth}{
\centering
\includegraphics[width=0.49\textwidth]{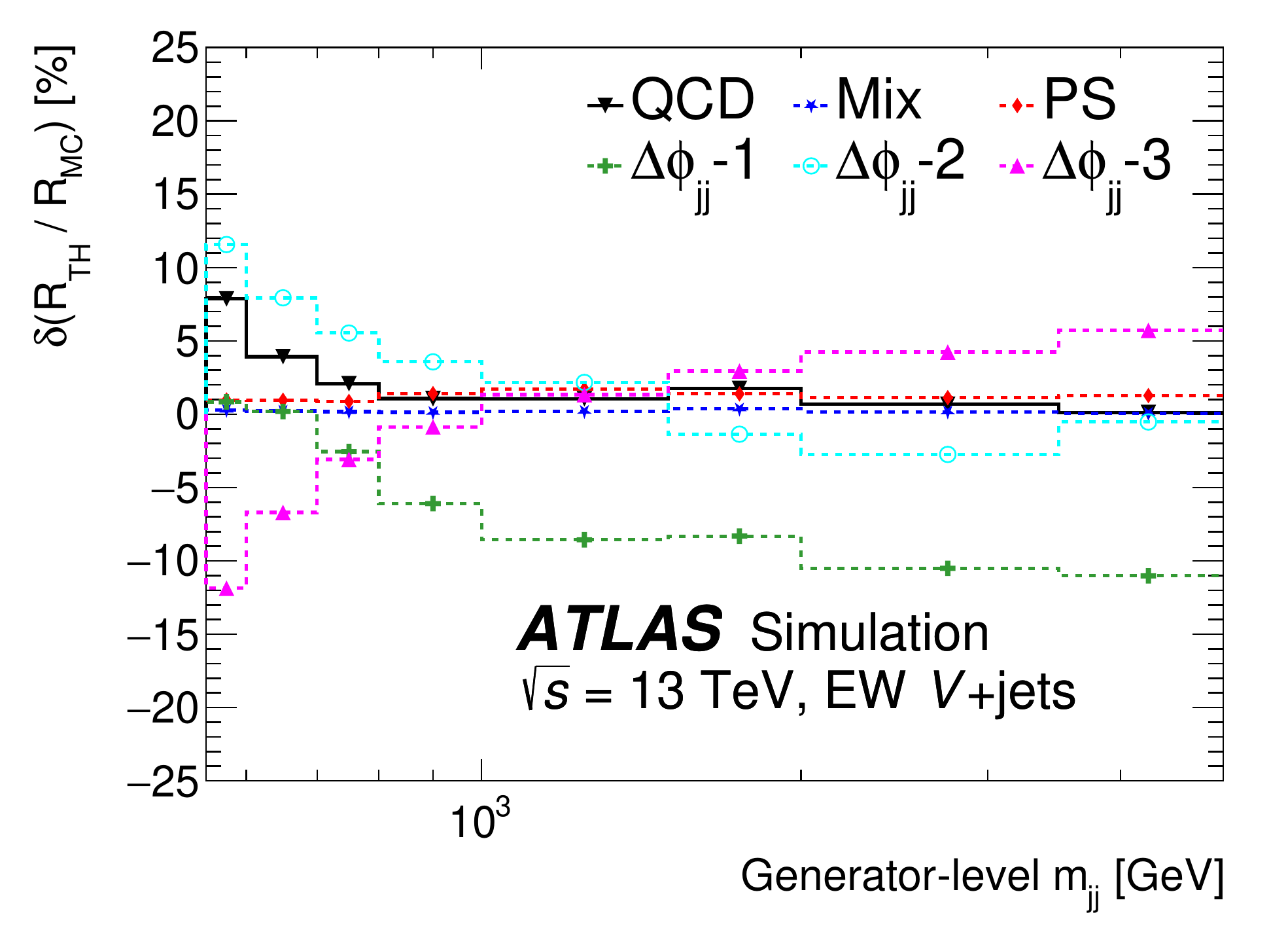}
}
\label{fig:wz_unc_b}
}
\caption{Fractional uncertainties of the double ratio $\mathcal{R}^{Z/W}_{\text{TH}}{\,\big/\,}\mathcal{R}^{Z/W}_{\text{MC}}$.
`QCD' refers to the change in the ratio induced by the NLO QCD corrections, `Mix' shows the difference between additive and multiplicative application of the NLO QCD and NLO electroweak corrections, and `PS' stands for the uncertainties attributed to the parton shower.
For the strong $V\text{+\,jets}$ processes \protect\subref{fig:wz_unc_a}, `Sherpa jet veto' denotes the difference between the ratios with and without a jet veto being applied.
For the electroweak $V\text{+\,jets}$ processes \protect\subref{fig:wz_unc_b}, the three `$\Delta\phi_{\text{jj}}$-[1,2,3]' uncertainty components quantify the impact of the $\Delta\phi_{\text{jj}}$ dependence on the correction and are evaluated in the ranges 0--1, 1--2, and ${>}2$.}
\end{figure}
 
\subsubsection{Signal uncertainties}
 
The inclusive VBF and ggF Higgs boson production cross sections and uncertainties are provided by the LHC Higgs working group~\cite{deFlorian:2227475}.
For the VBF process, there is an uncertainty due to the $\pT$-dependent NLO electroweak correction.
This is estimated from \hawk to be 2\%.
The \mjj-dependent effects of the renormalisation and factorisation scale uncertainties and their correlations were computed by the LHC Higgs working group~\cite{deFlorian:2016spz} for the \POWPY signal sample.
The factorisation and renormalisation scales are varied independently.
Additionally, the Stewart--Tackmann technique~\cite{Stewart:2011cf} is used to compute the uncertainty due to the jet veto.
These uncertainties are in the range from 1\% to 3\%.
By comparison with a \POWHER[7] sample, parton shower uncertainties are estimated to have a 2\%--4\% effect on the VBF signal.
PDF uncertainties assessed from the \NNPDF set affect the VBF signal by 1\%--2\%.
 
Uncertainties associated with the ggF process are also evaluated by renormalisation and factorisation scale variations.
A Stewart--Tackmann procedure is used to estimate the jet bin migration uncertainty.
This gives a 45\% uncertainty in bins with $N_\text{jet}{\,=\,}2$ and 41\% in the bins with $N_\text{jet}{\,=\,}3$ or 4.
The smaller PDF and parton shower uncertainties are also included.
 
\subsection{Experimental uncertainties}
 
Several experimental uncertainties impact the sensitivity of the analysis.
They can be grouped into three categories: uncertainties in the luminosity, uncertainties in the trigger efficiencies, and uncertainties related to the used physics objects such as electrons, muons, jets, and the \MET.
 
The uncertainty in the luminosity is 1.7\%~\cite{ATLAS-CONF-2019-021} and impacts only the signal yield because its effect on the dominant \Vjets backgrounds cancels out in the chosen approach (see Section~\ref{sec:fitsetup}).
 
To maximise the sensitivity of the analysis, the \MET requirement of at least either 160 or 200\,\GeV{} for the signal selection is chosen.
But since at these values the \MET triggers are yet not fully efficient, systematic uncertainties have to be considered to account for possible trigger efficiency differences between data and simulation.
This is done by comparing the combined L1+HLT efficiency turn-on curve for simulated signal events with the turn-on curve derived from data using a sample recorded with single-muon triggers.
They agree within the statistical uncertainty, which is below 1\%.
To cover the difference between the trigger efficiencies for \Wmn{} and \Znn{}, a further systematic uncertainty is taken from the difference found in simulated samples, which is smaller than 0.5\% for all signal region bins.
For the $W$ and $Z$ control regions, scale factors and uncertainties of the corresponding single-lepton and dilepton triggers are applied~\cite{TRIG-2018-01,TRIG-2018-05}.
 
For electrons and muons, uncertainties in the reconstruction and isolation efficiencies~\cite{PERF-2017-01,PERF-2015-10}, as well as in energy scale and resolution~\cite{EGAM-2018-01,ATL-PHYS-PUB-2019-018}, are considered.
For the electron veto, an uncertainty in the electron inefficiency is taken into account.
For jets, uncertainties in the energy scale and resolution~\cite{ATL-PHYS-PUB-2015-015} and in the pile-up tagging efficiencies are used~\cite{ATLAS-CONF-2014-018,ATL-PHYS-PUB-2019-026}.
The above uncertainties associated with the reconstructed objects are propagated to the calculation of \MET to evaluate their impact on the predicted yields.
The uncertainties in the scale and resolution of the \MET soft term are also propagated to the overall \MET uncertainty~\cite{ATLAS-CONF-2018-023}.
% End of text imported from the .//sections/uncertainties.tex input file
 
\section{Statistical model}
\label{sec:fitsetup}
% The next lines are included from the .//sections/fitsetup.tex input file
A simultaneous maximum-likelihood fit is used to extract the signal yield from the observed event counts in the signal region and control regions.
The likelihood is a product of terms for the Poisson probabilities to observe each data count of bin $i$, $N^\text{CR/SR}_{i}$, from a signal-plus-background model.
Furthermore, it depends on a set of parameters representing the yields from multijet processes in the fake-$e$, fake-$\mu$ and pile-up CRs, and on a set of nuisance parameters $\theta_j$ that represent experimental and theoretical uncertainties.
The strong and electroweak components of the \Vjets estimates in each SR bin $i$ and the corresponding CR bins are scaled by one dedicated free parameter $\beta_i$.
Since the \Vjets CR is enriched in events from \Vjets processes, the $\beta_i$ factors are almost exclusively determined so as to match the data counts in these CR bins, and these normalisation constraints are appropriately transferred to the \Vjets prediction in the corresponding SR bin.
 
The dependence of all background ($B$) and signal ($S$) yields on the nuisance parameters $\vec{\theta}$ is omitted to simplify the notation.
The signal region yield likelihood is
\begin{equation}\nonumber
\mathcal{L}^\text{SR}=\prod_{i}\mathcal{P}\left( N^\text{SR}_{i} \mid \beta_{i}\cdot B_{Z,i}^\text{SR}+\beta_i\cdot B_\textit{W,i}^\text{SR}+B_\text{MJ,$i$}^\text{SR}+B^\text{SR}_{\text{other},i}+\mu\cdot S^\text{SR}_{i} \right),
\end{equation}
where $B_{W,i}^\text{SR}$ and $B_{Z,i}^\text{SR}$ are the \Wjets and \Zjets yields in the simulation, with the latter already weighted by the double ratio $\RZW{\mathrm{TH}}\,\big/\,\RZW{\textsc{MC}}$ as discussed in Section~\ref{sec:WZbg}.
The $B_\text{MJ,$i$}^\text{SR}$ are the multijet background estimates detailed below;
$B^\text{SR}_{\text{other},i}$ are sums of the simulated $t\bar{t}$, $VV$, $VVV$ yields and VBF $H$ with $H\to\tau^+\tau^-$ or $H\to W^\ast W$;
$S^\text{SR}_{i}$ is the simulated yield for an invisible Higgs boson decay branching fraction of 100\%; and
$\mu$ is the signal strength.
Since the same $\beta_i$ is used for \Wjets and \Zjets contributions, the larger number of events in the $W_{\ell\nu}$ CRs reduces the uncertainty of the \Zjets background, particularly in bins with high \mjj{} requirements.
 
The background from multijet events, $B^\text{SR}_{\text{MJ},i}$, is determined as detailed in Section~\ref{sec:MJ} by two independent methods.
The pile-up-CR method uses a multijet-enriched control region that is binned according to the SR to constrain the free fit parameters $n_{\text{PU-MJ},i}$.
Those are scaled by the ratio $R_{\text{MJ},i}$ to give the final multijet estimates.
In bins 6--10, which have high \MET and $1{\,<\,}\dphijj{\,<\,}2$, the total event yield is evaluated using the merged bins of the PU-CR, while the shape, i.e.\ the fraction $f_{\text{R+S},i}$ of events in each bin, is predicted by the R+S method.
For bins 11--13, with $N_\text{jet}{\,>\,}2$, the prediction is obtained solely by using the R+S method.
The corresponding likelihood is
\begin{equation}\nonumber
\mathcal{L}^\text{PU-CR}=\prod_{i\in[1,5]\text{ or }\{6-10\}\text{ or }[14,16]}\mathcal{P}\left( N_{i}^{\text{PU-CR}} \mid B_{\text{non-MJ,}\textit{i}}^{\text{PU-CR}} + n_{\text{PU-MJ},i}\right),
\end{equation}\nonumber
where $B_{\text{non-MJ,}\textit{i}}^{\text{PU-CR}}$ is the non-multijet contribution determined from simulation, and $\{$6--10$\}$ represents the index of the merged bins.
The full multijet estimate that enters the signal region likelihood is
\begin{equation}\nonumber
B_\text{MJ,$i$}^\text{SR} =
\begin{cases}
n_{\text{PU-MJ},i}\cdot R_{\text{MJ},i} & \text{for } i\in[1,5]\text{ or }[14,16] \\
n_{\text{PU-MJ},\{6-10\}}\cdot R_{\text{MJ},\{6-10\}}\cdot f_{\text{R+S},i} & \text{for } i\in[6,10] \\
B_{\text{MJ-R+S},i}^\text{SR} & \text{for } i\in[11,13] \\
\end{cases}.
\end{equation}
 
The \Vjets backgrounds are constrained by dedicated $W$ and $Z$ control regions and the associated likelihood is:
\begin{eqnarray*}
\mathcal{L}^{V\!+\text{jets-CR}} &=& \prod_{i}\mathcal{P}\left( N_{i}^{Z\text{CR}} \mid \beta_\textit{i}\cdot B_\textit{Z,i}^{Z\text{CR}} + B_{\text{non-}\textit{Z,i}}^{Z\text{CR}} \right) \nonumber \\
&& \prod_{i}\mathcal{P}\left( N_{i}^{\textit{W$\mu\nu$}\text{CR}} \mid \beta_\textit{i}\cdot B_\textit{W,i}^{\textit{W$\mu\nu$}\text{CR}} + B_{\text{non-}\textit{W,i}}^{\textit{W$\mu\nu$}\text{CR}} + R_M\cdot n_{\text{fake-$\mu$},i}\right) \nonumber \\
&& \prod_{i}\mathcal{P}\left( N_{i}^{\textit{We$\nu$}\text{CR}} \mid \beta_\textit{i}\cdot B_\textit{W,i}^{\textit{We$\nu$}\text{CR}} + B_{\text{non-}\textit{W,i}}^{\textit{We$\nu$}\text{CR}} + R_{S,i}\cdot n_{\text{fake-$e$},i} \right),
\end{eqnarray*}
where $\beta_i$ are the same free parameters present in the signal region yield likelihood to propagate the control region constraints to the expected $W$ and $Z$ background yields in the SR.
The
$B_\textit{W,i}^{\textit{W$\mu\nu$}\text{CR}}$ and
$B_\textit{W,i}^{\textit{We$\nu$}\text{CR}}$ terms are the simulated
\Wjets yields separated by lepton flavour, and
$B_\textit{Z,i}^{Z\text{CR}}$ are the \Zjets yields in simulation scaled by
the same theoretical model for the ratio of $Z/W$ as in the SR.
The
$B_{\text{non-}\textit{Z,i}}^{Z\text{CR}}$,
$B_{\text{non-}\textit{W,i}}^{\textit{W$\mu\nu$}\text{CR}}$, and
$B_{\text{non-}\textit{W,i}}^{\textit{We$\nu$}\text{CR}}$ are the simulated
yields due to non-$W$, non-$Z$, and non-fake contributions to these control regions.
The `fake' electrons (muons) are modelled with free parameters $n_{\text{fake-$e$},i}$ ($n_{\text{fake-$\mu$},i}$) multiplied by the $R_{S,i}$ ($R_M$) scale factors introduced in Section \ref{sec:WZbg}.
 
The last set of yield probabilities in the likelihood are for the fake-$e$ and fake-$\mu$ control regions
\begin{eqnarray*}
\mathcal{L}^\text{fake-CR} &=& \prod_{i}\mathcal{P}\left( N_{i}^{\text{fake-$\mu$CR}} \mid \beta_\textit{i}\cdot B_\textit{W,i}^{\text{fake-$\mu$CR}} + B_{\text{non-}\textit{W,i}}^{\text{fake-$\mu$CR}} + n_{\text{fake-$\mu$},i} \right) \nonumber \\
&& \prod_{i}\mathcal{P}\left( N_{i}^{\text{fake-$e$CR}} \mid \beta_\textit{i}\cdot B_\textit{W,i}^{\text{fake-$e$CR}} + B_{\text{non-}\textit{W,i}}^{\text{fake-$e$CR}} + n_{\text{fake-$e$},i} \right),
\end{eqnarray*}
where the first two terms in each correspond to the real-lepton contributions calculated in the same way as in the $W$ and $Z$ control regions, and the last terms, $n_{\text{fake-$\mu$},i}$ and $n_{\text{fake-$e$},i}$, are the contributions from the fake leptons.
 
To constrain the nuisance parameters within their uncertainties the following likelihood term is considered:
\begin{equation}\nonumber
\mathcal{L}^\text{NP}=\prod_{j}\mathcal{G}\left(0\rvert\theta_{j}\right),
\end{equation}
where $\mathcal{G}$ is the Gaussian probability density function.
All systematic uncertainties are implemented as nuisance parameters $\vec\theta$ modifying the $B$, $S$, and $R$ parameters.
Their nominal values are zero and they are constrained by Gaussian probability distributions with widths corresponding to the size of their uncertainties, which are normalised to one.
The effect of the nuisance parameters on the yields is through an exponential response function $(1+\varepsilon_{ij})^{\theta_j}$ where $\varepsilon_{ij}$ is the fractional uncertainty amplitude of bin $i$ from the uncertainty source $j$.
This treatment prevents the fitted yields from becoming negative in cases with large uncertainties.
 
Each experimental uncertainty source is taken to be fully correlated across all signal and control regions.
For the correlation of theory systematic uncertainties the following assumptions are made.
PDF uncertainties are treated as fully correlated across bins.
The perturbative uncertainties of each of the four \Vjets samples (strong and electroweak components of $W$ and $Z$ production) are split into 17 components: one for each of the 16 SR (CR) bins, and one component that is fully correlated across all bins.
Each of the 16 uncorrelated uncertainty sources only affects one SR (CR) bin, with an uncertainty amplitude obtained from the seven-point scale variation as discussed in Section~\ref{sec:WZjetTheorySyst}.
The last uncertainty component, which affects all bins for a particular \Vjets sample, is assigned such that the combined uncertainty is equal to the systematic uncertainty for the inclusive SR (CR).
 
Combining all the terms gives the full likelihood expression
\begin{equation}\nonumber
\mathcal{L}(\mu,\vec{\beta}_{V}, \vec{n}_\text{PU-MJ/fake-$e$/fake-$\mu$}, \vec{\theta})=\mathcal{L}^\text{SR} \cdot \mathcal{L}^\text{PU-CR} \cdot \mathcal{L}^{V\!+\text{jets-CR}} \cdot \mathcal{L}^\text{fake-CR} \cdot \mathcal{L}^\text{NP},
\end{equation}
which contains a total of 58 free parameters and 105 observed event counts.
Asymptotic formulae for the CL$_\text{s}$ frequentist approach~\cite{Junk:1999kv,Read:2002hq,Cowan:2010js} are used to set an upper limit on $\mu$ at 95\%~CL.
This corresponds to the observed upper limit on the branching fraction of Higgs boson decays into invisible particles.
% End of text imported from the .//sections/fitsetup.tex input file
 
\section{Validation}
% The next lines are included from the .//sections/validation.tex input file
Since the event selection requires $\dphijj{\,<\,}2$, events with $2{\,<\,}\dphijj{\,<\,}2.5$ are not part of the analysis but are instead used to define a validation region (VR).
The VR data set is used to validate the background estimation techniques, and in particular to test the used ratio of the \Wjets and \Zjets cross sections and the assigned uncertainties (see Section \ref{sec:WZjetTheorySyst}).
The validation parallels the full analysis chain with five VR bins in \mjj with the same boundaries as the SR bins 1--5 or 6--10.
The expected signal from invisibly decaying Higgs bosons at an assumed branching fraction $\mathcal{B}_\text{inv}$ of 0.15 is less than 5\% of the background in all VR bins corresponding to the SR.
Because of the larger multijet background for bins with $160\,\GeV{\,<\,}\MET{\,<\,}200\,\GeV$ or $N_\text{jets}{\,=\,}3$ or 4, only events with $\MET{\,>\,}200\,\GeV$ with $N_\text{jets}{\,=\,}2$ are considered.
Corresponding validation regions, each with five bins, are defined for \Wjets, \Zjets, fake-$e$, and fake-$\mu$.
The multijet background is estimated with the R+S method.
The likelihood function is adjusted to the reduced number of VR bins.
A fit is performed in the VR bins, excluding the zero-lepton bins, using only the background contributions and with $\mu$ set to zero.
In total, the likelihood function for the VR has 15 free parameters and 30 observed event counts.
The fitted \Vjets background scaling factors $\beta_i$ are then applied to the corresponding VR bins with zero leptons, and the resulting background expectations are compared with the observed yields as shown in Figure~\ref{fig:highdphijj}.
Good agreement between the total background prediction and data yields is obtained across all bins, including the zero-lepton validation region, which justifies the application of this strategy to the SR bins with $\dphijj{\,<\,}2$.
 
\begin{figure}
\centering
\includegraphics[width=1.\textwidth]{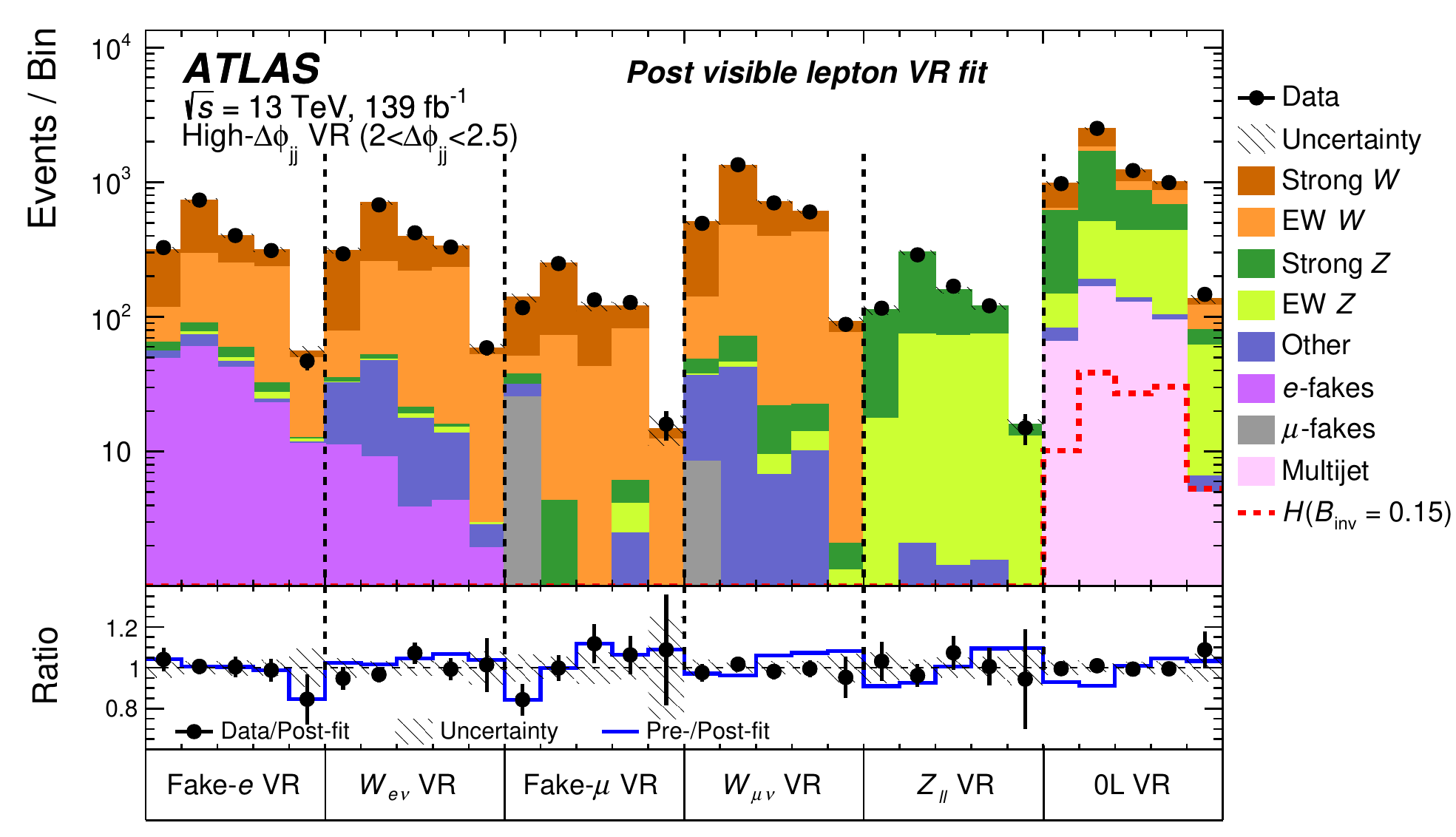}
\caption{
Post-fit yields and their uncertainties for all VR bins with $2{\,<\,}\Delta\phi_\text{jj}{\,<\,}2.5$ and with $\mu$ set to zero.
The five bins in each region correspond to the $m_\text{jj}$ bins of the SR bins 1--5 and 6--10.
The fit is performed only in the VR bins with one or two leptons, and the fitted background scaling factors $\beta_i$ are applied to the corresponding VR bins with zero leptons.
Minor contributions from $t\bar{t}$, $VV$, $VVV$, and VBF $H$ with $H\to\tau^+\tau^-$ or $H\to W^\ast W$ are combined and labelled `other'.
The resulting background expectations are compared with the observed yield.
The signal is scaled to a branching ratio of 15\% for Higgs boson decays into invisible particles.
The ratio of the observed yields to the prediction of the model post-fit is shown by the points in the bottom plot and the relative change of the prediction with respect to the pre-fit yields is shown by the blue line.
}
\label{fig:highdphijj}
\end{figure}
% End of text imported from the .//sections/validation.tex input file
 
\section{Results}
% The next lines are included from the .//sections/results.tex input file
The fit strategy as described in Section \ref{sec:fitsetup} is applied to the data set. Overall, good agreement between the predicted and observed yields is observed.
The expected background yields in the inclusive signal region and the control regions prior to the likelihood fit are summarised in Table~\ref{tab:yieldsprefitAll}.
The event yields in the 16 signal selection bins after a background-only likelihood fit, i.e.\ with the signal strength $\mu$ set to zero, are compared with the observed numbers of events in Tables~\ref{tab:yieldspostfitSR} and~\ref{tab:yieldspostfitSR2}, and a comparison in CR and SR bins is presented in Figure~\ref{fig:ResultsPostFit}, excluding the fake-lepton and pile-up CR bins.
Figure~\ref{fig:Dist_PostFit} shows the \mjj and \dphijj distributions after the likelihood fit in the inclusive signal region.
The ratio of yields before and after the likelihood fit shows a clear trend in \mjj{}.
This is in agreement with the observation that \Sherpa[2.2.1] tends to overestimate the strong \Zjets processes at large values of \mjj{}~\cite{STDM-2017-27}.
 
\begin{table}[tb]
\caption{Observed and expected background event yields with associated uncertainties in the signal region (SR) and control regions ($Z_{\ell\ell}$~CR, $W_{e\nu}$~CR, $W_{\mu\nu}$~CR, fake-$e$~CR, fake-$\mu$~CR, pile-up~CR) prior to the likelihood fit.
Minor backgrounds from $t\bar{t}$, $VV$, $VVV$, and VBF $H\to W^\ast W\, /\, \tau^+\tau^-$ are combined and labelled `other'.
The uncertainties in the backgrounds include the statistical, experimental, and theoretical uncertainties, taking into account the correlations between the individual SR and CR bins.
The predicted signal yields (VBF, ggF, and $VH$) for $\mathcal{B}_\mathrm{inv}{\,=\,}15\%$ (the observed limit) are presented for comparison.
For all CRs with leptons, the contribution from the predicted signal is negligible.
}
\label{tab:yieldsprefitAll}
\centering
\resizebox{\textwidth}{!}{
\begin{tabular}{lr@{$~\pm~$}lr@{$~\pm~$}lr@{$~\pm~$}lr@{$~\pm~$}lr@{$~\pm~$}lr@{$~\pm~$}lr@{$~\pm~$}lr@{$~\pm~$}lr@{$~\pm~$}lr@{$~\pm~$}l}
\toprule
Process	        &  \multicolumn{2}{c}{SR}	& \multicolumn{2}{c}{$Z_{\ell\ell}$~CR}	& \multicolumn{2}{c}{ $W_{e\nu}$~CR}	& \multicolumn{2}{c}{ $W_{\mu\nu}$~CR} & \multicolumn{2}{c}{Fake-$e$ CR}& \multicolumn{2}{c}{Fake-$\mu$ CR} & \multicolumn{2}{c}{Pile-up CR}\\
\midrule
$Z$ strong	&  6030 &  2050 & 1220 &  440 & 42 &  14 & 143 &  21  & 160 &  57 & 40 & 14 & 42 & $^{59}_{42}$ \\
$Z$ EWK	    &  2630 &  260 & 618 &  74 & 12.1 &  1.8 & 28.3 &  3.3 & 23.5 &  3.3 & 15.3 & 2.8 & 7.2 & 3.1 \\
$W$ strong	&  3710 &  1300 & \multicolumn{2}{c}{-} & 3260 &  1180 & 5170 &  1850  & 1810 &  650 & 1010 & 370 & 49 & $^{92}_{49}$ \\
$W$ EWK	    &  1610 &  150 & \multicolumn{2}{c}{-} & 2360 &  190 & 3410 &  240 & 1400 &  140 & 822 & 69 & 22.9 & 7.7 \\
Fake-$e$	    &  \multicolumn{2}{c}{-} & \multicolumn{2}{c}{-} & 191 & 70 & \multicolumn{2}{c}{-} &  1100 & 330 & \multicolumn{2}{c}{-}& \multicolumn{2}{c}{-} \\
Fake-$\mu$	&  \multicolumn{2}{c}{-} & \multicolumn{2}{c}{-} & \multicolumn{2}{c}{-}  & 43 & 15 & \multicolumn{2}{c}{-} &  130 & 51 & \multicolumn{2}{c}{-} \\
Multijet	    &  830 & 190 & \multicolumn{2}{c}{-} & \multicolumn{2}{c}{-} & \multicolumn{2}{c}{-} & \multicolumn{2}{c}{-} & \multicolumn{2}{c}{-} & 1890 & 110 \\
Other	    &  180 & 46 & 46 &  25 & 346 &  82 & 351 &  71	& 67 &  13 & 89 & 27 & 19.5 & 9.0 \\
\midrule
Total bkg.		&  14\,990 &  2990 & 1880 &  510 & 6210 &  1260 & 9150 &  1890 & 4560 &  760 & 2110 & 390 & 2030 & 110 \\
\midrule
$H$ (VBF)	&  886 & 81 & \multicolumn{10}{c}{} & 3.9 & 1.3 \\
$H$ (ggF)	&  106 & 41 & \multicolumn{10}{c}{Predicted signal for $\mathcal{B}_\mathrm{inv}=15\%$} & 1.0  &  $^{1.5}_{1.0}$ \\
$H$ $(VH)$		&  0.9 & 0.2 & \multicolumn{10}{c}{} & \multicolumn{2}{c}{-}\\
\midrule
Data	        &  \multicolumn{2}{c}{16\,490} & \multicolumn{2}{c}{2051} & \multicolumn{2}{c}{6361} & \multicolumn{2}{c}{9294} & \multicolumn{2}{c}{4563} & \multicolumn{2}{c}{2110} & \multicolumn{2}{c}{2033} \\
\bottomrule
\end{tabular}
}
\end{table}
 
\begin{table}
\caption{Yields of data, signal and major backgrounds in the signal regions for each bin after the likelihood fit with $\mu$ set to zero.
Minor backgrounds from $t\bar{t}$, $VV$, $VVV$, and VBF $H\to W^\ast W\, /\, \tau^+\tau^-$ are combined and labelled `other'.
The uncertainties in the backgrounds are derived by the fit and take into account the correlations of the background uncertainties.
The predicted signal yields (VBF, ggF, and $VH$) for $\mathcal{B}_\mathrm{inv}{\,=\,}15\%$ (the observed limit) are presented for comparison, along with their associated uncertainties, assuming that the experimental uncertainties are fully correlated between the ggF and VBF contributions.
The $\beta_V$ are the background normalisation factors.
}
\label{tab:yieldspostfitSR}
\centering
\begin{tabular}{l|r@{$~\pm~$}lr@{$~\pm~$}lr@{$~\pm~$}lr@{$~\pm~$}lr@{$~\pm~$}l}
\toprule
& \multicolumn{10}{c}{$\njet{}=2,|\dphijj|<1$, \mjj bins } \\
& \multicolumn{2}{c}{ 0.8--1.0\,\TeV} & \multicolumn{2}{c}{ 1.0--1.5\,\TeV} & \multicolumn{2}{c}{ 1.5--2.0 TeV} & \multicolumn{2}{c}{ 2.0--3.5\,\TeV} & \multicolumn{2}{c}{>3.5\,\TeV} \\
Process	& \multicolumn{2}{c}{Bin 1} & \multicolumn{2}{c}{Bin 2} & \multicolumn{2}{c}{Bin 3} & \multicolumn{2}{c}{Bin 4} & \multicolumn{2}{c}{Bin 5} \\
\midrule
$Z$ strong 	& 1110 & 57   & 1289 & 79   & 362 & 41 & 169 & 29 & 10.6 & 3.4   \\
$Z$ EWK		& 132 & 27    & 318 & 69    & 190 & 32 & 186 & 27 & 28.9 & 4.8   \\
$W$ strong	& 632 & 41    & 735 & 77    & 179 & 24 & 82 & 19  & 2.6 & 1.8    \\
$W$ EWK		& 78 & 16     & 157 & 33    & 90 & 16  & 120 & 17 & 25.2 & 4.4   \\
Multijet		& 76 & 49     & 124 & 41    & 81 & 25  & 82 & 27  & 9.1 & 4.4    \\
Other		& 15.6 & 2.8      & 21.4 & 3.3      & 12.5 & 2.3   & 11.6 & 2.8   & 3.4 & 1.0    \\
\midrule
Total bkg.	& 2044 & 43   & 2646 & 48   & 915 & 28 & 650 & 23 & 79.8 & 7.1   \\
$H$ ($\mathcal{B}_\text{inv}=0.15$) & 74 & 11 & 154 & 16 & 98.9 & 9.9 & 102 & 11 & 17.9 & 2.7  \\ 
\midrule
Data			& \multicolumn{2}{c}{2059}          & \multicolumn{2}{c}{2640}         & \multicolumn{2}{c}{905}        & \multicolumn{2}{c}{647}       & \multicolumn{2}{c}{77} \\
\midrule
Data/bkg.		& 1.01 & 0.03     & 1.00 & 0.03     & 0.99 & 0.04  & 1.00 & 0.05  & 0.97 & 0.14  \\
\midrule
$\beta_V$	& 1.18  &  0.23  &  1.07  &  0.21  &  1.01  &  0.15  &  0.97  &  0.12  &  0.88  &  0.12 \\
\bottomrule
\toprule
& \multicolumn{10}{c}{$\njet{}=2,1<|\dphijj|<2$, \mjj bins } \\
& \multicolumn{2}{c}{ 0.8--1.0\,\TeV} & \multicolumn{2}{c}{ 1.0--1.5\,\TeV} & \multicolumn{2}{c}{ 1.5--2.0\,\TeV} & \multicolumn{2}{c}{ 2.0--3.5\,\TeV} & \multicolumn{2}{c}{>3.5\,\TeV} \\
Process	& \multicolumn{2}{c}{Bin 6} & \multicolumn{2}{c}{Bin 7} & \multicolumn{2}{c}{Bin 8} & \multicolumn{2}{c}{Bin 9} & \multicolumn{2}{c}{Bin 10} \\
\midrule
$Z$ strong	& 768 & 39    & 1357 & 80   & 424 & 50    & 222 & 39    & 14.5 & 4.8  \\
$Z$ EWK		& 89 & 19     & 418 & 77    & 301 & 45    & 313 & 38    & 55.5 & 6.1  \\
$W$ strong	& 416 & 39    & 715 & 63    & 228 & 29    & 121 & 34    & 4.3 & 1.4   \\
$W$ EWK		& 56 & 13     & 215 & 41    & 145 & 24    & 165 & 21    & 41.8 & 5.3  \\
Multijet		& 9.8 & 5.3       & 21 & 12     & 8.0 & 4.4       & 6.2 & 3.5       & 0.5 & 0.3   \\
Other		& 9.4 & 1.6       & 19.1 & 3.1      & 7.8 & 1.4       & 10.6 & 2.5      & 3.7 & 1.2   \\
\midrule
Total bkg.	& 1348 & 35   & 2745 & 49   & 1115 & 29   & 838 & 26    & 120.3 & 8.4 \\
$H$ ($\mathcal{B}_\text{inv}=0.15$) & 24.9 & 3.0 & 86.4 & 7.1 & 58.4 & 4.7 & 64.2 & 5.9 & 12.2 & 1.7  \\ 
\midrule
Data			& \multicolumn{2}{c}{1354} &  \multicolumn{2}{c}{2745} &  \multicolumn{2}{c}{1131} &  \multicolumn{2}{c}{841} &  \multicolumn{2}{c}{134} \\
\midrule
Data/bkg.		& 1.00 & 0.04     & 1.00 & 0.03     & 1.01 & 0.04     & 1.00 & 0.05     & 1.11 & 0.12 \\
\midrule
$\beta_V$	& 1.20  &  0.24  &  1.14  &  0.20  &  0.99  &  0.12  &  0.94  &  0.11  &  0.96  &  0.12 \\
\bottomrule
\end{tabular}
\end{table}
 
\begin{table}
\caption{Yields of data, signal and major backgrounds in the signal regions for each bin after the likelihood fit with $\mu$ set to zero.
Minor backgrounds from $t\bar{t}$, $VV$, $VVV$, and VBF $H\to W^\ast W\, /\, \tau^+\tau^-$ are combined and labelled `other'.
The uncertainties in the backgrounds are derived by the fit and take into account the correlations of the background uncertainties.
The predicted signal yields (VBF, ggF, and $VH$) for $\mathcal{B}_\mathrm{inv}{\,=\,}15\%$ (the observed limit) are presented for comparison, along with their associated uncertainties, assuming that the experimental uncertainties are fully correlated between the ggF and VBF contributions.
The $\beta_V$ are the background normalisation factors.
}
\label{tab:yieldspostfitSR2}
\centering
\resizebox{\textwidth}{!}{
\begin{tabular}{l|r@{$~\pm~$}lr@{$~\pm~$}lr@{$~\pm~$}l|r@{$~\pm~$}lr@{$~\pm~$}lr@{$~\pm~$}l}
\toprule
& \multicolumn{6}{c|}{$3\le \njet{} \le 4$, \mjj bins } & \multicolumn{6}{c}{$160\,\GeV{} < \MET{} < 200\,\GeV{}$, \mjj bins } \\
& \multicolumn{2}{c}{ 1.5--2.0 TeV} & \multicolumn{2}{c}{ 2.0--3.5 TeV} & \multicolumn{2}{c|}{>3.5 TeV} & \multicolumn{2}{c}{ 1.5--2.0 TeV} & \multicolumn{2}{c}{ 2.0--3.5 TeV} & \multicolumn{2}{c}{>3.5 TeV}\\
Process	& \multicolumn{2}{c}{Bin 11} & \multicolumn{2}{c}{Bin 12} & \multicolumn{2}{c|}{Bin 13} & \multicolumn{2}{c}{Bin 14} & \multicolumn{2}{c}{Bin 15} & \multicolumn{2}{c}{Bin 16}\\
\midrule
$Z$ strong	& 275 & 41 & 224 & 44 & 19.3 & 6.7 & 530 & 47 & 228 & 32 & 8.4 & 3.5 \\
$Z$ EWK		& 121 & 39 & 148 & 44 & 28.4 & 6.0 & 230 & 38 & 199 & 28 & 18.7 & 4.0 \\
$W$ strong	& 178 & 30 & 147 & 30 & 10.0 & 3.8 & 404 & 52 & 143 & 29 & 8.6 & 3.9 \\
$W$ EWK		& 76 & 19 & 104 & 23 & 29.7 & 5.4 & 169 & 29 & 164 & 23 & 23.6 & 5.6 \\
Multijet		& 33 & 21 & 33 & 15 & 3.2 & 2.4 & 218 & 61 & 97 & 32 & 9.6 & 3.5 \\
Other		& 10.2 & 1.8 & 14.4 & 2.8 & 3.0 & 0.9 & 15.8 & 2.6 & 16.8 & 3.6 & 2.2 & 0.8\\
\midrule
Total bkg.	& 694 & 25 & 671 & 23 & 93.6 & 8.2 & 1\,568 & 38 & 848 & 27 & 71.0 & 7.3 \\
$H$ ($\mathcal{B}_\text{inv}=0.15$) & 38.7 & 6.2 & 54.9 & 8.0 & 9.2 & 2.4 & 95.6 & 9.6 & 90.3 & 7.8 & 15.3 & 3.2 \\ 
\midrule
Data			& \multicolumn{2}{c}{688} &  \multicolumn{2}{c}{681} &  \multicolumn{2}{c|}{90} &  \multicolumn{2}{c}{1561} &  \multicolumn{2}{c}{861} &  \multicolumn{2}{c}{76} \\
Data/bkg.		& 0.99 & 0.05 & 1.01 & 0.05 & 0.96 & 0.13 & 1.00 & 0.03 & 1.01 & 0.05 & 1.07 & 0.16 \\
\midrule
$\beta_V$	& 0.98  &  0.23  &  0.84  &  0.16  &  0.92  &  0.14  &  1.22  &  0.19  &  1.06  &  0.14 &  0.64  &  0.11 \\
\bottomrule
\end{tabular}
}
\end{table}
 
\begin{figure}
\centering
\includegraphics[width=1.\linewidth]{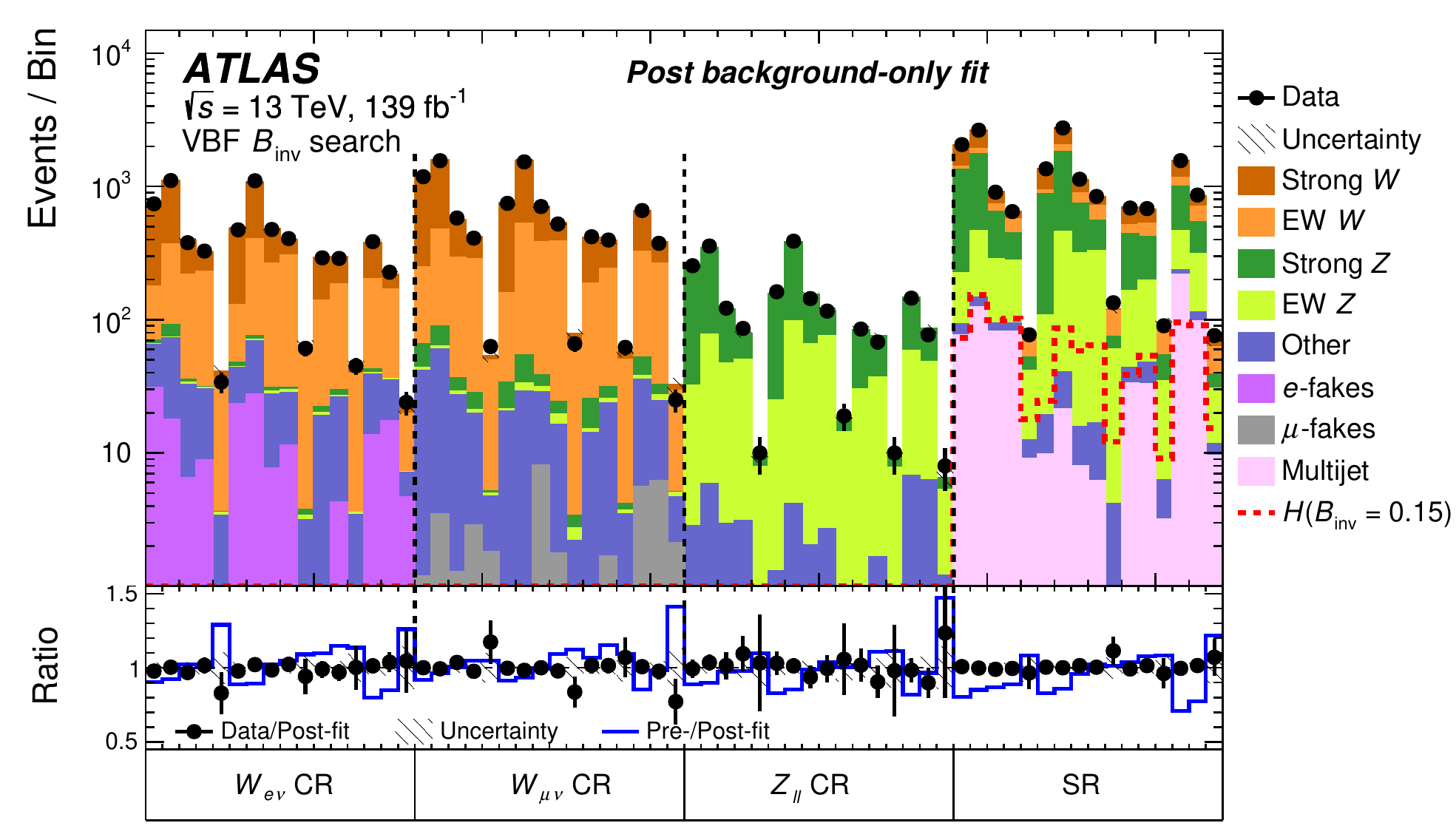}
\caption{Post-fit results of all SR and CR bins with $\mu$ set to zero.
Minor contributions from $t\bar{t}$, $VV$, $VVV$, and VBF $H$ with $H\to\tau^+\tau^-$ or $H\to W^\ast W$ are combined and labelled `other'.
The signal (dashed, red line) is scaled to a branching ratio of 15\% for Higgs boson decays into invisible particles.
The 16 bins in the SR and each of the three CRs are defined in Figure~\ref{fig:SRbinning}.
The fake-lepton and pile-up CRs, which contain an additional 41 bins that also enter the likelihood fit, are not shown, as this improves the readability.
The ratio of the observed yields to the prediction of the model post-fit is shown by the points in the bottom plot and the relative change of the prediction with respect to the pre-fit yields is shown by the blue line.
}
\label{fig:ResultsPostFit}
\end{figure}
 
\begin{figure}
\centering
\subfloat[]{
\parbox{0.75\textwidth}{
\centering
\includegraphics[width=0.75\textwidth]{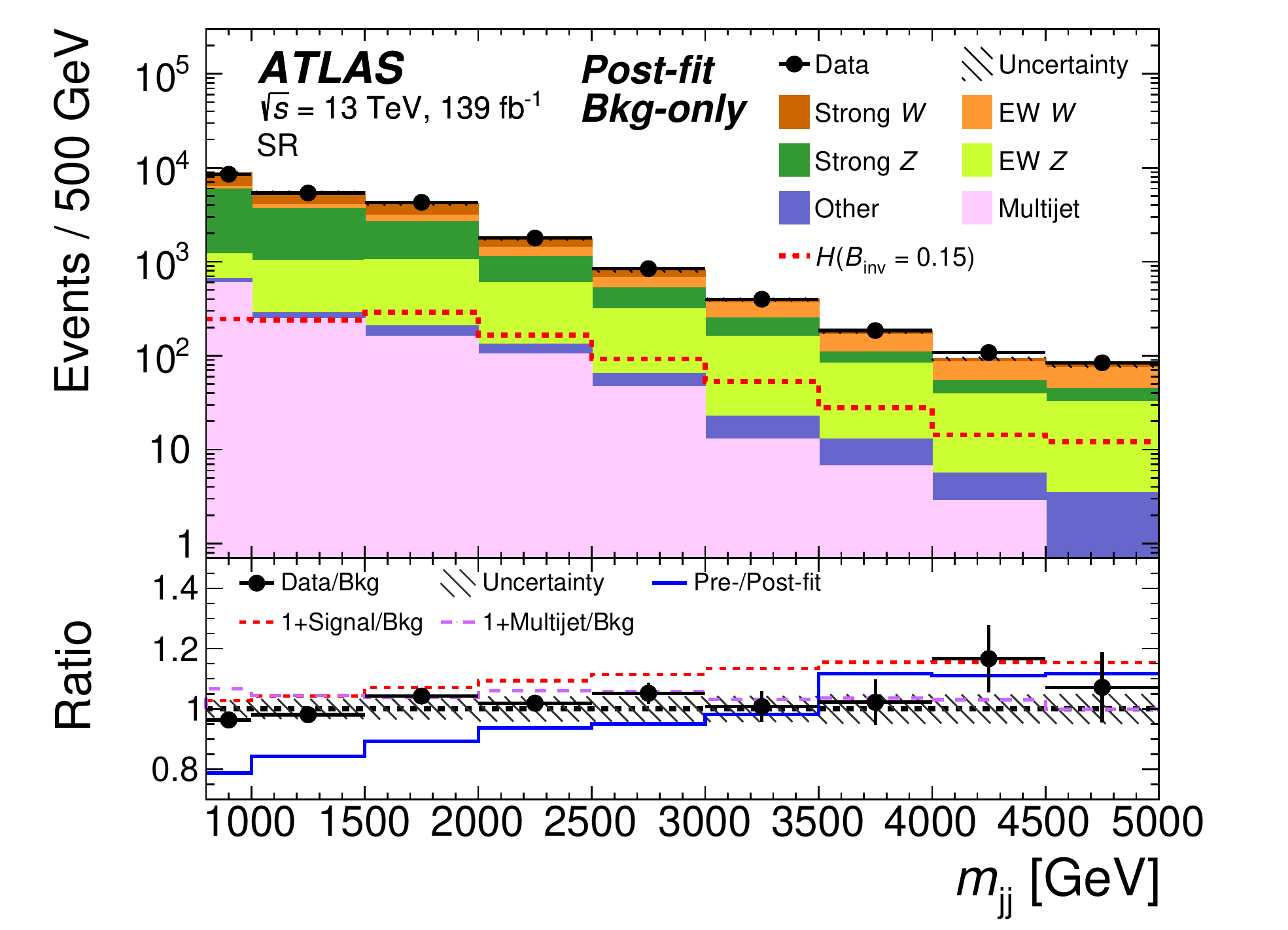}
}
\label{fig:Dist_PostFit_a}
}
 
\subfloat[]{
\parbox{0.75\textwidth}{
\centering
\includegraphics[width=0.75\textwidth]{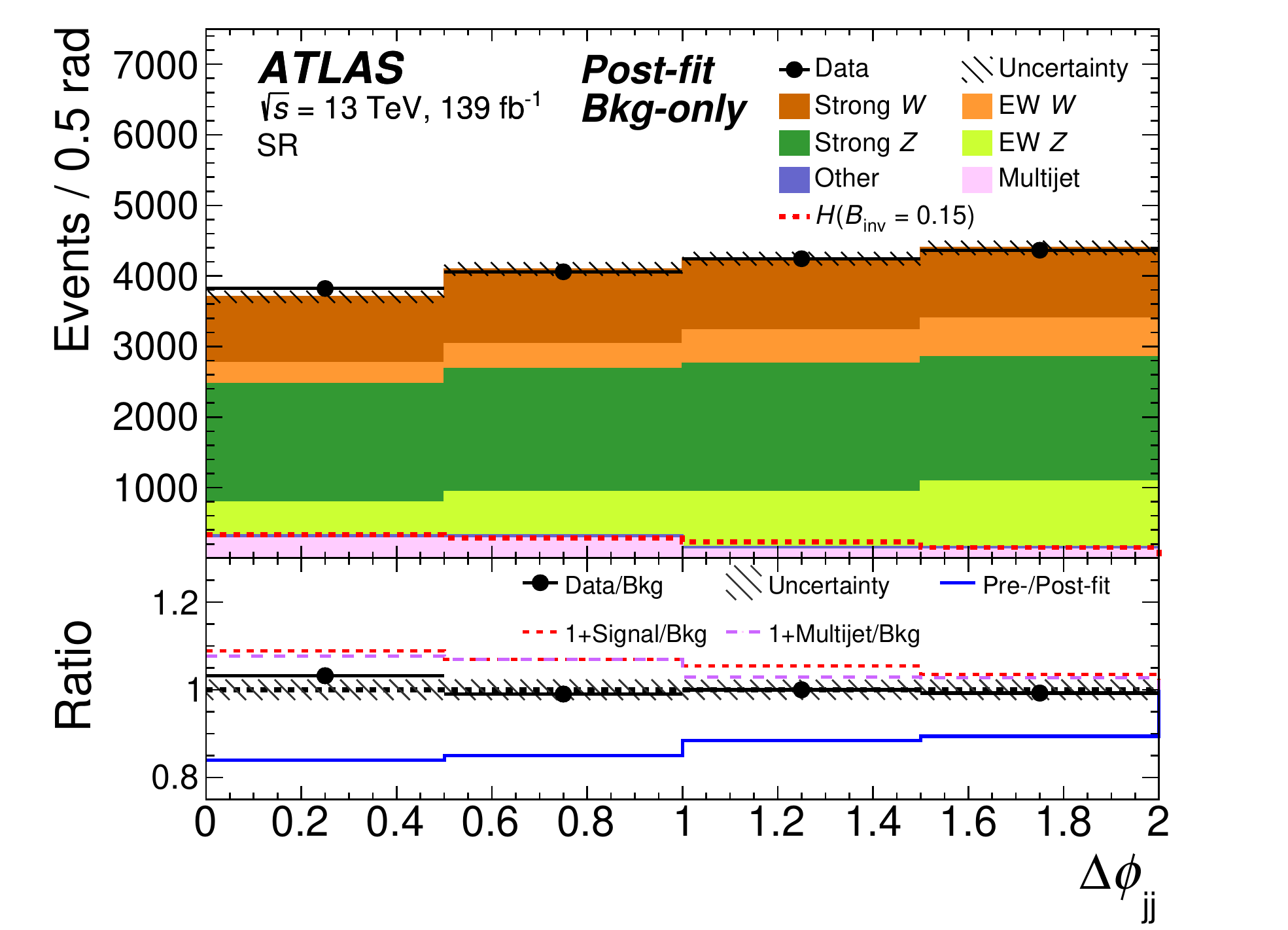}
}
\label{fig:Dist_PostFit_b}
}
\caption{Post-fit \protect\subref{fig:Dist_PostFit_a} $m_\text{jj}$ and \protect\subref{fig:Dist_PostFit_b} $\Delta\phi_\text{jj}$ distributions in the inclusive signal region.
Minor contributions from $t\bar{t}$, $VV$, $VVV$, and VBF $H$ with $H\to\tau^+\tau^-$ or $H\to W^\ast W$ are combined and labelled `other'.
The signal (dashed red line) is scaled to a branching ratio of 15\% for Higgs boson decays into invisible particles.
In addition to the data-to-background ratio, the lower panels show the ratio of the background expectation before and after the likelihood fit, the relative size of the multijet background, and the signal-to-background ratio.
The latter two quantities are shifted by one to use the same $y$-axis scaling.
}
\label{fig:Dist_PostFit}
\end{figure}
 
\newpage
 
The fitted value of the signal strength $\mu$ is $0.053{\,\pm\,}0.052$.
The contributions of the systematic uncertainties to the 68\% confidence interval of the fitted $\mu$ are shown in Table~\ref{tab:unc_impact}.
They are evaluated in a fit with the data yields set to the total background expectation, as is done to determine the expected limits.
Groups of systematic uncertainties (Section~\ref{sec:systematics}) are subsequently switched off, i.e.\ the corresponding nuisance parameters are fixed to their best-fit values.
The reduction of the 68\% confidence interval of $\mu$ is obtained by quadratically subtracting the resulting size of the confidence interval from the nominal case, which includes all systematic uncertainties.
The statistical uncertainties of the data yields in the \Vjets CR bins and the remaining SR and CR bins have an impact of ${\sim}42\%$ and ${\sim}54\%$, respectively.
For the experimental uncertainties, the dominant two groups are the uncertainties associated with the leptons (reconstruction and isolation efficiencies, lepton energy scale and resolution) and the uncertainties in the JER, having an impact of ${\sim}32\%$ and ${\sim}29\%$, respectively.
The uncertainties in the multijet background (Section~\ref{sec:MJ}) contribute with ${\sim}40\%$ and the contributions of fake leptons to the $W$ CRs have an impact of ${\sim}37\%$.
The dominant theoretical uncertainties are in the \Vjets background estimates and amount to ${\sim}28\%$.
Given, that the fit does not show any significant signal contribution, an upper limit of 0.145 is set on the branching fraction of Higgs boson decays into invisible particles at 95\% confidence level.
This is in agreement with the expected value of $0.103^{+0.041}_{-0.028}$ as shown in Table~\ref{tab:limits}.
In comparison with the observed (expected) limit on $\mathcal{B}_\text{inv}$ of $0.37$ ($0.28$) from the previous search~\cite{EXOT-2016-37}, this analysis shows a significant improvement.
 
\begin{table}
\caption{The contributions to the 68\% confidence interval of the fitted signal strength $\mu$ from different sources of systematic uncertainty.
They are evaluated in a fit with the data yields set to the total background expectation.
Due to residual correlations between categories, the sum in quadrature of the individual contributions only approximately equals the total uncertainty.
The uncertainty due to the number of data events (`Data stats.') is obtained by fixing all systematic uncertainties to their best-fit values.
The entry `$V\text{+\,jets}$ data stats.' is the statistical uncertainty originating from data yields for the $W$ and $Z$ normalisation.
The experimental uncertainties and the uncertainty related to the size of MC simulated samples (`MC stats.') are treated as separate categories.
The `Remaining' category contains $E_\text{T}^\text{miss}$, luminosity, and pile-up uncertainties, and a diboson uncertainty in the $Z_{\ell\ell}$ CR.
}
\label{tab:unc_impact}
\centering
\begin{tabular}{cc}
\toprule
Source  & Contribution to $\pm 1\sigma$ ($=0.052$) \\
\midrule
Data stats. & 0.029 \\
$V\text{+\,jets}$ data stats. & 0.022 \\
MC stats. & 0.014 \\
\midrule
Multijet & 0.021 \\
$\mu/e$-fakes & 0.019 \\
Lepton & 0.017 \\
(Lepton -- muon only) & ~~0.0049 \\
(Lepton -- electron only) & 0.016 \\
JER & 0.015\\
JES & 0.011 \\
Remaining & 0.012 \\
\midrule
$V\text{+\,jets}$ -- theory & 0.015 \\
Signal -- theory & ~~0.0056 \\
\bottomrule
\end{tabular}
\end{table}
 
\begin{table}
\caption{Observed and expected limits on $\mathcal{B}_\text{inv}$ for a Higgs boson with a mass of 125~\GeV{} calculated at the 95\%~CL for a $139\,\text{fb}^{-1}$ data set. The $\pm 1\sigma$ and $\pm 2\sigma$ variations of the expected limit are also shown.}
\label{tab:limits}
\centering
\begin{tabular}{cccccc}
\toprule
Observed & Expected & $+1 \sigma$ & $-1 \sigma$ & $+2 \sigma$ & $-2 \sigma$ \\
\midrule
0.145 & 0.103 & 0.144 & 0.075 & 0.196 & 0.055\\
\bottomrule
\end{tabular}
\end{table}

The limit on $\mathcal{B}_\text{inv}$ is related to a limit on the spin-independent WIMP--nucleon cross section~\cite{Kanemura:2010sh,Djouadi:2011aa,Fox:2011pm,deSimone:2014pda}.
This allows the result obtained here to be compared with limits from a selection of the most sensitive direct-detection experiments~\cite{CRESST:2019jnq,Agnes:2018ves,PandaX-4T:2021bab} in Figure~\ref{fig:DirectDMInterplay}.
The relation between $\mathcal{B}_\text{inv}$ and $\sigma_{\text{WIMP-nucleon}}$ is presented in the effective field theory framework assuming that the new-physics scale is $\mathcal{O}(1)\,\TeV$, well above the scale probed at the Higgs boson mass~\cite{Djouadi:2011aa}.
To enable comparisons with the 90\%~CL direct-detection limits, the observed $90\%$~CL upper limit of $12.7\%$ on $\mathcal{B}_\text{inv}$ from this analysis is used, the expected limit being $8.7\%$.
For the scalar WIMP interpretation, cross sections exceeding values ranging from $3{\,\cdot\,}10^{-43}\,\text{cm}^2$ to $1{\,\cdot\,}10^{-45}\,\text{cm}^2$ are excluded for masses between $1\,\GeV$ and $60\,\GeV$.
For the Majorana fermion WIMP interpretation, cross sections exceeding values ranging from $4{\,\cdot\,}10^{-47}\,\text{cm}^2$ to $7{\,\cdot\,}10^{-45}\,\text{cm}^2$ are excluded for the same mass range, and for a vector-like WIMP the exclusion limit ranges from $5{\,\cdot\,}10^{-51}\,\text{cm}^2$ to $3{\,\cdot\,}10^{-46}\,\text{cm}^2$.
Adding a renormalisable mechanism for generating the vector-like WIMP masses could modify the above-mentioned correlation substantially~\cite{Baek:2014jga,Arcadi:2020jqf,Zaazoua:2021xls}.
Many UV-complete models predict a new scalar particle that mixes with the Higgs boson.
This adds at least two free parameters to the model, for example its mass $m_2$ and the mixing angle $\alpha$.
The dependence of the exclusion limit for the vector-like WIMP hypothesis on the mass $m_2$ is shown in Figure~\ref{fig:DirectDMInterplay}.
The uncertainty band in the plot uses the latest computation of the nucleon form factors~\cite{Hoferichter:2017olk}.
The overlay shows the complementarity in coverage by the direct-detection experiments and the searches at colliders, such as the presented analysis.
 
\begin{figure}
\centering
\includegraphics[width=1.0\linewidth]{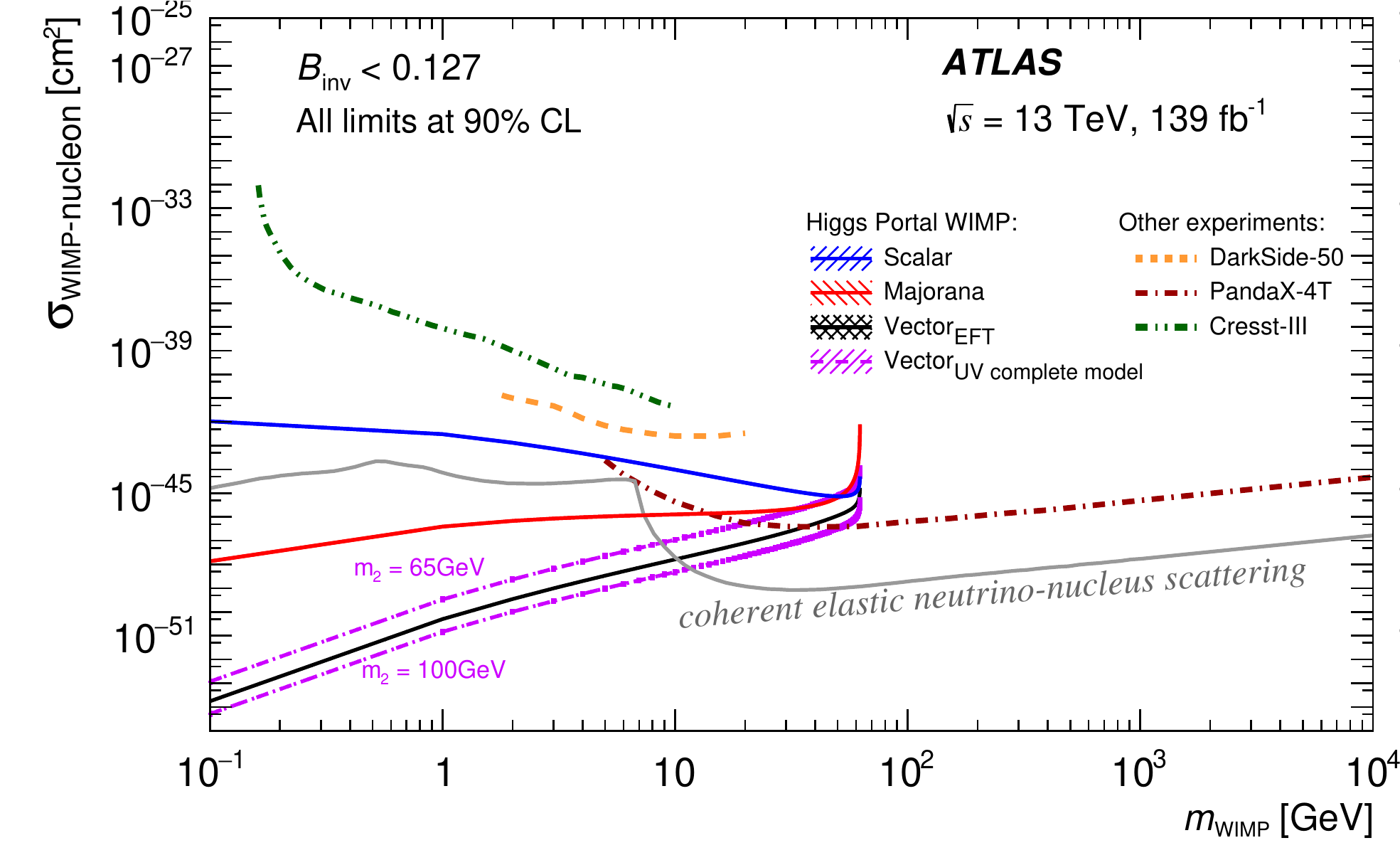}
\caption{Upper limits on the spin-independent WIMP--nucleon cross section using Higgs portal interpretations of $\mathcal{B}_\text{inv}$ at $90\%$ CL vs $m_\text{WIMP}$. For the vector-like WIMP hypothesis, the dependence on the mass $m_2$ of the new scalar particle, which is often predicted by renormalisable models, is shown for two different values taken from Ref.~\cite{Zaazoua:2021xls}.
For comparison with direct searches for DM, the plot shows results from Refs.~\cite{CRESST:2019jnq,Agnes:2018ves,PandaX-4T:2021bab}.
The neutrino floor for coherent elastic neutrino--nucleus scattering is taken from Refs.~\cite{Billard:2013qya,Ruppin:2014bra} and assumes germanium as the target over the whole WIMP mass range.
The dependence on the choice of target nucleus is relatively small, given the large range of cross sections shown.}
\label{fig:DirectDMInterplay}
\end{figure}
 
The results are further interpreted as a search for invisible decays of heavy scalar particles acting as mediators to dark matter.
The considered masses range from $50\,\GeV{}$ to $2\,\TeV$, and the upper limit on the product of cross section and branching ratio to invisible particles ($\sigma^\text{VBF}{\,\cdot\,}\mathcal{B}_\text{inv}$) is shown in Figure~\ref{fig:HeavyHiggsLimits}.
The derived limits become stronger for heavier mediator masses due to an accumulation of the signal events at larger values of \mjj, where the background yields are smaller.
The 95\% CL upper limit on $\sigma^\text{VBF}{\,\cdot\,}\mathcal{B}_\text{inv}$ is $1.0\,\text{pb}$ at a mediator mass of $50\,\GeV{}$ and strengthens to  $0.1\,\text{pb}$ for a mediator mass of $2\,\TeV{}$.
 
\begin{figure}
\centering
\includegraphics[width=0.8\linewidth]{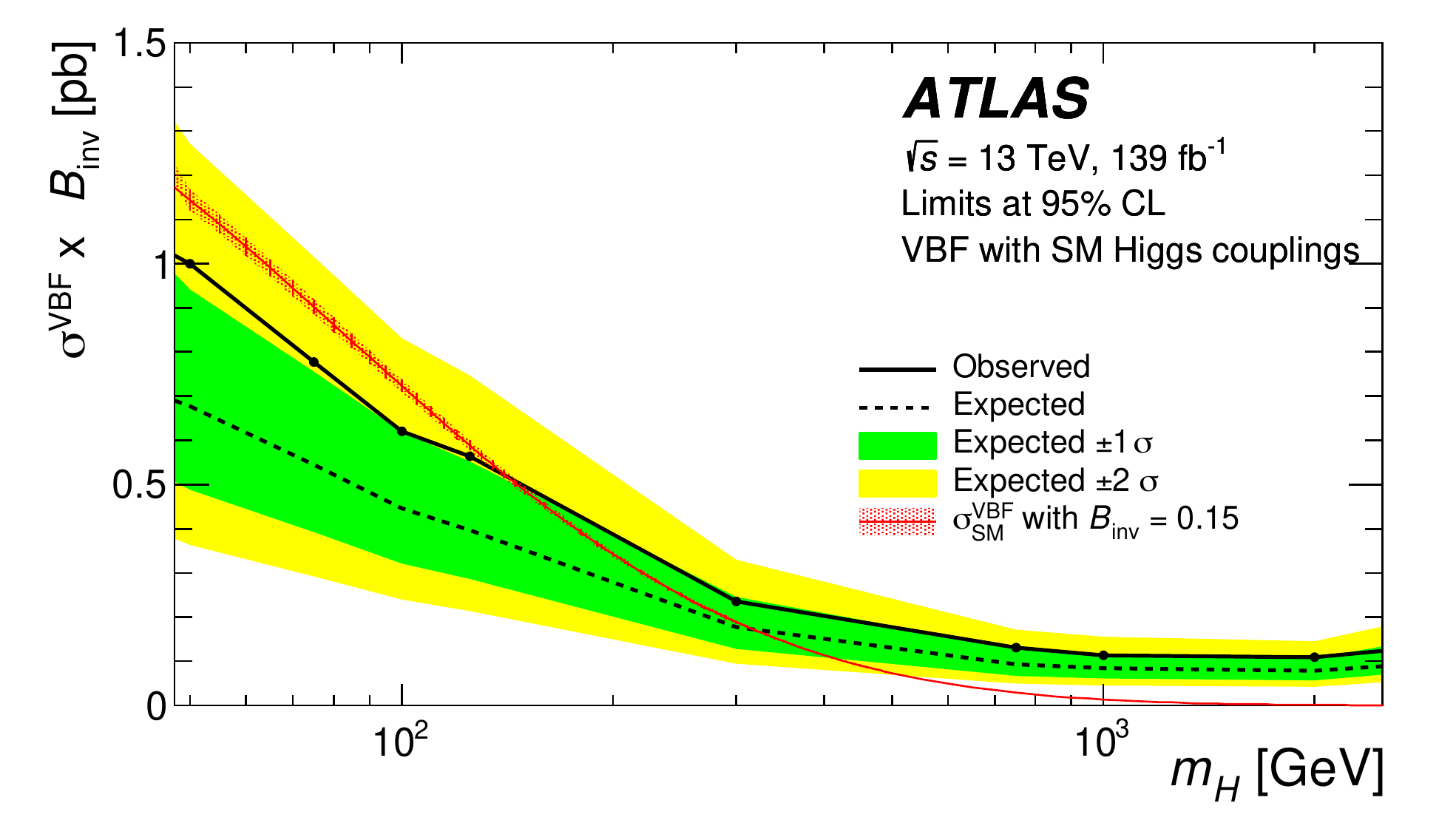}
\caption{Upper limit on cross section times branching ratio to invisible particles for a scalar mediator as a function of its mass. For comparison the VBF cross section at NLO in QCD, i.e.\ without the electroweak corrections, for a particle with SM Higgs boson couplings, multiplied by a $\mathcal{B}_\text{inv}$ value of 15\%, is overlaid.}
\label{fig:HeavyHiggsLimits}
\end{figure}
% End of text imported from the .//sections/results.tex input file

\FloatBarrier
 
\section{Conclusion}
% The next lines are included from the .//sections/conclusion.tex input file
A search for Higgs boson decays into invisible particles is presented using $139\,\ifb$ of proton--proton collision data with a centre-of-mass energy of $\sqrt{s}{\,=\,}13\,\TeV$ collected between 2015 and 2018 by the ATLAS detector at the LHC.
The VBF production mode is targeted by requiring two energetic jets that are not back-to-back in the azimuthal plane and exhibit a large rapidity separation, in addition to large values of the two-jet invariant mass and missing transverse momentum.
Events with three or four jets are also considered if the additional jets are compatible with the hypothesis that they originate from final-state radiation from one of the VBF jets.
The dominant \Zjets and \Wjets backgrounds are evaluated using common control regions, which is made possible due to the use of a dedicated theoretical calculation at next-to-leading order in the phase space that is relevant for this analysis.
Two independent data-driven techniques are used to estimate the multijet background.
 
Assuming the Standard Model cross section for the Higgs boson, an upper limit of $0.145$ is set on the branching fraction $\mathcal{B}_\text{inv}$ into invisible particles at $95\%$ CL, with an expected limit of $0.103$.
This is a significant improvement on the previously published result using $36\,\ifb$ of Run-2 data and originates from various changes.
The increased size of the data sample and simulated samples, together with refinements of the lepton identification and acceptance results in a relative improvement of the expected limit by ${\sim}36\%$.
The extension of the requirement that both leading jets are not pile-up tagged to forward jets with $|\eta|{\,>\,}2.5$ improves the sensitivity by ${\sim}17\%$.
Considering events with three or four jets and events with lower values of missing transverse momentum, along with the finer binning of the signal region reduces the upper limit by ${\sim}19\%$.
The changes to the multijet and \Vjets background estimation lead to further improvements of ${\sim}10\%$ and ${\sim}6\%$, respectively.
 
The result is interpreted using Higgs portal models to exclude regions in the parameter space of $\sigma_{\text{WIMP}\text{-nucleon}}$ and $m_\text{WIMP}$ for various WIMP models.
The obtained results are further interpreted as a search for invisible decays of new scalar particles with masses of up to $2\,\TeV$, resulting in an upper limit on $\sigma^\text{VBF}{\,\cdot\,}\mathcal{B}_\text{inv}$ of $1.0\,\text{pb}$ for a mediator mass of $50\,\GeV$ and of $0.1\,\text{pb}$ for $2\,\TeV$.
% End of text imported from the .//sections/conclusion.tex input file
 
\section*{Acknowledgements}
 
% The next lines are included from the .//acknowledgements/Acknowledgements.tex input file

We thank CERN for the very successful operation of the LHC, as well as the
support staff from our institutions without whom ATLAS could not be
operated efficiently.
 
We acknowledge the support of
ANPCyT, Argentina;
YerPhI, Armenia;
ARC, Australia;
BMWFW and FWF, Austria;
ANAS, Azerbaijan;
SSTC, Belarus;
CNPq and FAPESP, Brazil;
NSERC, NRC and CFI, Canada;
CERN;
ANID, Chile;
CAS, MOST and NSFC, China;
Minciencias, Colombia;
MEYS CR, Czech Republic;
DNRF and DNSRC, Denmark;
IN2P3-CNRS and CEA-DRF/IRFU, France;
SRNSFG, Georgia;
BMBF, HGF and MPG, Germany;
GSRI, Greece;
RGC and Hong Kong SAR, China;
ISF and Benoziyo Center, Israel;
INFN, Italy;
MEXT and JSPS, Japan;
CNRST, Morocco;
NWO, Netherlands;
RCN, Norway;
MEiN, Poland;
FCT, Portugal;
MNE/IFA, Romania;
JINR;
MES of Russia and NRC KI, Russian Federation;
MESTD, Serbia;
MSSR, Slovakia;
ARRS and MIZ\v{S}, Slovenia;
DSI/NRF, South Africa;
MICINN, Spain;
SRC and Wallenberg Foundation, Sweden;
SERI, SNSF and Cantons of Bern and Geneva, Switzerland;
MOST, Taiwan;
TAEK, Turkey;
STFC, United Kingdom;
DOE and NSF, United States of America.
In addition, individual groups and members have received support from
BCKDF, CANARIE, Compute Canada and CRC, Canada;
COST, ERC, ERDF, Horizon 2020 and Marie Sk{\l}odowska-Curie Actions, European Union;
Investissements d'Avenir Labex, Investissements d'Avenir Idex and ANR, France;
DFG and AvH Foundation, Germany;
Herakleitos, Thales and Aristeia programmes co-financed by EU-ESF and the Greek NSRF, Greece;
BSF-NSF and GIF, Israel;
Norwegian Financial Mechanism 2014-2021, Norway;
NCN and NAWA, Poland;
La Caixa Banking Foundation, CERCA Programme Generalitat de Catalunya and PROMETEO and GenT Programmes Generalitat Valenciana, Spain;
G\"{o}ran Gustafssons Stiftelse, Sweden;
The Royal Society and Leverhulme Trust, United Kingdom.
 
The crucial computing support from all WLCG partners is acknowledged gratefully, in particular from CERN, the ATLAS Tier-1 facilities at TRIUMF (Canada), NDGF (Denmark, Norway, Sweden), CC-IN2P3 (France), KIT/GridKA (Germany), INFN-CNAF (Italy), NL-T1 (Netherlands), PIC (Spain), ASGC (Taiwan), RAL (UK) and BNL (USA), the Tier-2 facilities worldwide and large non-WLCG resource providers. Major contributors of computing resources are listed in Ref.~\cite{ATL-SOFT-PUB-2021-003}.
 
% End of text imported from the .//acknowledgements/Acknowledgements.tex input file

We thank Jonas Lindert, Marek Sch\"onherr, and Stefano Augusto Pozzorini for providing the full next-to-leading-order corrections to the \Vjets cross sections in the vector-boson-fusion signature while being associated with ATLAS.

\printbibliography

\clearpage
% ATLAS Collaboration author list
% Reference date of EXOT-2020-11 is 2021-11-03
% Author list last updated on date 21-JUN-22
% Data extracted on 21-Jun-2022 for paper reference EXOT-2020-11
% at 12:01pm
 
\begin{flushleft}
\hypersetup{urlcolor=black}
{\Large The ATLAS Collaboration}

\bigskip

\AtlasOrcid[0000-0002-6665-4934]{G.~Aad}$^\textrm{\scriptsize 100}$,    
\AtlasOrcid[0000-0002-5888-2734]{B.~Abbott}$^\textrm{\scriptsize 126}$,    
\AtlasOrcid[0000-0002-7248-3203]{D.C.~Abbott}$^\textrm{\scriptsize 101}$,    
\AtlasOrcid[0000-0002-2788-3822]{A.~Abed~Abud}$^\textrm{\scriptsize 36}$,    
\AtlasOrcid[0000-0002-1002-1652]{K.~Abeling}$^\textrm{\scriptsize 53}$,    
\AtlasOrcid[0000-0002-2987-4006]{D.K.~Abhayasinghe}$^\textrm{\scriptsize 93}$,    
\AtlasOrcid[0000-0002-8496-9294]{S.H.~Abidi}$^\textrm{\scriptsize 29}$,    
\AtlasOrcid[0000-0002-9987-2292]{A.~Aboulhorma}$^\textrm{\scriptsize 35e}$,    
\AtlasOrcid[0000-0001-5329-6640]{H.~Abramowicz}$^\textrm{\scriptsize 158}$,    
\AtlasOrcid[0000-0002-1599-2896]{H.~Abreu}$^\textrm{\scriptsize 157}$
% ...
\bigskip
\\

% ...
$^{29}$Physics Department, Brookhaven National Laboratory, Upton NY; United States of America.\\
% ...
$^{35}$$^{(a)}$Facult\'e des Sciences Ain Chock, R\'eseau Universitaire de Physique des Hautes Energies - Universit\'e Hassan II, Casablanca;$^{(b)}$Facult\'{e} des Sciences, Universit\'{e} Ibn-Tofail, K\'{e}nitra;$^{(c)}$Facult\'e des Sciences Semlalia, Universit\'e Cadi Ayyad, LPHEA-Marrakech;$^{(d)}$LPMR, Facult\'e des Sciences, Universit\'e Mohamed Premier, Oujda;$^{(e)}$Facult\'e des sciences, Universit\'e Mohammed V, Rabat;$^{(f)}$Mohammed VI Polytechnic University, Ben Guerir; Morocco.\\
$^{36}$CERN, Geneva; Switzerland.\\
% ...
$^{53}$II. Physikalisches Institut, Georg-August-Universit\"{a}t G\"ottingen, G\"ottingen; Germany.\\
% ...
$^{93}$Department of Physics, Royal Holloway University of London, Egham; United Kingdom.\\
% ...
$^{100}$CPPM, Aix-Marseille Universit\'e, CNRS/IN2P3, Marseille; France.\\
$^{101}$Department of Physics, University of Massachusetts, Amherst MA; United States of America.\\
% ...
$^{126}$Homer L. Dodge Department of Physics and Astronomy, University of Oklahoma, Norman OK; United States of America.\\
% ...
$^{157}$Department of Physics, Technion, Israel Institute of Technology, Haifa; Israel.\\
$^{158}$Raymond and Beverly Sackler School of Physics and Astronomy, Tel Aviv University, Tel Aviv; Israel.\\
% ...
$^{a}$ Also at Borough of Manhattan Community College, City University of New York, New York NY; United States of America.\\
$^{b}$ Also at Bruno Kessler Foundation, Trento; Italy.\\
$^{c}$ Also at Center for High Energy Physics, Peking University; China.\\
$^{d}$ Also at Centro Studi e Ricerche Enrico Fermi; Italy.\\
$^{e}$ Also at CERN, Geneva; Switzerland.\\
$^{f}$ Also at D\'epartement de Physique Nucl\'eaire et Corpusculaire, Universit\'e de Gen\`eve, Gen\`eve; Switzerland.\\
% ...

\end{flushleft}

% Created with Glance <Atlas.Glance@cern.ch>

\end{document}